%% file: Second_Submit_TASK_arxiv_ready.tex
\documentclass[11pt]{fizik}

\usepackage{hyperref}
\hypersetup{
colorlinks=true,
urlcolor=blue,
citecolor=blue}
\usepackage[all]{xy,xypic}

\usepackage{eufrak,amscd,bezier,latexsym,mathrsfs,enumerate,multirow}
\usepackage[utf8]{inputenc}
\usepackage[english]{babel}
\usepackage[dvipsnames]{xcolor}
\usepackage{academicons}
\definecolor{orcidlogocol}{HTML}{A6CE39}
\usepackage[pagewise]{lineno}

\usepackage{xspace}
\usepackage{color}
\usepackage{subcaption}
\usepackage{graphicx}
\usepackage{float}
\usepackage{booktabs}
\usepackage{xcolor}
\usepackage{setspace}
\usepackage{physics}
\usepackage{nicefrac}
\usepackage{bbold}
\usepackage{accents}
\usepackage{mathtools}

\def\bt{\begin{equation}}
\def\bea{\begin{eqnarray}}
\def\ee{\end{equation}}
\def\eea{\end{eqnarray}}

\yil{}
\vol{}
\fpage{}
\lpage{}
\doi{}

\title{Evolution of Wormholes under $f(R,T)$ Theory, the Karmarkar Condition and the Casimir Energy}            
\author[Murat Metehan TURKOGLU]{                                                                                     

\textbf{Murat Metehan TURKOGLU$^{1}$\thanks{mmturkoglu@gelisim.edu.tr}}\\ 
$^{1}$Department of Aeronautical Engineering, Istanbul Gelisim University, Avcılar, Istanbul \\Graduate School, Defense Technologies, Aeronautics and Astronautics Engineering Programme, \\ Istanbul Technical University, Istanbul    
\\ [1.8em]

\rec{.. 2022}
\acc{.. 2022}
\finv{.. 2022}
}

\input{fizik.tex}

\begin{document}

\maketitle

\begin{abstract} In this study, both the evolution of wormholes (by examining both the energy conditions and using the TOV equations) and the effects of the Karmarkar condition on the solutions obtained under certain specific cases were examined in the light of the $f(R,T)$ gravity theory, using two $f(R,T)$ functions predicted to describe the accelerated expansion of the universe. In this context, for the first time in the literature, a generalized shape function was obtained using the Karmarkar condition. It was observed that solutions of the type $R-a_{1}^2/R+a_{2}g(T)$ satisfy the energy conditions (with the dominant energy condition being partially satisfied), whereas solutions of the type $R+a_{1}^2R^2+a_{2}g(T)$ require the presence of exotic matter. In both cases, stable, static, and traversable wormhole solutions were obtained. By applying the Karmarkar condition to the $R+a_{1}^2R^2+a_{2}g(T)$ type solutions, which violate the energy conditions, the relationship between wormhole geometry and energy conditions was investigated. The study examined whether the Karmarkar condition eliminates the need for exotic matter, and it was found that the solutions do not remove the necessity of exotic matter. Additionally, it was demonstrated that a specific value of the parameter, ${\beta}$, which determines the radial variation of the shape function, could ensure the stability of the wormhole throat with the aid of Casimir energy. In other words, it is considered possible that the geometric evolution of the wormhole throat could trigger the transition from positive energy (baryonic matter) to negative energy (dark matter, dark energy, or other exotic matter) by inducing Casimir forces.

\end{abstract}

\section{Introduction}
\label{intro}
Wormholes can be thought of as topological bridges that are hypothesized to connect distant regions of the universe or different universes, emerging as solutions to Einstein’s Field Equations. The first scientific analysis of wormholes was conducted in $1916$ by Flamm, who examined the Schwarzschild solution, one of the newly discovered solutions of general relativity at that time \cite{Flamm}. Subsequently, in 1935, modern wormhole solutions were introduced by Einstein and Rosen (ER), who proposed a model in which different regions of space are connected through a "bridge." This model is now known as the Einstein-Rosen bridge \cite{ER}. 
The reason for the popularity of wormholes is that they have been conceptualized as a type of "time machine." In this sense, Morris and Thorne \cite{MorrisThorne} analyzed wormholes as traversable time machines, contributing to their modern popularity.
The most important feature that wormholes must satisfy to be traversable is that the gravitational tidal forces acting on a person traveling in a spacecraft remain within reasonable limits. Additionally, geometrically, wormholes must satisfy the condition known as flaring-out, which imposes a boundary condition on the minimum throat radius. However, this condition is incompatible with the Null Energy Condition (NEC), which states that an observer must measure a non-negative average energy density in spacetime. The violation of the NEC requires the existence of exotic matter, and modified theories of gravity have been invoked to obtain physically consistent wormhole solutions. For example, Armendariz-Picon showed that the NEC is satisfied in the presence of a massless scalar field and that the Einstein equations provide nonsingular, traversable wormhole geometries \cite{Armendariz}. Sushkov obtained exact solutions for a static, spherically symmetric wormhole with phantom energy and demonstrated that the spatial distribution of the phantom energy is confined to the throat region of the wormhole \cite{SushkovPha}, (Additionally, other studies in the literature where similar results have been obtained can be reviewed, \cite{SNLobo}, \cite{Zaslavskii}). Furthermore, exact solutions of spherically symmetric wormholes supported by Generalized Chaplygin Gas (GCG) have also been explored in the literature \cite{Lobo_2006}, (for a more comprehensive review, the following references are recommended for further study, \cite{peter,jamil1,jamil2,cataldo,parsaei,Kuhfittig:2018hts}).

The question of whether exotic matter fields are necessary for the traversability of wormholes is significant. In this context, it is examined whether the solutions of the field equations, obtained by adding the trace of the energy-momentum tensor to the gravitational Lagrangian in $f\left(R,T\right)$ gravity, eliminate the need for exotic matter \cite{zubair0,zubair}. In their study, Deng and Meng, \cite{wang}, examined the structure and topology of wormholes in the presence of dark energy. In the Einstein-Dirac-Maxwell theory, traversable, asymptotically flat wormhole solutions that are free from singularities and exotic matter were obtained using massive fermions \cite{blaz}. Additionally, another method for obtaining wormholes without exotic matter is the application of the thin-shell approximation \cite{israel}, \cite{darmois}. Moreover, it has been demonstrated in Brans-Dicke gravity that exotic matter is not required to support the throat geometry of wormholes \cite{garciaa,papan}.

There are many studies examining wormholes within the framework of modified theories of gravity, $f\left(R,T\right)$ gravity. \cite{camera1,lobo55,capozziello1,rosa1,rosa22,lobo1,garattini1,lobo2,garattini2,lobo3,garattini3,harko1,Anchordoqu1997,Bhawal1992,dotti1,bronnikov1,lobo6,agnese1,nandi1,benedictis,eiroa,mazhar,godani,pavlovic,Chanda:2021dvc,Harko:2011kv,BarrientosO:2014mys,moraes1,mishra1,sahoo1,SALEEM,CHANDRA2021}. Additionally, recent studies have yielded highly significant results. For example, Ganiyeva et al. have presented a wormhole solution that satisfies all energy conditions without requiring fine-tuning \cite{ganiyeva}. Roy et al. have demonstrated that within the framework of the f(R,T) theory, evolving and traversable wormhole solutions that violate the NEC condition but do not require exotic matter are possible \cite{Royet}. Yousaf and Asad used Visser's cut-and-paste technique to construct thin-shell wormholes and revealed that minimally coupled f(R,T) gravity models support various wormhole configurations \cite{yousaf}. Bhatti et al. discuss the junction using distribution formalism and have examined the isotropic perfect fluid as well as the polytropic equation of state that supports exotic matter at the throat of the shell \cite{Bhatti}. Rastgoo and Parsaei have presented wormhole solutions with an asymptotically linear equation of state and have proven that their solutions satisfy all energy conditions \cite{Rastpars}. Chaudhary et al. have obtained stable and traversable wormhole solutions supported by exotic matter with the help of specific shape functions \cite{chaudaryet}. Lu et al. have obtained wormhole solutions that satisfy all energy conditions without the need for exotic matter or any specific type of matter, using a specially chosen f(R,T) function \cite{Lu}. Tangphati et al. have shown that within the framework of f(R,T) theory, the wormhole energy density is always positive, while the radial pressure is negative. They have stated that this result indicates the necessity of exotic matter for the existence of wormholes \cite{Tangphati}. Yashwanth et al. have demonstrated that within the framework of f(R,T) theory, Finslerian wormhole models require exotic matter to maintain the stability of the wormhole throat \cite{Yashwanth}. Mondal and Rahaman have studied wormholes within the framework of f(R,T) gravity theory using the Navarro–Frenk–White (NFW) density profile, the Universal Rotation Curve (URC) dark matter profile, and the mass density profile. They have shown that some special wormhole solutions may exist without requiring the presence of exotic matter \cite{Mondal}. Azmat et al. have examined the Casimir energy densities generated between two parallel plates, a cylinder, and a sphere, and compared them with the  f(R,T) field equations. They have demonstrated that exotic matter is required for the stability of wormholes \cite{Azmat}. Chaudhary et al. have examined various cases of wormholes supported by phantom fluid within the framework of f(R,T) theory. They have shown that the first type of phantom solutions violate the radial null energy condition (NEC), while the tangential NEC is satisfied. On the other hand, they have proven that the second type of wormhole solutions violate the NEC \cite{Chadtwo}.

The most important feature of the $f\left(R,T\right)$ theory is that it explains the observed expansion of the universe without resorting to dark matter and dark energy \cite{Harko:2011kv}. In this study, cosmological models of the type $R-a_{1}^2/R+a_{2}g(T)$ and $R+a_{1}^2R^2+a_{2}g(T)$ are discussed in detail. The main motivation of the paper is to predict the behavior of fundamental physical quantities (energy density, radial and tangential pressure) describing the evolution of a wormhole using f(R,T) functions, which are proposed to characterize the late-time acceleration of the universe, and to determine under which conditions wormholes require the presence of exotic matter. Based on the assumption that the wormhole geometry influences its evolution, the study investigates whether changes in the geometry can trigger transitions from baryonic matter to exotic matter (or vice versa).

We  fix the speed of light and the gravitational constant via $c=G=1$. Throughout this work, a unitless value $r/M$ is assigned by normalizing the radial coordinate r with respect to M (mass of wormhole), so that valid results can be produced for any value of M (see \cite{rosa1}).

\section{Wormhole Geometry}\label{sec:WomGeo}

In this study, we focus on a spherically symmetric and static wormhole model. The general spherically symmetric and static wormhole metric is given in spherical coordinates, $\left(t,r,\theta,\phi\right)$, as follows,
\begin{equation}\label{eq:metric}
ds^2=-e^{\delta\left(r\right)}dt^2+\left[1-\frac{b\left(r\right)}{r}\right]^{-1}dr^2+r^2d\Omega^2,
\end{equation}
where $\delta\left(r\right)$ is the redshift function, $b\left(r\right)$ is the shape function, and $d\Omega^2=d\theta^2+\sin^2\theta d\phi^2$ is the surface-element on the two-sphere.
The traversability of the wormhole depends on the functions $\delta\left(r\right)$ and $b\left(r\right)$ satisfying several conditions. For instance, a spacetime traveler must be able to pass through the wormhole throat at $r=r_0$. For this, there should be no event horizon in the spacetime. This condition is fulfilled if the redshift function remains finite throughout the entire spacetime, in other words, $|\delta\left(r\right)|<\infty$. The second condition is known as the flaring-out condition, which ensures that the wormhole throat has a minimum size.
\begin{equation}\label{eq:flaring_out}
b\left(r_0\right)=r_0,\qquad b'\left(r_0\right)<1.
\end{equation}
The redshift function and shape function that satisfy the conditions above have been chosen as follows \cite{rosa22},
\begin{equation}\label{eq:def_zeta}
\delta\left(r\right)=\delta_0\left(\frac{r_0}{r}\right)^\alpha,
\end{equation}
\begin{equation}\label{eq:def_shape}
b\left(r\right)=b_0\left(\frac{r_0}{r}\right)^\beta,
\end{equation}
where $\delta_0$, $\alpha$, and $\beta$ are constants, and for the solutions to be physically acceptable ( to ensure asymptotic flatness), $\alpha$ and $\beta$ must be chosen as positive.
We can analyze the topology of a traversable wormhole through its embedding in three-dimensional Euclidean space. To comply with spherical symmetry, we can choose $\theta=\pi/2$, and since we assume the wormhole to be static, we can set the time as constant. In this case, the line element is obtained as follows,
\begin{eqnarray}\label{embe}
ds^{2}=\frac{dr^{2}}{1-b/r}+r^{2}d\phi^{2}.
\end{eqnarray}
The surface described by the equation (\ref{embe}) can be represented using the cylindrical metric defined in terms of the three-dimensional cylindrical coordinates $(r,\phi,z)$.
\begin{eqnarray}\label{cyl}
ds^{2}=dz^{2}+dr^{2}+r^{2}d\phi^{2},
\end{eqnarray}
The equation \ref{cyl} can be rewritten as,
\begin{eqnarray}\label{embe2}
ds^{2}=\left[1+\left(\frac{dz}{dr}\right)^{2}\right]dr^{2}
+r^{2}d\phi^{2}.
\end{eqnarray}
where
\begin{eqnarray}
dz=\frac{dz}{dr}dr.
\end{eqnarray}
Using (\ref{embe}) and (\ref{embe2}) we get
\begin{eqnarray}\label{emb3}
\frac{dz}{dr}=\pm 
\left(\frac{r}{b}-1\right)^{-1/2},
\end{eqnarray}
Here, since the wormhole has a minimum radius, as $r \to b_{0}$, we find that $dz/dr \to \infty$. Additionally, to obtain an asymptotically flat solution, as $r \to \infty$, both $b/r \to 0$ and $\delta \to 0$ must hold. 

\begin{figure}[H]
    \centering
    \begin{subfigure}[b]{0.23\textwidth}
        \includegraphics[width=\textwidth,height=60mm]{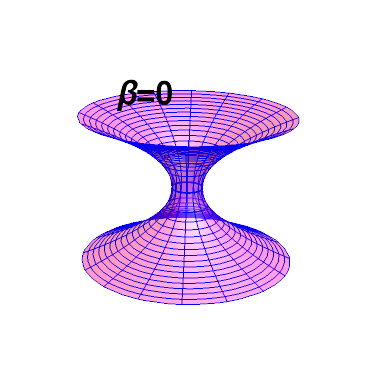}
        \caption*{(a)}
    \end{subfigure}
    \begin{subfigure}[b]{0.23\textwidth}
        \includegraphics[width=\textwidth,height=60mm]{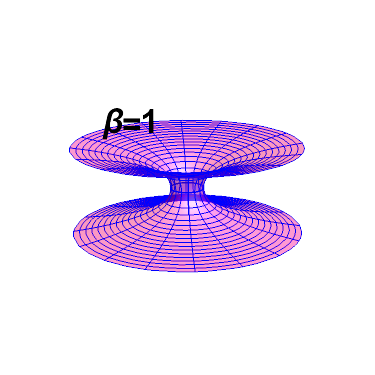}
        \caption*{(b)}
    \end{subfigure}
    \begin{subfigure}[b]{0.23\textwidth}
        \includegraphics[width=\textwidth,height=60mm]{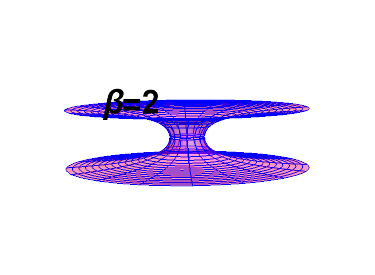}
        \caption*{(c)}
    \end{subfigure}
    \begin{subfigure}[b]{0.23\textwidth}
        \includegraphics[width=\textwidth,height=60mm]{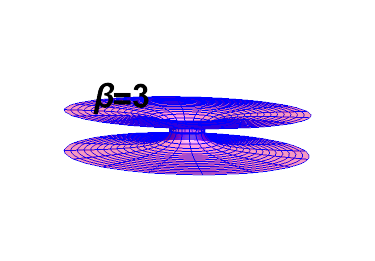}
        \caption*{(d)}
    \end{subfigure}
    \caption{Embedding diagram for different $\beta$ values according to eq.~\ref{eq:embedding}. The throat radius has been set to $r_{0}=1$.}
    \label{fig:embed1}
\end{figure}

\begin{figure}[H]
\includegraphics[width=9cm]{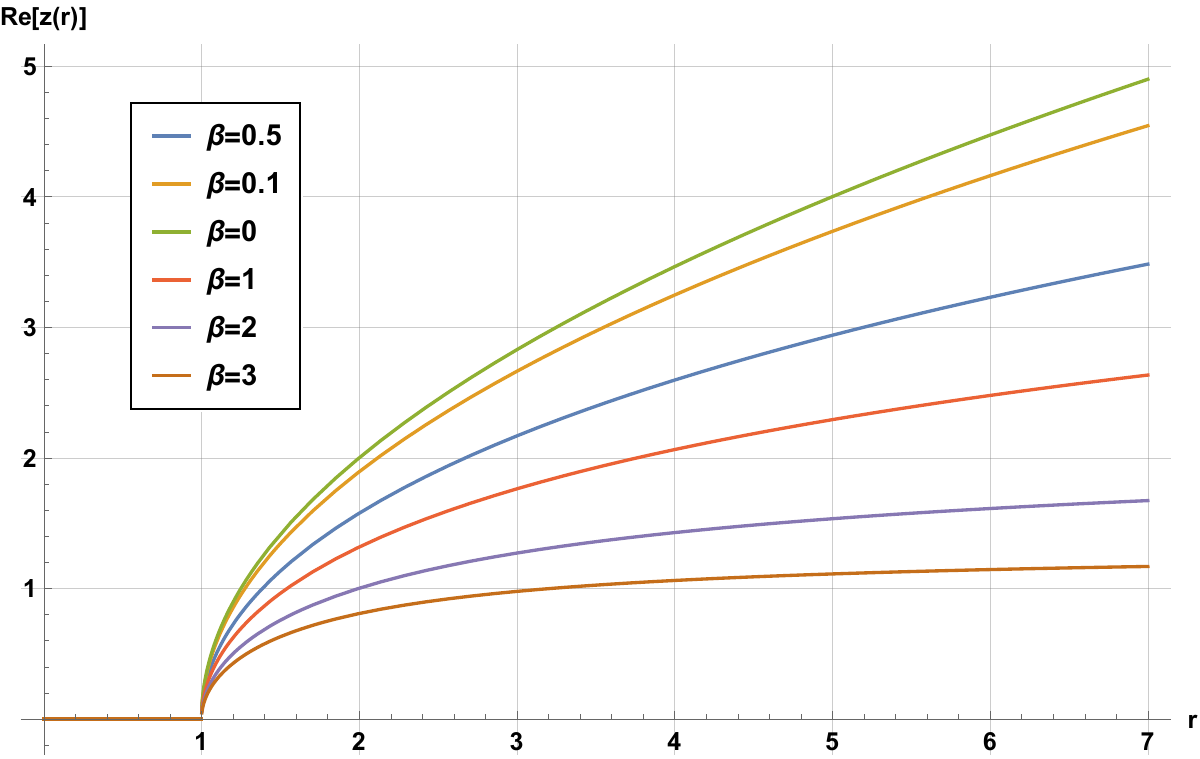}
\centering
\caption{The behavior of the function $z(r)$ for different values of $\beta$ with $b_{0}=1$. }
\label{fig:b02beta}       
\end{figure}

Using the shape function, the embedding function can be find as as
\begin{eqnarray}
\label{eq:embedding}
z(r)=i r_{2}F_{1}\left(\frac{1}{2},\frac{1}{1+\beta},1+\frac{1}{1+\beta},(r/b_{0})^{1+\beta}\right) \nonumber \\ 
-i/b_{0}\sqrt{\pi}\frac{\Gamma(1+\frac{1}{1+\beta})}{\Gamma(\frac{1}{2}+\frac{1}{1+\beta})},
\end{eqnarray}
where ${}_{2}F_{1}(\alpha,\beta,\gamma,t)$ is the Hypergeometric function. The evolution of the wormhole topology depending on different values of $\beta$ for $r_{0}=1$ is shown in figure \ref{fig:embed1}. The variation of the function $z(r)$ for different $\beta$ values is shown in figure \ref{fig:b02beta}.

According to Morris and Thorne \cite{MorrisThorne}, for wormholes to be traversable, the shape function must satisfy the following conditions:\\

\begin{figure}[H]
\includegraphics[width=9cm]{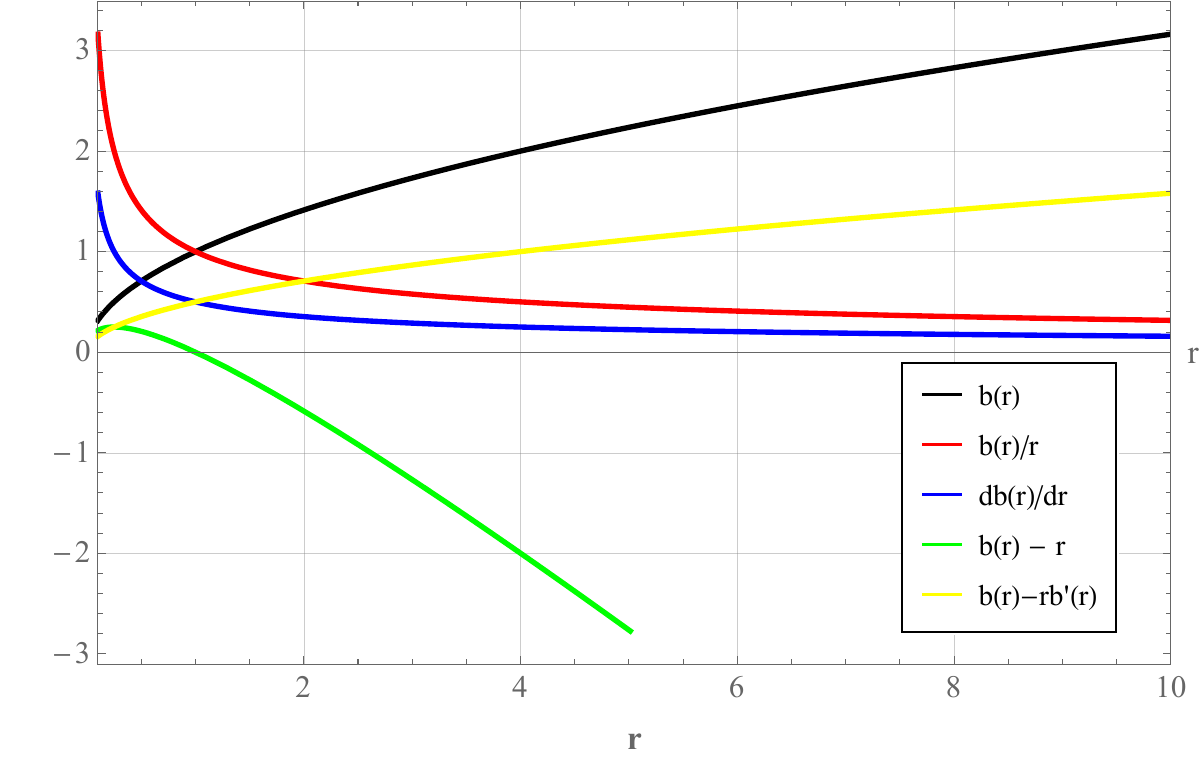}
\centering
\caption{Evaluation of wormhole shape function $b(r)$ for $r_{0}=1$.}
\label{sha}       
\end{figure}

\begin{itemize}
\item At the throat: $r=b(r)$ at $r=r_o$ .
\item The essential condition $\frac{b(r)-rb'(r)}{b(r)^2} > 0$ must be satisfy at $r=r_o$.
\item $b(r)$ must satisfy $b'(r)<1$.
\item asymptotically flat space-time geometry condition, $\frac{b(r)}{r}\to 0$ as $r \to \infty.$
\end{itemize}
The behavior of the conditions that the shape function must satisfy is shown in figure \ref{sha}.

\section{$f\left(R,T\right)$ field equations}\label{sec:field}

The $f\left(R,T\right)$ action is given as \cite{Harko:2011kv}

\begin{equation}\label{frteq1}
S=\int d^4x\sqrt{-g}\left[\frac{1}{16\pi}f(R,T)+L_{m}\right].
\end{equation}
where $g$ is the determinant of the metric $g_{\mu\nu}$ and $L_{m}$ the matter lagrangian.

If we take the variation of the above action with respect to the metric, we obtain the field equations as shown below

\begin{eqnarray}\label{frt2}
f_RR_{\mu\nu}-\frac{1}{2}f(R,T)g_{\mu\nu}+(g_{\mu\nu}\nabla^\mu\nabla_{\nu}-\nabla_\mu\nabla_\nu)f_R= 
8\pi T_{\mu\nu}+f_T(T_{\mu\nu}-L_{m}g_{\mu\nu}).
\end{eqnarray}
where $f_R\equiv\partial f(R,T)/\partial R$, $R_{\mu\nu}$ is the Ricci tensor, $T_{\mu\nu}$ is the energy-momentum tensor and $f_T\equiv\partial f(R,T)/\partial T$.

The covariant derivative of equation \ref{frt2} yields

\begin{eqnarray}\label{frt3}
\nabla^{\mu}T_{\mu\nu}=\frac{f_T}{8\pi+f_T}\times 
 \left[(L_{m}g_{\mu\nu}-T_{\mu\nu})\nabla^{\mu}\ln f_T+\nabla^{\mu}\left(L_{m}-\frac{1}{2}T\right)g_{\mu\nu}\right].
\end{eqnarray}

As seen, the energy-momentum tensor is not conserved in the $f(R,T)$ theory (For discussions on the violation of the conservation law of the energy momentum tensor, see references \cite{harko-2014,Harko:2011kv}).
In this study, the matter forming the wormholes is assumed to be described by an anisotropic energy-momentum tensor

\begin{equation}\label{wh9}
 T_{\mu\nu}=-p_{t}(r)g_{\mu\nu}+(p_{t}(r)+\rho(r))U_{\mu}U_{\nu}+(p_{r}(r)-p_{t}(r))N_{\mu}N_{\nu}
\end{equation}
where $\rho(r)$, $p_{r}(r)$ and $p_{t}(r)$ are the energy density, the radial pressure and the tangential pressure of the fluid, respectively, and the matter Lagrangian is assumed to be defined as $L_{m}=\frac{1}{3}(p_{r}+2p_{t})$.  the four velocity  $U^{\mu}$ and the radial unit vector $N^{\mu}$ satisfy the conditions $U_{\nu}U^{\nu}=1$, $N_{\nu}N^{\nu}=-1$ and $U_{\nu}N^{\nu}=0$.

\subsection{ Field equations for the case $R-a_{1}^2/R+a_{2}T$ }\label{subsec:field1}

For the first case, by substituting the $R-a_{1}^2/R+a_{2}T$ in the equation \ref{frt2} yields,

\begin{eqnarray}\label{case1}
R_{\mu\nu}\left( 1+\frac{a_{1}^{2}}{R^2}\right)+\frac{1}{2}Rg_{\mu\nu}(\frac{a_{1}^{2}}{R^2}-1)=
(8\pi+a_{2}) T_{\mu\nu}+a_{2}g_{\mu\nu}\left(\frac{1}{2}T-L_{m}\right)+
a_{1}^{2}(\nabla_\mu\nabla_\nu-g_{\mu\nu}\nabla^\mu\nabla_{\nu})(R^{-2})
\end{eqnarray}

Also, by substituting the above form for $f(R,T)$ in the equation \ref{frt3} yields (the same result also applies to the second case \ref{subsec:field2}),

\begin{eqnarray}\label{koru1}
\nabla^{\mu}T_{\mu\nu}=\frac{a_{2}}{8\pi+a_{2}}\left( \nabla_\nu\left(L_{m}-\frac{1}{2}T\right)\right).
\end{eqnarray}

As seen in the equation \ref{koru1}, when $a_{2}\to0$, the conservation condition is satisfied.

The values of $r_{0}=1, \delta_{0}=1, \alpha=1, M=1$ were used in solving the field equations determined by the equation \ref{case1}, and sets of equations were obtained for $\beta = 1, \beta = 2, \beta = 3$. From these sets of equations, density and pressure values were obtained separately.

The expressions for density and pressure for $\beta = 1$ have been obtained as follows:
\begin{eqnarray}\label{case1rho}
\rho=-\frac {1} {24 (a_{2} + 4\pi) (a_{2} + 8\pi) r^6\left (-5 r^2 + 2 r + 1 \right)^4}\times \nonumber \\
(96\pi r^2 (1 + 8 r + 4 r^2 - 88 r^3 - 74 r^4 + 440 r^5 + \nonumber \\
     100 r^6 - 1000 r^7 + 625 r^8 + 2 a_{1}^{2} r^{10} + 12a_{1}^{2}r^{11} - \nonumber \\
     578 a_{1}^{2} r^{12} - 1992 a_{1}^{2} r^{13} + 2710a_{1}^{2} r^{14} + \nonumber \\
     8620 a_{1}^{2}r^{15} - 8870a_{1}^{2} r^{16} - 7040a_{1}^{2} r^{17} + 7200 a_{1}^{2} r^{18}) + \nonumber \\ 
  a_{2} (1 + 10 r + 35 r^{2} + 40 r^{3} - 190 r^4 - 1028 r^5 - 130 r^6 + \nonumber \\
     5800 r^7 + 125 r^8 - 13750 r^9 + 9375 r^{10} + 28 a_{1}^{2} r^{12} + \nonumber \\
     24 a_{1}^{2} r^{13} - 8596a_{1}^{2} r^{14} - 26928 a_{1}^{2} r^{15} + \nonumber \\
     40484a_{1}^{2} r^{16} + 117080a_{1}^{2} r^{17} - 125740a_{1}^{2} r^{18} - \nonumber \\
     96160 a_{1}^{2} r^{19} + 100800a_{1}^{2} r^{20}))\nonumber \\
\end{eqnarray}
\begin{eqnarray}\label{case1pr}
p_{r}=\frac {1} {24 (a_{2} + 4\pi) (a_{2} + 8\pi) r^6\left (-5 r^2 + 2 r + 1 \right)^4}\times \nonumber \\
(96\pi r (-1 - 2 r + 5 r^2) (-1 - 5 r + 10 r^2 + 55 r^3 - 70 r^4 -  \nonumber \\
     187 r^5 + 290 r^6 + 25 r^7 - 125 r^8 - 2 a_{1}^{2} r^{11} -  \nonumber \\
     36 a_{1}^{2} r^{12}+ 64 a_{1}^{2} r^{13} + 336 a_{1}^{2} r^{14} - 438a_{1}^{2} r^{15} -  \nonumber \\
     260 a_{1}^{2} r^{16} + 320 a_{1}^{2} r^{17}) +  \nonumber \\
  a_{2} (1 + 34 r + 179 r^2 - 272 r^3 - 2686 r^4 + 1324 r^5 +  \nonumber \\
     16094 r^6 - 11144 r^7 - 39235 r^8 + 46850 r^9 + 3375 r^{10} -  \nonumber \\
     15000 r^{11} + 28a_{1}^{2} r^{12} + 696a_{1}^{2} r^{13} + 5228a_{1}^{2} r^{14} +  \nonumber \\
     5424 a_{1}^{2} r^{15} - 22492a_{1}^{2} r^{16} - 22216a_{1}^{2} r^{17} +  \nonumber \\
     39380a_{1}^{2} r^{18} + 26240a_{1}^{2} r^{19} - 33600a_{1}^{2} r^{20})) \nonumber \\
\end{eqnarray}
\begin{eqnarray}\label{case1pt}
p_{t}=\frac {1} {24 (a_{2} + 4\pi) (a_{2} + 8\pi) r^6\left (-5 r^2 + 2 r + 1 \right)^4}\times \nonumber \\
(24\pi(-1 - 12 r - 31 r^2 + 114 r^3 + 462 r^4 - 576 r^5 - \nonumber \\
     2406 r^6 + 2652 r^7 + 4755 r^8 - 7300 r^9 + 1125 r^{10} + \nonumber \\
     1250 r^{11} + 4 a1^2 r^{12} + 112a_{1}^{2} r^{13} - 1804 a_{1}^{2} r^{14} - \nonumber \\
     7640  a_{1}^{2} r^{15} + 8124  a_{1}^{2} r^{16} + 32448 a_{1}^{2} r^{17} - \nonumber \\
     29940 a_{1}^{2} r^{18} - 26680 a_{1}^{2} r^{19} + 25600a_{1}^{2} r^{20}) + \nonumber \\
  a_{2} (-5 - 62 r - 175 r^2 + 532 r^3 + 2486 r^4 - 2372 r^5 - \nonumber \\
     12790 r^6 + 11152 r^7 + 26255 r^8 - 33550 r^9 + 1125 r^{10} + \nonumber \\
     7500 r^{11} + 4  a_{1}^{2} r^{12} + 408 a_{1}^{2} r^{13} - 5548 a_{1}^{2} r^{14} - \nonumber \\
     24960 a_{1}^{2} r^{15} + 24188 a_{1}^{2} r^{16} + 104888  a_{1}^{2}r^{17} - \nonumber \\
     92500  a_{1}^{2} r^{18} - 87280 a_{1}^{2} r^{19} + 81600 a_{1}^{2} r^{20})) \nonumber \\
\end{eqnarray}

The expressions for density and pressure for $\beta = 2$ and  $\beta = 3$  have been obtained as follows:
\begin{eqnarray}\label{case1rho2}
\rho=-\frac {1} {24 (a_{2} + 4\pi) (a_{2} + 8\pi) r^7\left (-1-3r+8r^2+r^3 \right)^4}\times \nonumber \\
(192\pi r^2 (1 + 12 r + 22 r^2 - 184 r^3 - 435 r^4 + 1428 r^5 + \nonumber \\
     1882 r^6 - 6012 r^7 + 1750 r^8 + 1756 r^9 + 372 r^{10} + \nonumber \\
     32 r^{11} + (1 + a_{1}^{2}) r^{12} + 9 a_{1}^{2} r^{13} - 371 a_{1}^{2} r^{14} - \nonumber \\
     2022 a_{1}^{2} r^{15} + 1606 a_{1}^{2} r^{16} + 12103a_{1}^{2}r^{17} - \nonumber \\
     9089 a_{1}^{2} r^{18} - 4431a_{1}^{2} r^{19} - 13076a_{1}^{2} r^{20} + \nonumber \\
     12855a_{1}^{2} r^{21} + 2816a_{1}^{2} r^{22} + 144 a_{1}^{2} r^{23}) + \nonumber \\
  a_{2} (1 + 15 r + 90 r^2 + 265 r^3 - 295 r^4 - 5787 r^5 - 7570 r^6 + \nonumber \\
     45765 r^7 + 42510 r^8 - 187260 r^9 + 67652 r^{10} + 55590 r^{11} + \nonumber \\
     10245 r^{12} + 655 r^{13} + 28 a_{1}^{2} r^{14} + (-1 + 84a_{1}^{2}) r^{15} - \nonumber \\
     11276a_{1}^{2} r^{16} - 55320a_{1}^{2} r^{17} + 51400a_{1}^{2} r^{18} + \nonumber \\
     330460a_{1}^{2} r^{19} - 254444a_{1}^{2} r^{20} - 129996a_{1}^{2} r^{21} - \nonumber \\
     359120a_{1}^{2} r^{22} + 361668a_{1}^{2} r^{23} + 78944 a_{1}^{2}r^{24} + 4032a_{1}^{2} r^{25}))\nonumber \\
    \end{eqnarray}
\begin{eqnarray}\label{case1pr2}
p_{r}=\frac {1} {24 (a_{2} + 4\pi) (a_{2} + 8\pi) r^6\left (-5 r^2 + 2 r + 1 \right)^4}\times \nonumber \\
(96\pi r (-1 - 3 r + 8 r^2 + r^3) (-1 - 8 r + 6 r^2 + 124 r^3 - \nonumber \\
     69 r^4 - 636 r^5 + 842 r^6 - 224 r^7 + 438 r^8 - 388 r^9 - \nonumber \\
     184 r^{10} - 24 r^{11} - r^{12} - 2 a_{1}^{2} r^{13} - 44a_{1}^{2} r^{14} + \nonumber \\
     46a_{1}^{2} r^{15} + 592a_{1}^{2} r^{16} - 736 a_{1}^{2} r^{17} - 128 a_{1}^{2}r^{18} - \nonumber \\
     482 a_{1}^{2}r^{19} + 620 a_{1}^{2} r^{20} + 64a_{1}^{2} r^{21}) - \nonumber \\
  a_{2} (-1 - 39 r - 306 r^2 + 71 r^3 + 6319 r^4 + 3267 r^5 - \nonumber \\
     57494 r^6 + 7467 r^7 + 226842 r^8 - 253332 r^9 + 61372 r^{10} - \nonumber \\
     82374 r^{11} + 57771 r^{12} +  43769 r^{13} + (9000 - 28 a_{1}^{2}) r^{14} +\nonumber \\
 (769 - 852 a_{1}^{2}) r^{15} + (24 - 8212a_{1}^{2}) r^{16} - 15720 a_{1}^{2}r^{17} + \nonumber \\
     42872 a_{1}^{2} r^{18} + 79652a_{1}^{2} r^{19} - 75508 a_{1}^{2} r^{20} - \nonumber \\
     60276 a_{1}^{2} r^{21} - 126736a_{1}^{2} r^{22} + 152508 a_{1}^{2} r^{23} + \nonumber \\
     29056 a_{1}^{2} r^{24} + 1344 a_{1}^{2} r^{25})) \nonumber \\
\end{eqnarray}
\begin{eqnarray}\label{case1pt2}
p_{t}=\frac {1} {24 (a_{2} + 4\pi) (a_{2} + 8\pi) r^6\left (-5 r^2 + 2 r + 1 \right)^4}\times \nonumber \\
(24\pi (-1 - 17 r - 76 r^2 + 147 r^3 + 1501 r^4 - 311 r^5 - \nonumber \\
     11772 r^6 + 4367 r^7 + 40160 r^8 - 41840 r^9 - 900 r^{10} - \nonumber \\
     1630 r^{11} + 7327 r^{12} + 4071 r^{13} + (782 + 4 a_{1}^{2}) r^{14} + \nonumber \\
     5 (13 + 28 a_{1}^{2}) r^{15} + (2 - 2220 a_{1}^{2}) r^{16} - 15304 a_{1}^{2} r^{17} + \nonumber \\
     7848 a_{1}^{2} r^{18} + 91012 a_{1}^{2} r^{19} - 63900a_{1}^{2} r^{20} - \nonumber \\
     27652 a_{1}^{2} r^{21} - 99800 a_{1}^{2} r^{22} + 92540 a_{1}^{2} r^{23} + \nonumber \\
     20168a_{1}^{2} r^{24} + 1024 a1^2 r^{25}) + \nonumber \\
  a_{2} (-5 - 87 r - 414 r^2 + 571 r^3 + 7655 r^4 + 1179 r^5 - \nonumber \\
     57322 r^6 + 3423 r^7 + 193134 r^8 - 149724 r^9 - 21748 r^{10} - \nonumber \\
     38478 r^{11} + 36351 r^{12} + 23545 r^{13} + \nonumber \\
     4 (1161 + a_{1}^{2}) r^{14} + (389 + 492 a_{1}^{2}) r^{15} +\nonumber \\
 (12 - 6788 a_{1}^{2}) r^{16} - 50088 a_{1}^{2} r^{17} + 21400 a_{1}^{2} r^{18} + \nonumber \\
     295588 a_{1}^{2} r^{19} - 201572a_{1}^{2} r^{20} - 83220a_{1}^{2} r^{21} - \nonumber \\ 
     330272 a_{1}^{2} r^{22} + 299868 a_{1}^{2} r^{23} + 64784 a_{1}^{2} r^{24} + 3264 a_{1}^{2} r^{25})) \nonumber \\
\end{eqnarray}
\begin{eqnarray}\label{case1rho3}
\rho=-\frac {1} {24 (a_{2} + 4\pi) (a_{2} + 8\pi) r^8\left (-1-4r+12r^2+r^4 \right)^4}\times \nonumber \\
((96\pi r^2 (3 + 48 r + 144 r^2 - 960 r^3 - 3564 r^4 + 11376 r^5 + \nonumber \\
      20592 r^6 - 80256 r^7 + 63954 r^8 - 20592 r^9 + 20592 r^{10} - \nonumber \\
      1728 r^{11} + 2580 r^{12} - 48 r^{13} + 2 (72 +  a_{1}^{2}) r^{14} + \nonumber \\
      24  a_{1}^{2} r^{15} + (3 - 924  a_{1}^{2}) r^{16} - 7008  a_{1}^{2} r^{17} + \nonumber \\
      4666  a_{1}^{2} r^{18} + 61392 a_{1}^{2} r^{19} - 75848  a_{1}^{2} r^{20} + \nonumber \\
      11040  a_{1}^{2}r^{21} - 14394  a_{1}^{2} r^{22} - 69096  a_{1}^{2} r^{23} + \nonumber \\
      88580  a_{1}^{2} r^{24} - 3648 a_{1}^{2} r^{25} + 9790  a_{1}^{2} r^{26} + 288  a_{1}^{2} r^{28}) + \nonumber \\
   a_{2} (1 + 20 r + 160 r^2 + 640 r^3 - 165 r^4 - 16336 r^5 - \nonumber \\
      35040 r^6 + 183040 r^7 + 244970 r^8 - 1209480 r^9 + \nonumber \\
      995808 r^{10} - 275840 r^{11} + 306710 r^{12} - 17360 r^{13} + \nonumber \\
      34400 r^{14} + (1445 + 28  a_{1}^{2}) r^{16} + 4 (5 + 36  a_{1}^{2}) r^{17} - \nonumber \\
      14304  a_{1}^{2} r^{18} - 96192  a_{1}^{2} r^{19} + (-1 + 79724  a_{1}^{2}) r^{20} + \nonumber \\
      832320  a_{1}^{2} r^{21} - 1049152  a_{1}^{2} r^{22} + 149760 a_{1}^{2} r^{23} - \nonumber \\
      214764 a_{1}^{2} r^{24} - 946992  a_{1}^{2} r^{25} + 1239136 a_{1}^{2} r^{26} - \nonumber \\
      48192  a_{1}^{2} r^{27} + 137060  a_{1}^{2}2 r^{28} + 96  a_{1}^{2} r^{29} + 4032  a_{1}^{2} r^{30})) \nonumber \\
\end{eqnarray}
\begin{eqnarray}\label{case1pr3}
p_{r}=\frac {1} {24 (a_{2} + 4\pi) (a_{2} + 8\pi) r^8\left (-1-4r+12r^2+r^4 \right)^4}\times \nonumber \\
(96\pi r (-1 - 4 r + 12 r^2 + r^4) (-1 - 11 r + 236 r^3 - 76 r^4 - \nonumber \\
     1839 r^5 + 3420 r^6 - 2216 r^7 + 570 r^8 + 1263 r^9 - \nonumber \\
     1656 r^{10} + 252 r^{11} - 428 r^{12} + 11 r^{13} - 36 r^{14} - \nonumber \\
     2 a_{1}^{2} r^{15} - (1 + 52 a_{1}^{2}) r^{16} + 20 a_{1}^{2} r^{17} + \nonumber \\
     992 a_{1}^{2} r^{18} - 1484 a_{1}^{2} r^{19} + 72 a_{1}^{2} r^{20} - 116 a_{1}^{2} r^{21} - \nonumber \\
     784 a_{1}^{2} r^{22} + 1150 a_{1}^{2} r^{23} - 20 a_{1}^{2} r^{24} + 64 a_{1}^{2} r^{25}) + \nonumber \\
  a_{2} (1 + 44 r + 448 r^2 + 256 r^3 - 12453 r^4 - 14152 r^5 + \nonumber \\
     169632 r^6 - 17408 r^7 - 1048342 r^8 + 1898856 r^9 - \nonumber \\
     1306464 r^{10} + 383104 r^{11} + 275990 r^{12} - 453056 r^{13} + \nonumber \\
     116192 r^{14} - 162048 r^{15} + 7 (1523 + 4 a_{1}^{2}) r^{16} + \nonumber \\
     4 (-5149 + 252 a_{1}^{2}) r^{17} + 288 (1 + 41 a_{1}^{2}) r^{18} + \nonumber \\
     192 (-6 + 163 a_{1}^{2}) r^{19} - (1 + 86164 a_{1}^{2}) r^{20} - \nonumber \\
     24 (1 + 8996 a_{1}^{2}) r^{21} + 340160 a_{1}^{2} r^{22} - 40704 a_{1}^{2} r^{23} + \nonumber \\
     109332 a_{1}^{2} r^{24} + 377328 a_{1}^{2} r^{25} - 557984 a_{1}^{2} r^{26} + \nonumber \\
     8640 a_{1}^{2} r^{27} - 51868 a_{1}^{2} r^{28} - 384a_{1}^{2} r^{29} -1344 a_{1}^{2} r^{30}))\nonumber \\
\end{eqnarray}
\begin{eqnarray}\label{case1pt3}
p_{t}=\frac {1} {24 (a_{2} + 4\pi) (a_{2} + 8\pi) r^8\left (-1-4r+12r^2+r^4 \right)^4}\times \nonumber \\
(24\pi (-1 - 22 r - 136 r^2 + 160 r^3 + 3493 r^4 + 794 r^5 \nonumber \\
    - 39040 r^6 + 15680 r^7 + 192278 r^8 - 333644 r^9 + 219312 r^{10} - \nonumber \\
     108544 r^{11} + 25322 r^{12} + 26020 r^{13} + 64 r^{14} + \nonumber \\
     12736 r^{15} + (91 + 4 a_{1}^{2}) r^{16} + 6 (283 + 28a_{1}^{2}) r^{17} - \nonumber \\
     8 (-3 + 332 a_{1}^{2}) r^{18} - 32 (-3 + 830 a_{1}^{2}) r^{19} + (1 + 8756 a_{1}^{2}) r^{20} +\nonumber \\
    (2 + 237304 a_{1}^{2}) r^{21} - 280192 a_{1}^{2} r^{22} + 42752 a_{1}^{2} r^{23} - \nonumber \\
     42548 a_{1}^{2} r^{24} - 268904 a_{1}^{2} r^{25} + 327904 a_{1}^{2} r^{26} - \nonumber \\
     14912 a_{1}^{2} r^{27} + 35324 a_{1}^{2} r^{28} - 56 a_{1}^{2} r^{29} + 1024 a_{1}^{2} r^{30}) + \nonumber \\
  a_{-2} (-5 - 112 r - 728 r^2 + 448 r^3 + 17337 r^4 + 11468 r^5 - \nonumber \\
     183744 r^6 + 4096 r^7 + 904430 r^8 - 1285200 r^9 + 776784 r^{10} - \nonumber \\
     432896 r^{11} - 35566 r^{12} + 180232 r^{13} - 27136 r^{14} + \nonumber \\
     77568 r^{15} + (-1465 + 4 a_{1}^{2}) r^{16} + \nonumber \\
     32 (319 + 18 a_{1}^{2}) r^{17} + (72 - 8064 a_{1}^{2}) r^{18} - \nonumber \\
     192 (-3 + 455 a_{1}^{2}) r^{19} + (5 + 20276 a_{1}^{2}) r^{20} + \nonumber \\
     12 (1 + 65228 a_{1}^{2}) r^{21} - 909952 a_{1}^{2} r^{22} + 141312 a_{1}^{2} r^{23} - \nonumber \\
     124596 a_{1}^{2}r^{24} - 902112 a_{1}^{2} r^{25} + 1080640 a_{1}^{2} r^{26} - \nonumber \\
     50112 a_{1}^{2} r^{27} + 114044 a_{1}^{2} r^{28} - 240a_{1}^{2} r^{29} + 3264 a_{1}^{2} r^{30})) \nonumber \\
\end{eqnarray}

\subsection{ Field equations for the case $R+a_{1}^2R^{2}+a_{2}T$}\label{subsec:field2}

For the second case, by substituting the $R+a_{1}^2R^{2}+a_{2}T$  in the equation \ref{frt2} yields,

\begin{eqnarray}\label{case2}
R_{\mu\nu}\left( 1+2a_{1}^{2}R^2\right)-\frac{1}{2}Rg_{\mu\nu}(a_{1}^{2}R+1)=\nonumber \\
(8\pi+a_{2}) T_{\mu\nu}+a_{2}g_{\mu\nu}\left(\frac{1}{2}T-L_{m}\right)+\nonumber \\
2a_{1}^{2}(\nabla_\mu\nabla_\nu-g_{\mu\nu}\nabla^\mu\nabla_{\nu})R
\end{eqnarray}

The values of $r_{0}=1, \delta_{0}=1, \alpha=1, M=1$ were used in solving the field equations determined by the equation \ref{case2}, and sets of equations were obtained for $\beta = 1, \beta = 2, \beta = 3$. From these sets of equations, density and pressure values were obtained separately.

The expressions for density and pressure for $\beta = 1$ have been obtained as follows:

\begin{eqnarray}\label{case2rho1}
\rho=-\frac {1} {48 (a_{2} + 4\pi) (a_{2} + 8\pi) r^{12}}\times \nonumber \\
(-2 r^6 (a_{2} + 2 a_{2} r + 15 a_{2} r^2 + 96\pi r^2) +  \nonumber \\
 a_{1}^2 (24\pi (-1 - 4 r + 286 r^2 + 404 r^3 - 865 r^4 - 320 r^5 + 480 r^6) +  \nonumber \\
 a_{2} (-5 - 56 r + 938 r^2 + 1576 r^3 - 2945 r^4 - 1240 r^5 + 1680 r^6)))\nonumber \\
\end{eqnarray}
\begin{eqnarray}\label{case2pr1}
p_{r}=-\frac {1} {48 (a_{2} + 4\pi) (a_{2} + 8\pi) r^{12}}\times \nonumber \\
(-2 r^6 (96\pi r (-1 + r + r^2) + a_{2} (-1 - 26 r + 33 r^2 + 24 r^3)) + \nonumber \\
 a_{1}^2 (24 \pi (1 - 12 r + 58 r^2 + 180 r^3 - 327 r^4 - 200 r^5 + 320 r^6) + \nonumber \\
a_{2} (5 - 40 r + 1126 r^2 + 1928 r^3 - 4207 r^4 - 1880 r^5 + 3120 r^6)))\nonumber \\
\end{eqnarray}
\begin{eqnarray}\label{case2pt1}
p_{t}=\frac {1} {48 (a_{2} + 4\pi) (a_{2} + 8\pi) r^{12}}\times \nonumber \\
(2 r^6 (24 \pi (-1 - 4 r + 5 r^2 + 2 r^3) +  \nonumber \\
a_{2} (-5 - 22 r + 21 r^2 + 12 r^3)) - \nonumber \\
a_{1}^2 (24 \pi (1 + 32 r + 378 r^2 + 332 r^3 -  \nonumber \\
1111 r^4 - 300 r^5 + 640 r^6) + \nonumber \\
a_{2} (7 + 160 r + 1490 r^2 + 1144 r^3 -  \nonumber \\
4421 r^4 - 1120 r^5 + 2640 r^6))) \nonumber \\
\end{eqnarray}

The expressions for density and pressure for $\beta = 2$ have been obtained as follows:
\begin{eqnarray}\label{case2rho2}
\rho=\frac {1} {48 (a_{2} + 4\pi) (a_{2} + 8\pi) r^{14}}\times \nonumber \\
(-384 \pi r^9 + 2 a_{2} r^7 (-1 - 3 r - 32 r^2 + r^3) +  \nonumber \\
  a_{1}^2 (24 \pi (-1 - 6 r + 411 r^2 + 938 r^3 - 1690 r^4 - 480 r^5 -  \nonumber \\
  721 r^6 + 1280 r^7 + 96 r^8) +  \nonumber \\
  a_{2} (-5 - 72 r + 1317 r^2 + 3526 r^3 - 5744 r^4 - 1572 r^5 -  \nonumber \\
 2765 r^6 + 4456 r^7 + 336 r^8))) \nonumber \\
\end{eqnarray}
\begin{eqnarray}\label{case2pr2}
p_{r}=\frac {1} {48 (a_{2} + 4\pi) (a_{2} + 8\pi) r^{14}}\times \nonumber \\
(-2 r^7 (96 \pi r (-1 + r + r^3) +  \nonumber \\
 a_{2} (-1 - 27 r + 40 r^2 + r^3 + 24 r^4)) +  \nonumber \\
a_{1}^2 (24 \pi (1 - 14 r + 49 r^2 + 310 r^3 - 490 r^4  \nonumber \\
 - 104 r^5 - 383 r^6 + 632 r^7 + 64 r^8) +  \nonumber \\
 a_{2} (5 - 48 r + 1443 r^2 + 3962 r^3 - 7336 r^4 - 1932 r^5 -  \nonumber \\
 3859 r^6 + 7016 r^7 + 624 r^8)))  \nonumber \\
\end{eqnarray}
\begin{eqnarray}\label{case2pt2}
p_{t}=\frac {1} {48 (a_{2} + 4\pi) (a_{2} + 8\pi) r^{14}}\times \nonumber \\
(2 r^7 (24 \pi (-1 - 5 r + 6 r^2 + r^3 + 2 r^4) + \nonumber \\
  a_{2} (-5 - 27 r + 20 r^2 + 5 r^3 + 12 r^4)) - \nonumber \\
 a_{1}^2 (24 \pi (1 + 38 r + 557 r^2 + 850 r^3 - 2106 r^4 - 640 r^5 - \nonumber \\
  671 r^6 + 1620 r^7 + 128 r^8) + \nonumber \\
  a_{2} (7 + 192 r + 2193 r^2 + 2998 r^3 - 8240 r^4 - 2532 r^5 - \nonumber \\
  2465 r^6 + 6496 r^7 + 528 r^8)))\nonumber \\
\end{eqnarray}

The expressions for density and pressure for $\beta = 3$ have been obtained as follows:
\begin{eqnarray}\label{case2rho3}
\rho=\frac {1} {48 (a_{2} + 4\pi) (a_{2} + 8\pi) r^{16}}\times \nonumber \\
(-384 \pi r^9 + 2a_{2} r^7 (-1 - 3 r - 32 r^2 + r^3) +  \nonumber \\
(2 r^8 (-288 \pi r^2 +a_{2} (-1 - 4 r - 48 r^2 + r^4)) + \nonumber \\
a_{1}^2 (24 \pi (-1 - 8 r + 560 r^2 + 1792 r^3 - 3886 r^4 + 8 r^5 - \nonumber \\
608 r^6 - 1344 r^7 + 2879 r^8 + 96 r^{10}) + \nonumber \\
 a_{2} (-5 - 88 r + 1768 r^2 + 6704 r^3 - 13382 r^4 + 112 r^5 - \nonumber \\
 1960 r^6 - 5136 r^7 + 10075 r^8 - 24 r^9 + 336 r^{10})))\nonumber \\
\end{eqnarray}
\begin{eqnarray}\label{case2pr3}
p_{r}=\frac {1} {48 (a_{2} + 4\pi) (a_{2} + 8\pi) r^{16}}\times \nonumber \\
(-2 r^8 (96 \pi r (-1 + r + r^4) + \nonumber \\
  a_{2}(-1 - 28 r + 48 r^2 + r^4 + 24 r^5)) + \nonumber \\
 a_{1}^2 (24 \pi (1 - 16 r + 32 r^2 + 512 r^3 - 914 r^4 + 24 r^5 - \nonumber \\
 80 r^6 - 640 r^7 + 1153 r^8 - 8 r^9 + 64 r^{10}) + \nonumber \\
 a_{2} (5 - 56 r + 1784 r^2 + 7120 r^3 - 15418 r^4 + 80 r^5 - \nonumber \\
 2168 r^6 - 6768 r^7 + 14117 r^8 - 24 r^9 + 624 r^{10})))\nonumber \\
\end{eqnarray}
\begin{eqnarray}\label{case2pt3}
p_{t}=\frac {1} {48 (a_{2} + 4\pi) (a_{2} + 8\pi) r^{16}}\times \nonumber \\
(2 r^8 (24 \pi (-1 - 6 r + 8 r^2 + r^4 + 2 r^5) +  \nonumber \\
 a_{2} (-5 - 32 r + 24 r^2 + 5 r^4 + 12 r^5)) -  \nonumber \\
a_{1}^2 (24 \pi (1 + 44 r + 768 r^2 + 1616 r^3 - 4562 r^4 - 64 r^5 -  \nonumber \\
816 r^6 - 1232 r^7 + 3457 r^8 + 20 r^9 + 128 r^{10}) +  \nonumber \\
a_{2} (7 + 224 r + 3016 r^2 + 5648 r^3 - 17438 r^4 - 320 r^5 -  \nonumber \\
3208 r^6 - 4464 r^7 + 13543 r^8 + 96 r^9 + 528 r^{10}))) \nonumber \\
\end{eqnarray}

\section{Energy and Stability Conditions}\label{sec:ESC}

As discussed in detail in section \ref{intro}, in most cases, exotic matter that violates energy conditions is needed for wormholes to be stable and traversable. However, the existence of exotic matter is undesirable due to its physical properties. Therefore, whether traversable wormholes violate energy conditions is of significant theoretical importance. Energy conditions are derived from the Raychaudhuri equation, which is not directly related to any specific theory of gravity. However, they have been studied within the framework of various gravitational theories, \cite{bertolami2009,capozziello2018,santos2017,zubair2015,zubair2022,Mandal2022}.

The wormholes with anisotropic matter distribution, the energy conditions are given as follows \cite{Curiel};
\begin{enumerate}
\item {Null Energy Condition: $\rho+p_r \geq 0 ,  \rho+p_t \geq 0$},
\item{Weak Energy Condition: $\rho \geq 0, \rho+p_r \geq 0, \rho+p_t \geq 0$},
\item{Dominant Energy Condition: $\rho \geq 0, \rho \pm p_r \geq 0, \\ \rho \pm p_t \geq 0$},
\item{Strong Energy Condition: $\rho+p_r \geq 0, \rho+p_t \geq 0, \\ \rho+p_r+2 p_t \geq 0$}.
\end{enumerate}

In this study, we also considered the stability conditions of the wormhole solutions for the model we chose using the generalized Tolman-Oppenheimer-Volkoff (TOV) equation:

\begin{equation}\label{stability}
\frac{dp_r}{dr}+\frac{\delta'(\rho+p_r)}{2}+\frac{2}{r}(p_r-p_t)=0,
\end{equation}

This equation allows us to examine the stability of the wormhole with the help of gravitational force, $F_{gf}$, anisotropic force  $F_{af}$, and hydrostatic force, $F_{hf}$. The gravitational force is a result of the wormhole's gravitational attraction, while the anisotropic force stems from the system's anisotropy, and finally, the hydrostatic force arises as a result of the hydrostatic fluid, which we assume forms the wormhole. For the wormhole to be in equilibrium, the condition  $F_{gf}+F_{af}+F_{hf}=0$  must be satisfied. In this study, the stability conditions for the cases $\beta = 1$, $\beta = 2$, and $\beta = 3$ were examined, and the other parameter values were chosen as $r_{0}=1$, $\delta_{0}=1$, $\alpha=1$, and $M=1$. In \ref{subsec:sc1} and \ref{subsec:sc2}, where the energy and stability conditions are examined, the parameters $a_{1}$ and $a_{2}$ are chosen as $9\pi$ and $-9\pi$, respectively.

\subsection{Energy and Stability Conditions for the case $R-a_{1}^2/R+a_{2}T$ }\label{subsec:sc1}

\begin{figure}[H]
\includegraphics[width=8cm]{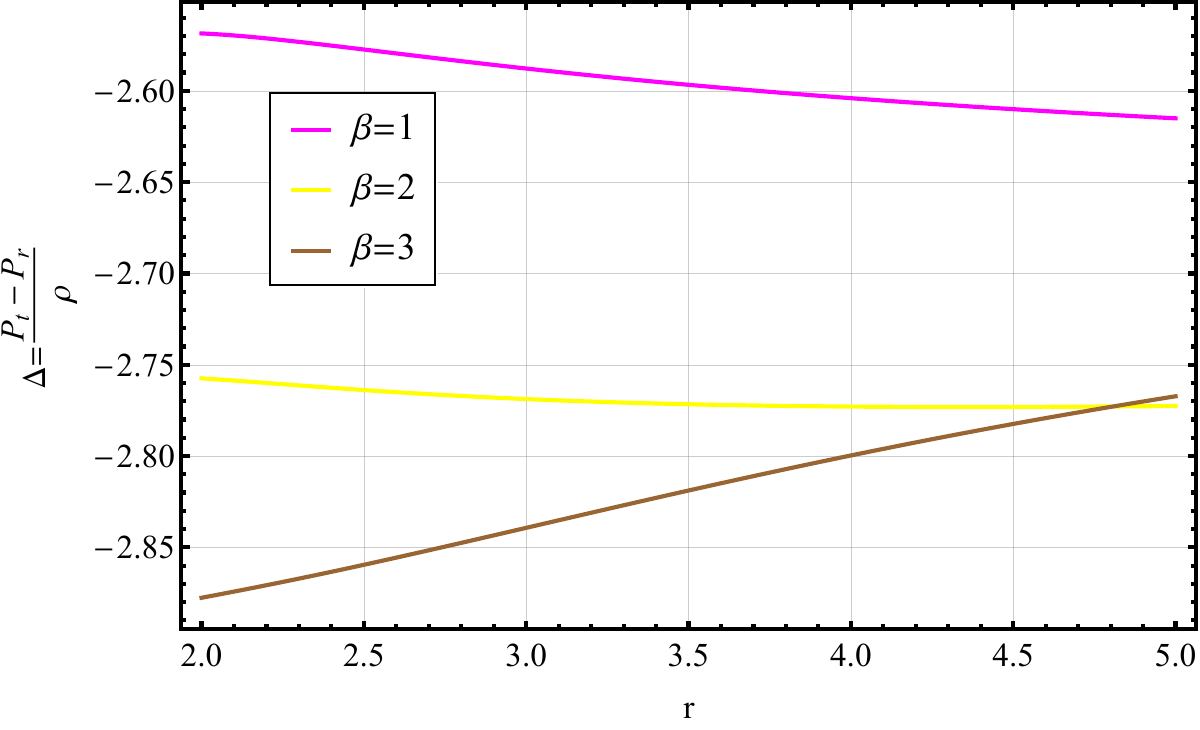}
\centering
\caption{Signs of the anisotropy parameter for the case $R-a_{1}^2/R+a_{2}T$. The curves are plotted for different values of the $\beta$ parameter, $\beta = 1$, $\beta = 2$, $\beta = 3$ and the values $r_{0}=1, \delta_{0}=1, \alpha=1, M=1$ are chosen. } 
\label{fig:delta1}
\end{figure}

\begin{figure}[H]
    \centering
    \begin{subfigure}[b]{0.45\textwidth}
        \centering
        \includegraphics[width=9cm]{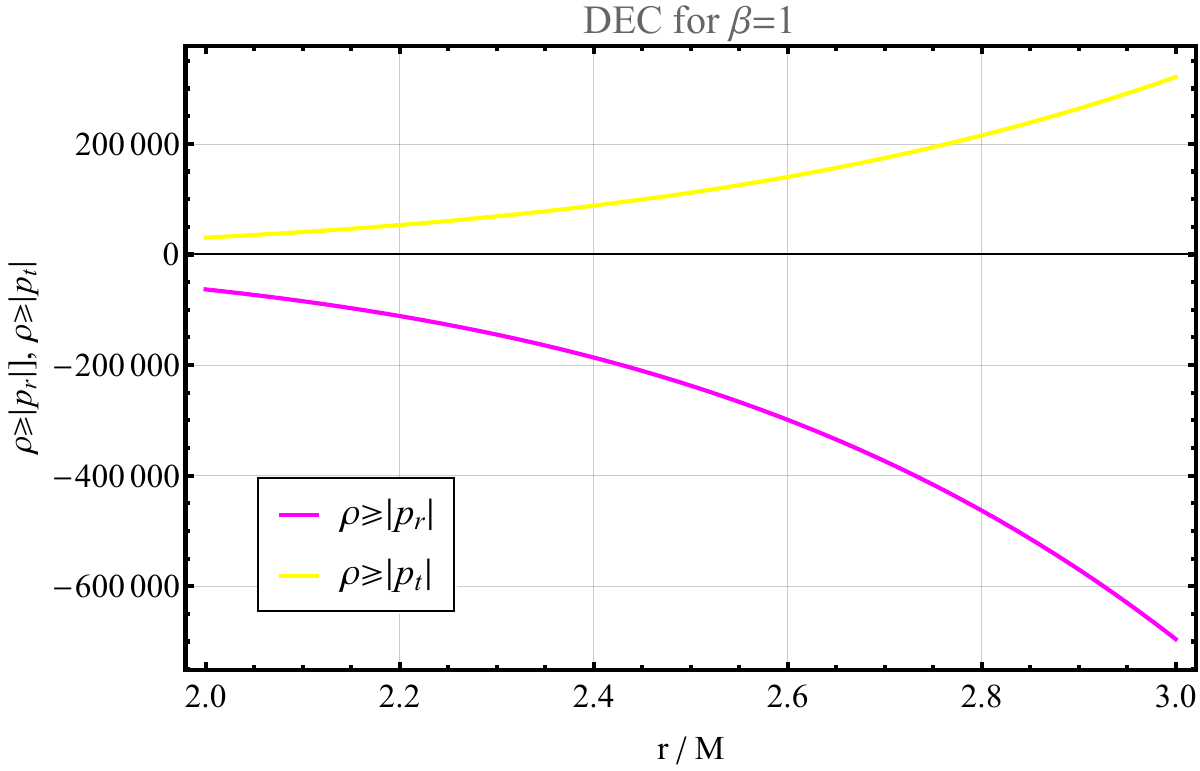}
        \caption*{(a)}
    \end{subfigure}
    \hfill
    \begin{subfigure}[b]{0.45\textwidth}
        \centering
        \includegraphics[width=9cm]{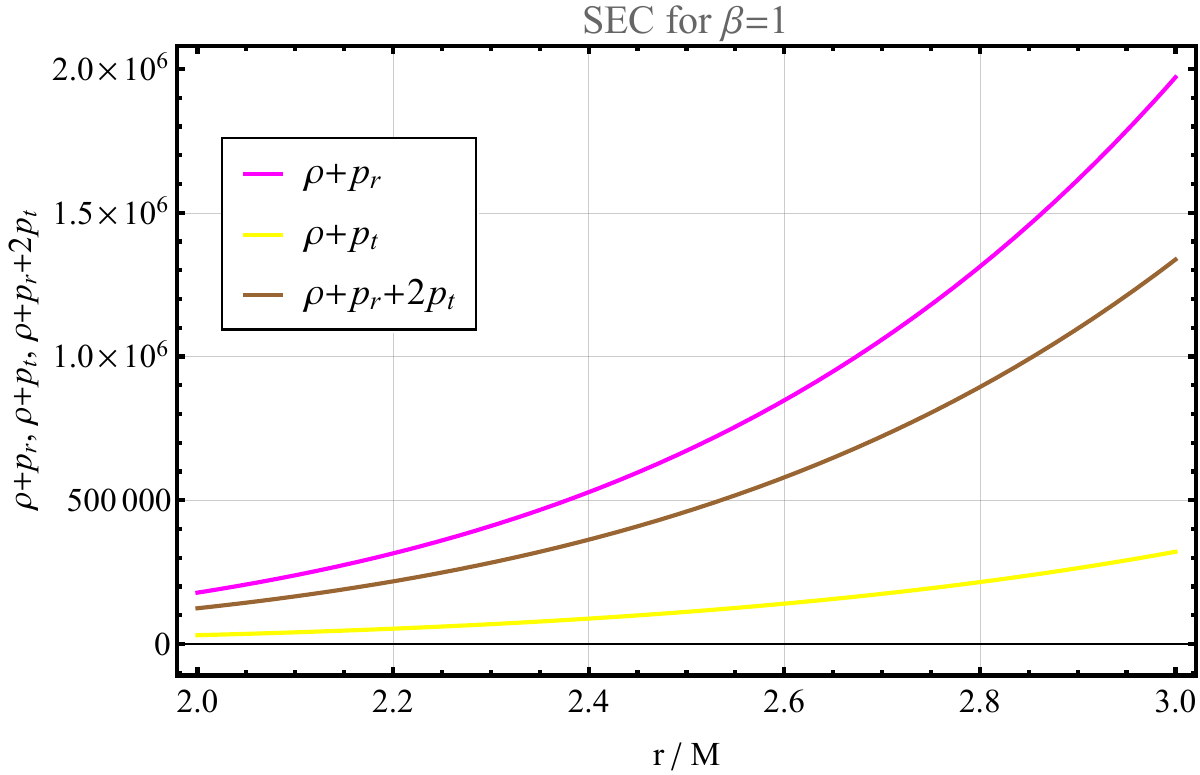}
        \caption*{(b)}
    \end{subfigure}

    \vspace{0.5cm} 

    \begin{subfigure}[b]{0.9\textwidth}
        \centering
        \includegraphics[width=9cm]{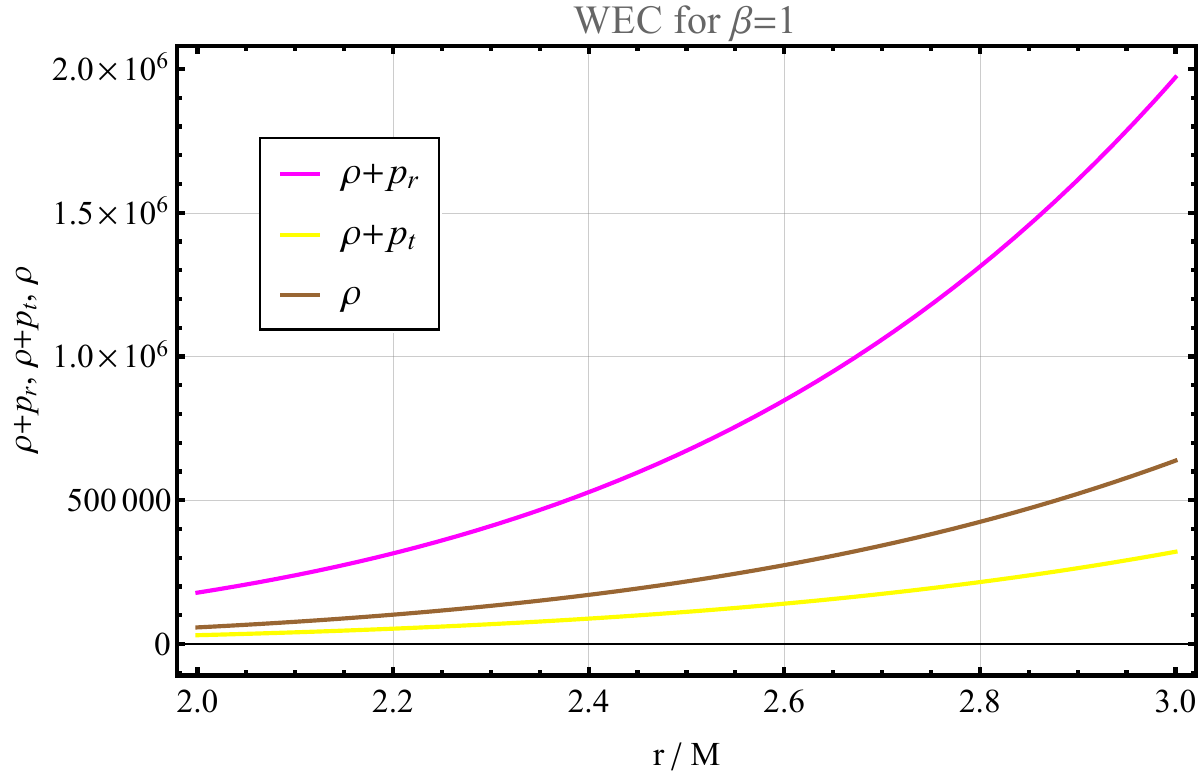}
        \caption*{(c)}
    \end{subfigure}

    \caption{Behaviour of energy conditions of the $R-a_{1}^2/R+a_{2}T$ case for $\beta=1$ and the values $r_{0}=1, \delta_{0}=1, \alpha=1, M=1$ are chosen.}
    \label{fig:decsecwecbeta1}
\end{figure}

\begin{figure}[H]
    \centering
    \begin{subfigure}[b]{0.45\textwidth}
        \centering
        \includegraphics[width=9cm]{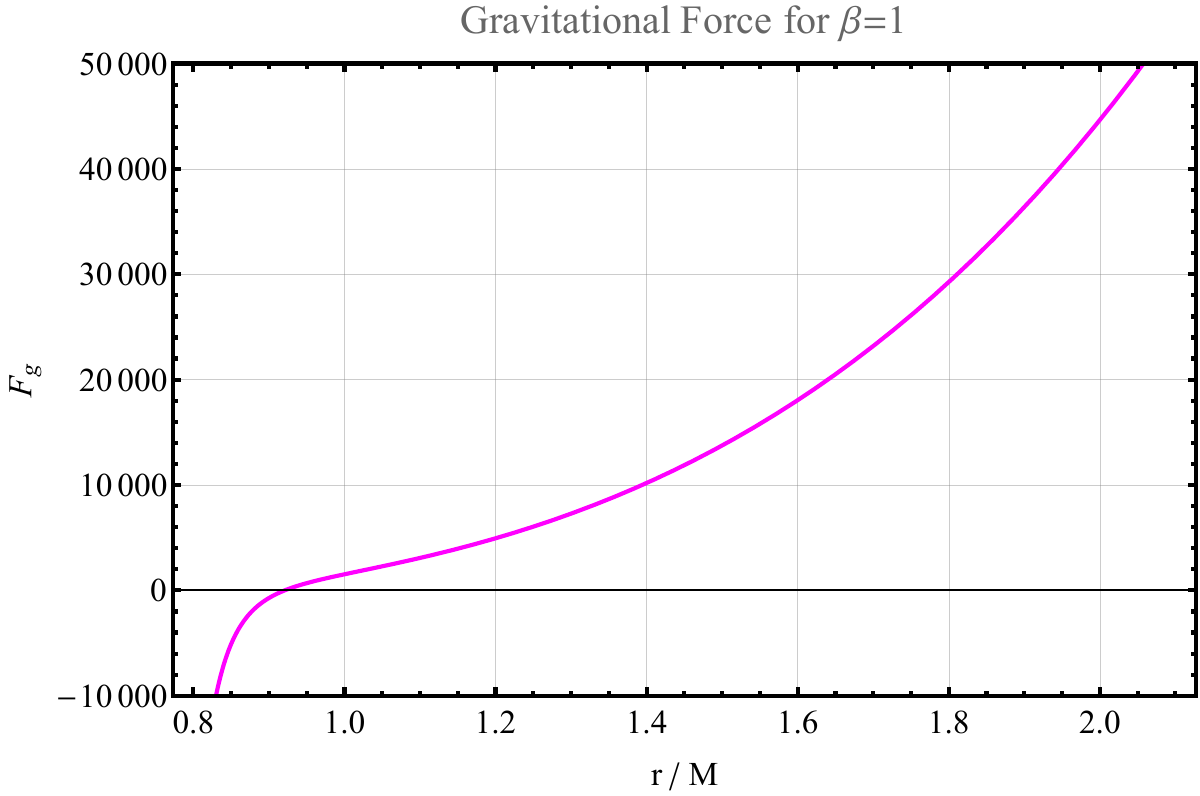}
        \caption*{(a)}
    \end{subfigure}
    \hfill
    \begin{subfigure}[b]{0.45\textwidth}
        \centering
        \includegraphics[width=9cm]{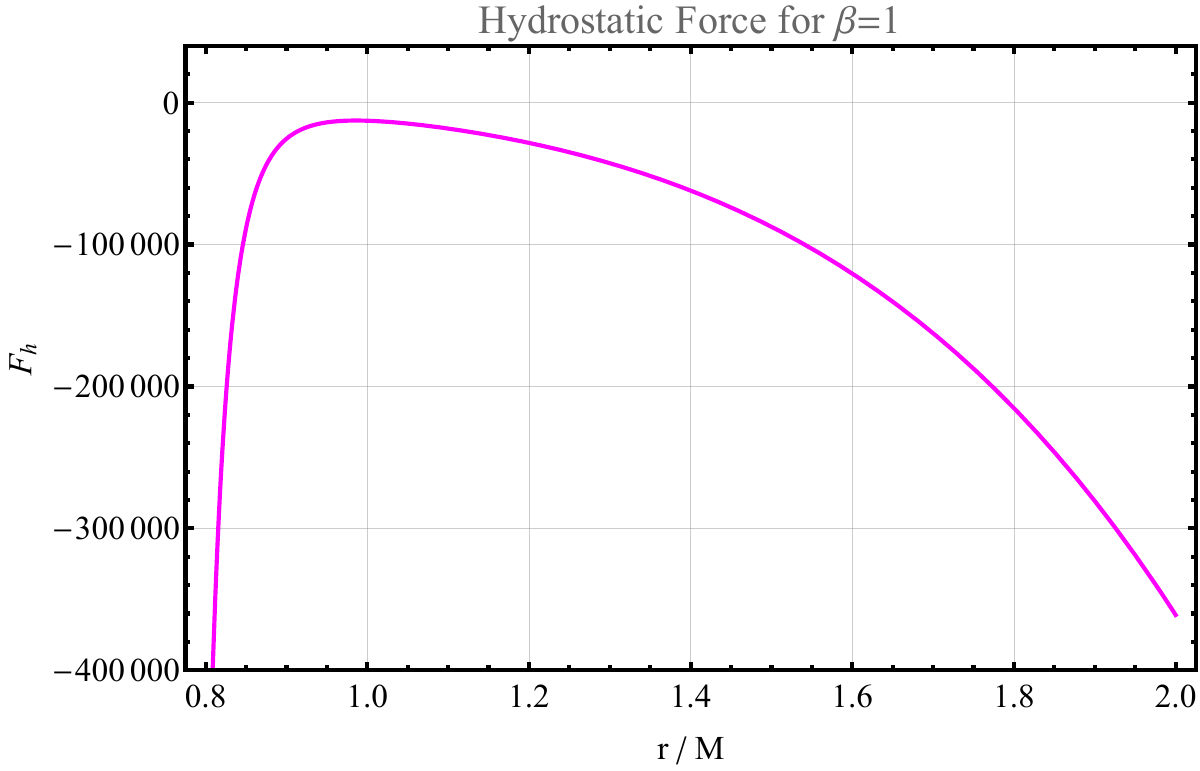}
        \caption*{(b)}
    \end{subfigure}

    \vspace{0.5cm}

    \begin{subfigure}[b]{0.45\textwidth}
        \centering
        \includegraphics[width=9cm]{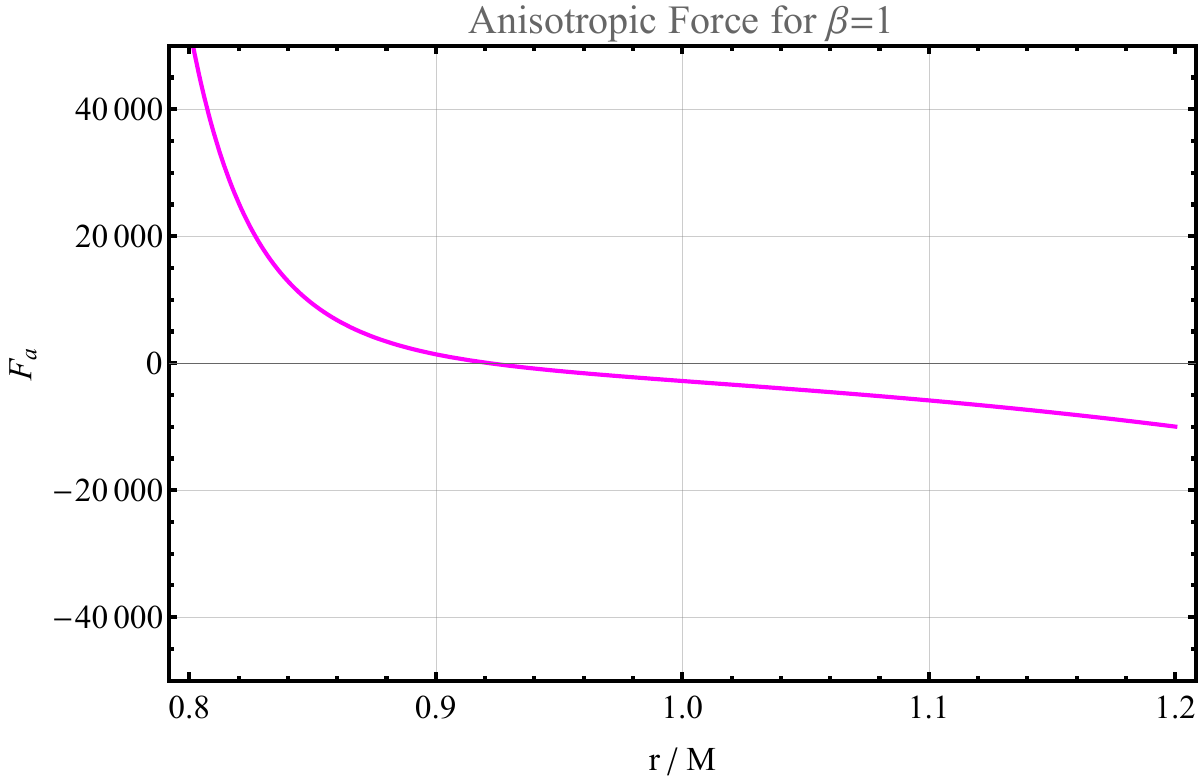}
        \caption*{(c)}
    \end{subfigure}
    \hfill
    \begin{subfigure}[b]{0.45\textwidth}
        \centering
        \includegraphics[width=9cm]{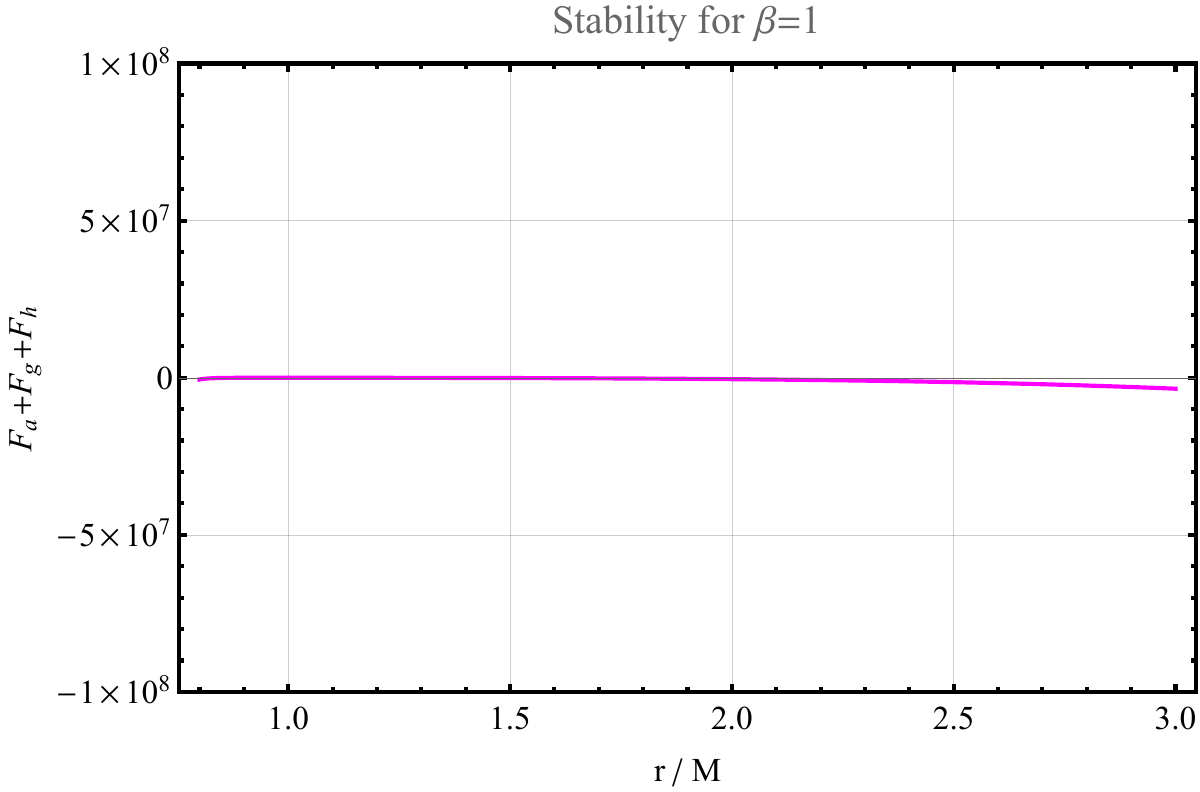}
        \caption*{(d)}
    \end{subfigure}

    \caption{Stability of the $R - a_{1}^{2}/R + a_{2}T$ case for $\beta=1$ and the values $r_{0}=1$, $\delta_{0}=1$, $\alpha=1$, $M=1$ are chosen. $F_{g}$, $F_{a}$ and $F_{h}$ represent gravitational force, anisotropic force and hydrostatic force, respectively.}
    \label{fig:gravhydroanisbeta1}
\end{figure}

\begin{figure}[H]
    \centering
    \begin{subfigure}[b]{0.45\textwidth}
        \centering
        \includegraphics[width=9cm]{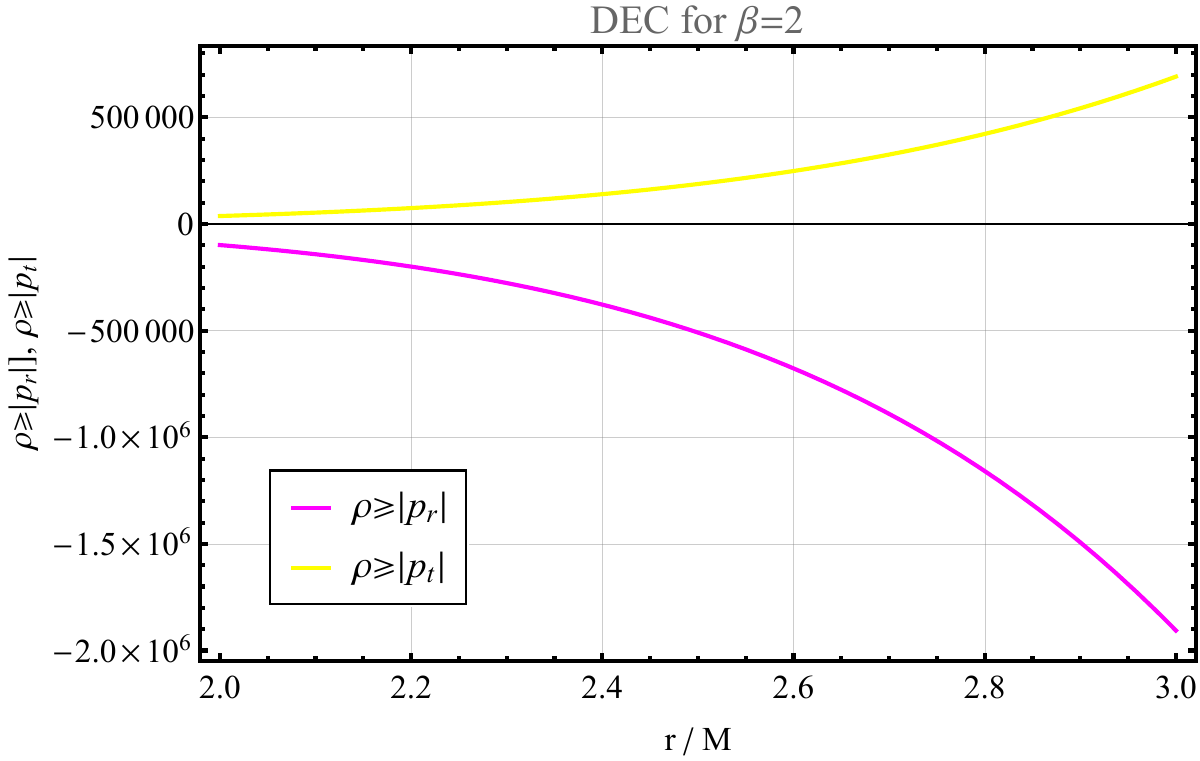}
        \caption*{(a)}
    \end{subfigure}
    \hfill
    \begin{subfigure}[b]{0.45\textwidth}
        \centering
        \includegraphics[width=9cm]{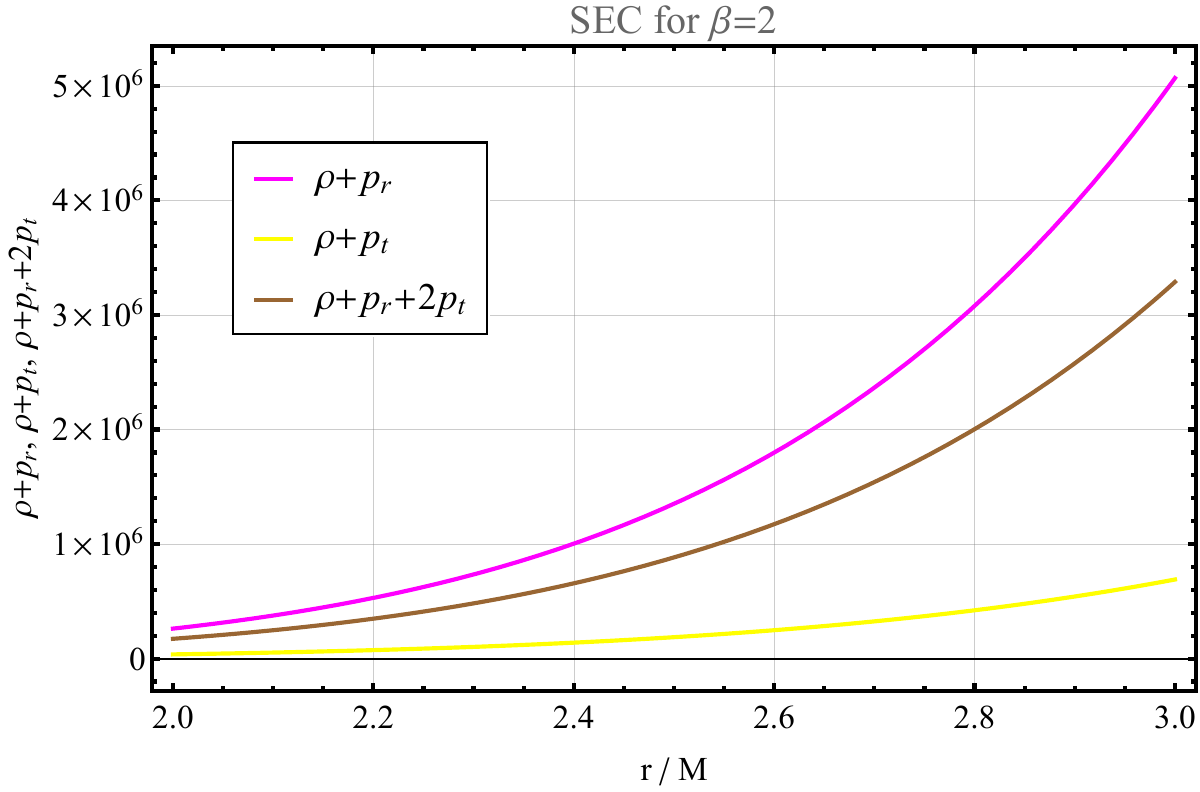}
        \caption*{(b)}
    \end{subfigure}

    \vspace{0.5cm}

    \begin{subfigure}[b]{0.9\textwidth}
        \centering
        \includegraphics[width=9cm]{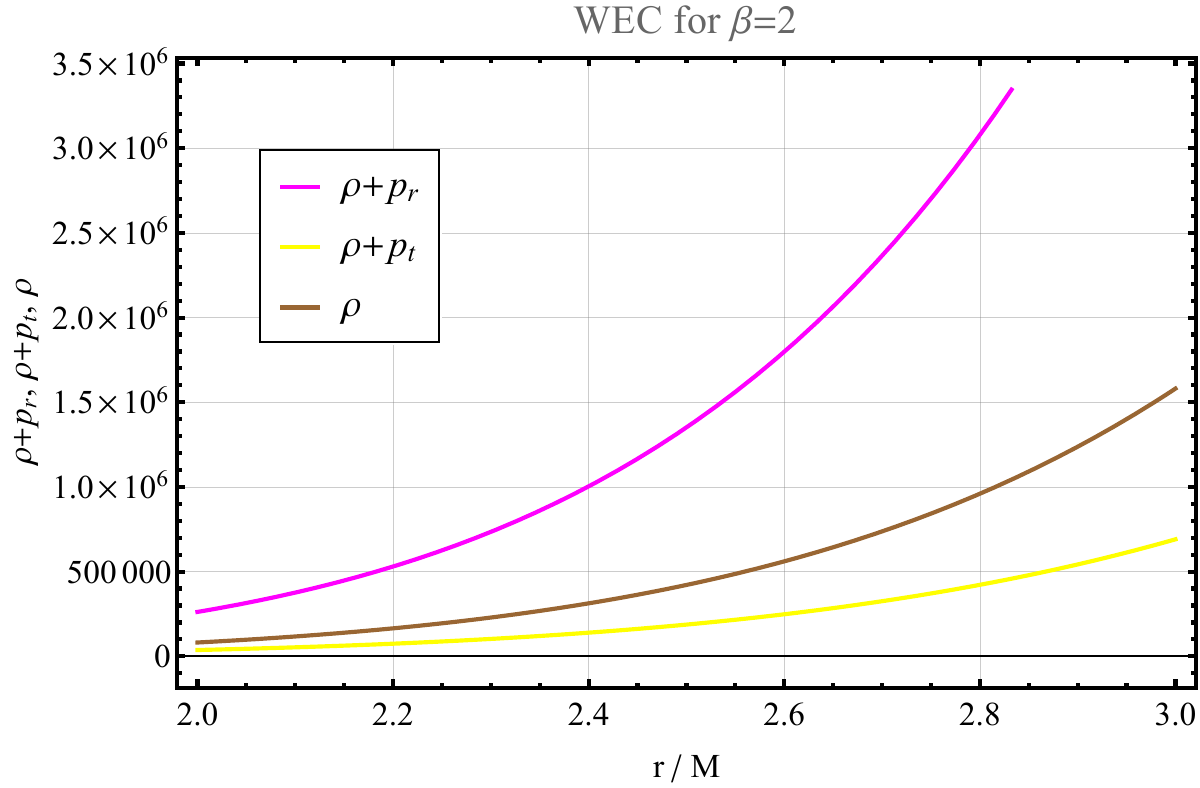}
        \caption*{(c)}
    \end{subfigure}

    \caption{Behaviour of energy conditions of the $R-a_{1}^2/R+a_{2}T$ case for $\beta=2$ and the values $r_{0}=1, \delta_{0}=1, \alpha=1, M=1$ are chosen.}
    \label{fig:decsecwecbeta2}
\end{figure}

\begin{figure}[H]
    \centering
    \begin{subfigure}[b]{0.45\textwidth}
        \centering
        \includegraphics[width=9cm]{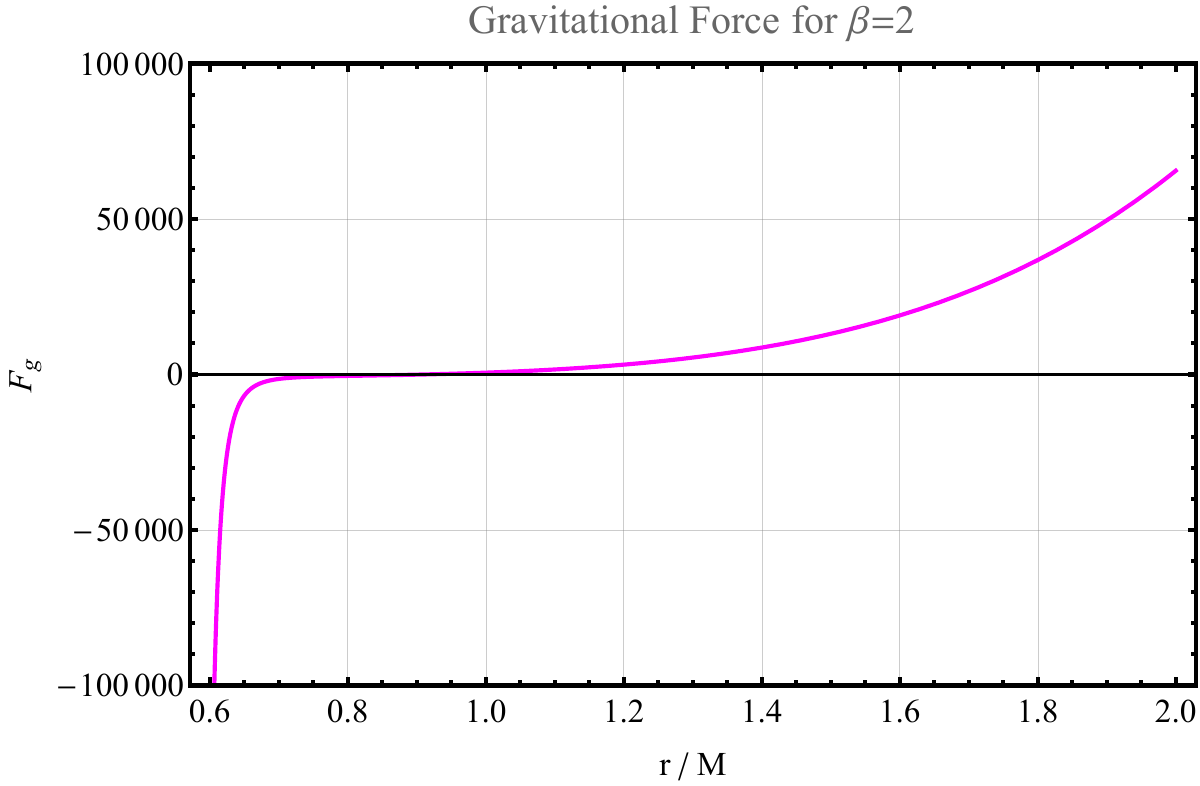}
        \caption*{(a)}
    \end{subfigure}
    \hfill
    \begin{subfigure}[b]{0.45\textwidth}
        \centering
        \includegraphics[width=9cm]{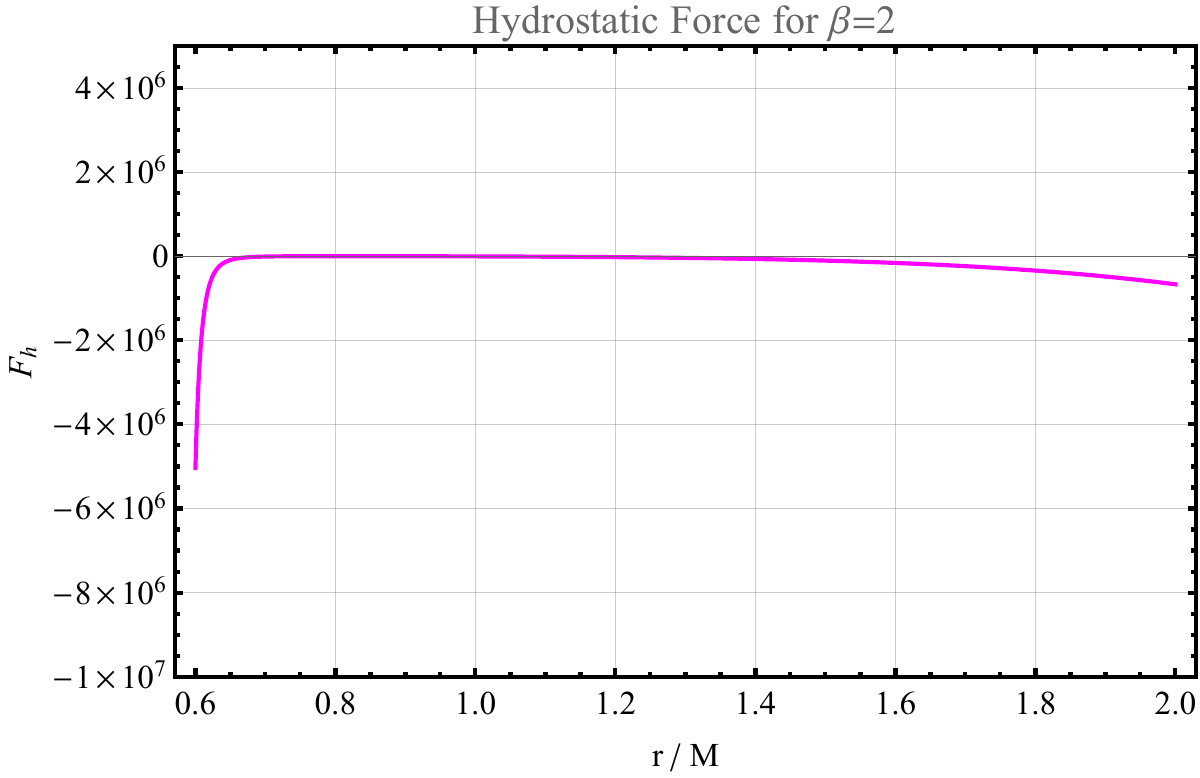}
        \caption*{(b)}
    \end{subfigure}

    \vspace{0.5cm}

    \begin{subfigure}[b]{0.45\textwidth}
        \centering
        \includegraphics[width=9cm]{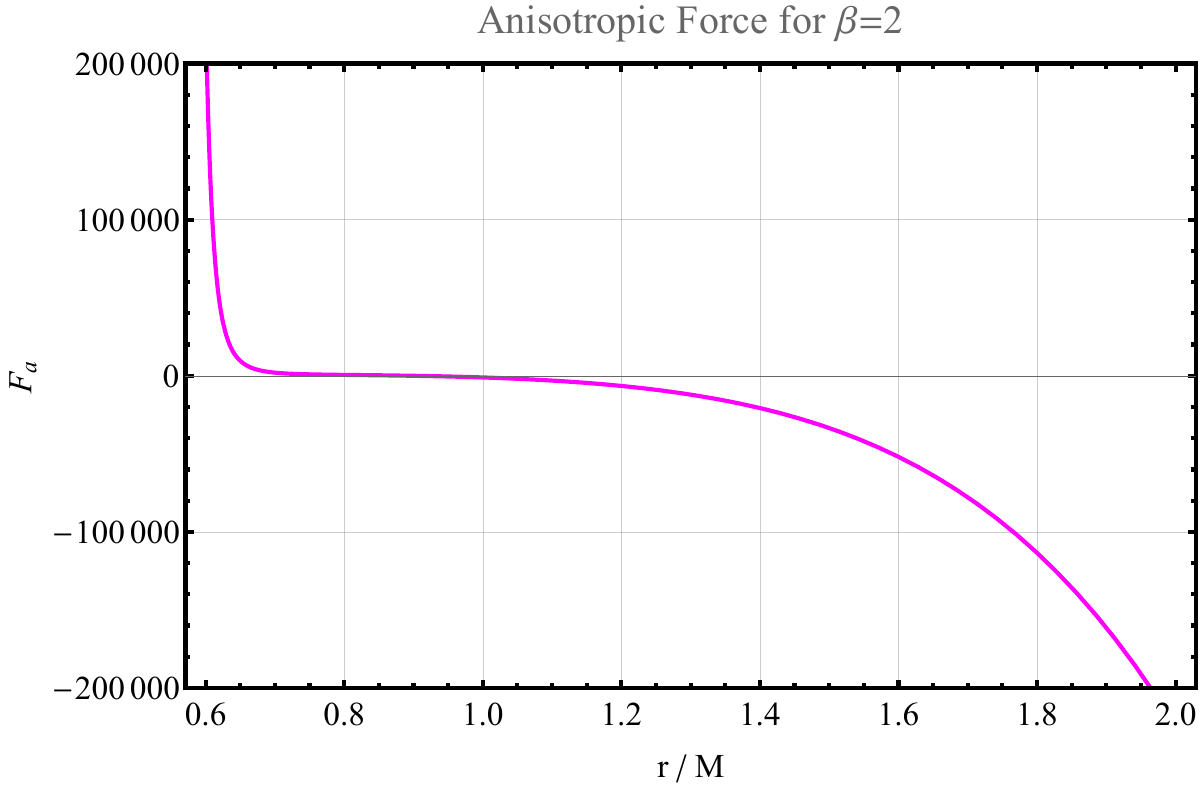}
        \caption*{(c)}
    \end{subfigure}
    \hfill
    \begin{subfigure}[b]{0.45\textwidth}
        \centering
        \includegraphics[width=9cm]{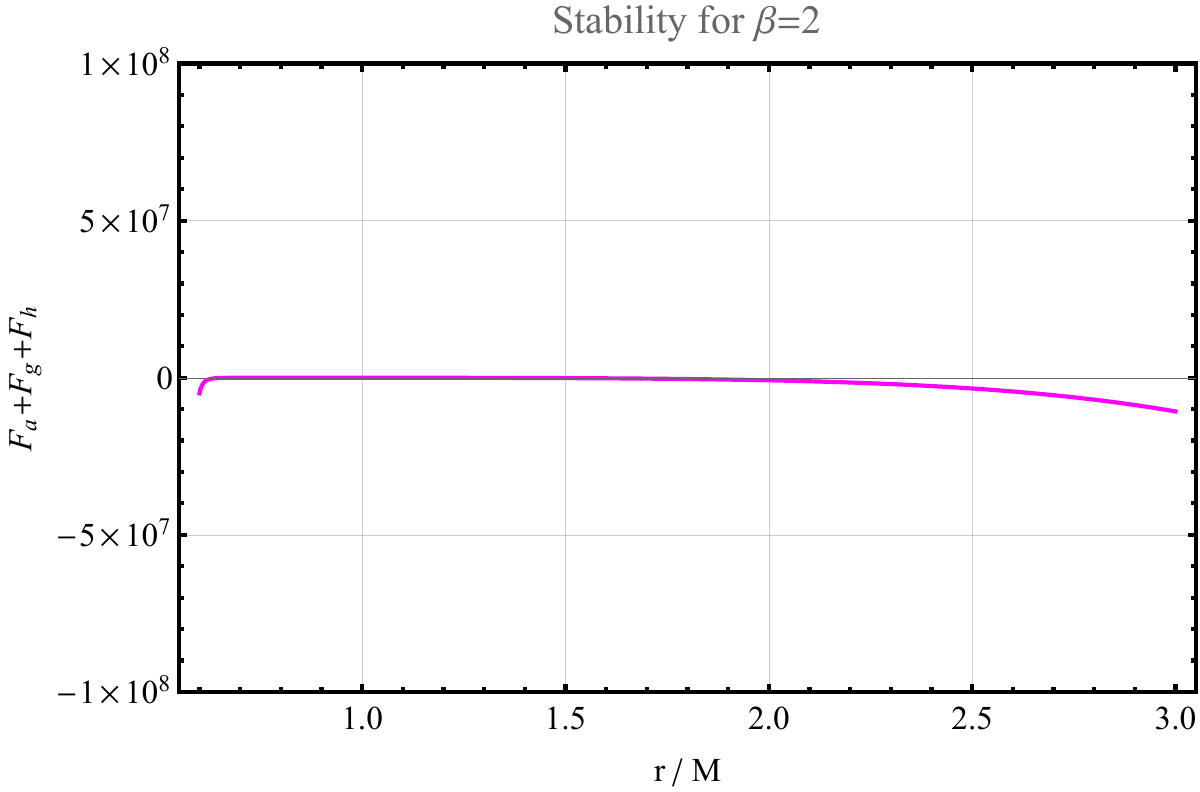}
        \caption*{(d)}
    \end{subfigure}

    \caption{Stability of the $R-a_{1}^2/R+a_{2}T$ case for $\beta=2$ and the values $r_{0}=1$, $\delta_{0}=1$, $\alpha=1$, $M=1$ are chosen. $F_{g}$, $F_{a}$ and $F_{h}$ represent gravitational force, anisotropic force and hydrostatic force, respectively.}
    \label{fig:gravhydroanisbeta2}
\end{figure}

\begin{figure}[H]
    \centering
    \begin{subfigure}[b]{0.45\textwidth}
        \centering
        \includegraphics[width=9cm]{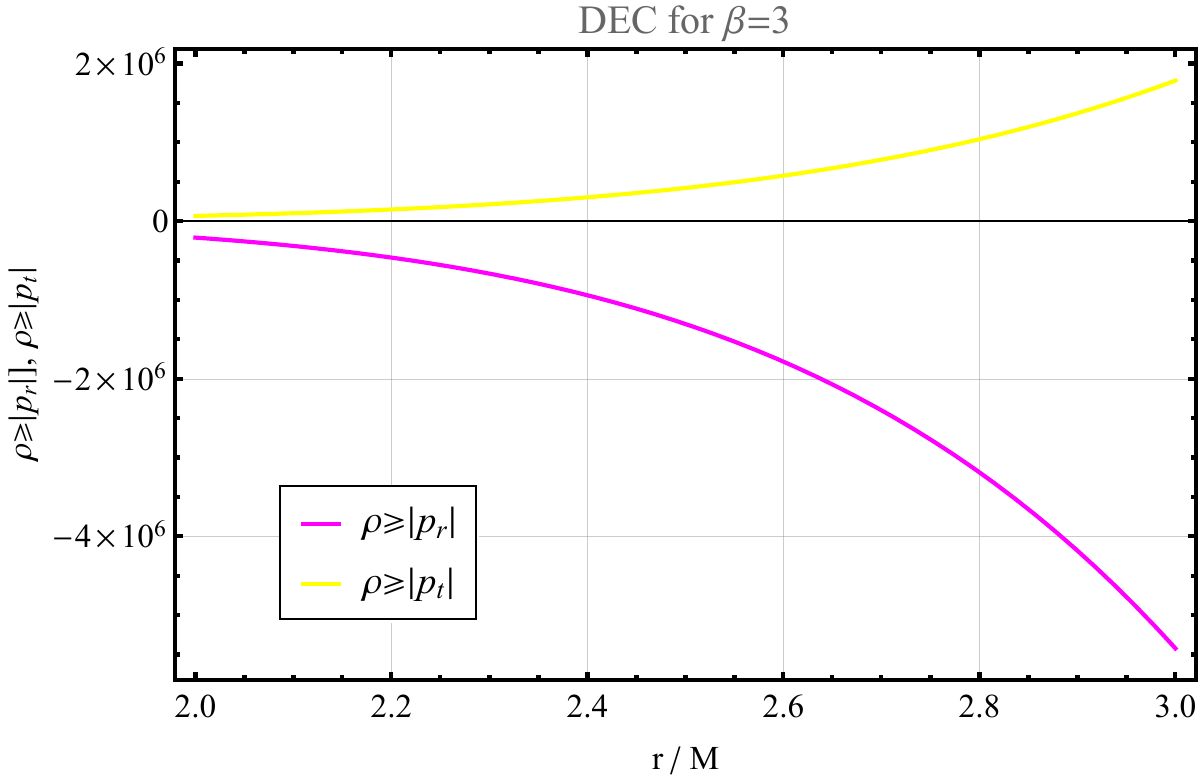}
        \caption*{(a)}
    \end{subfigure}
    \hfill
    \begin{subfigure}[b]{0.45\textwidth}
        \centering
        \includegraphics[width=9cm]{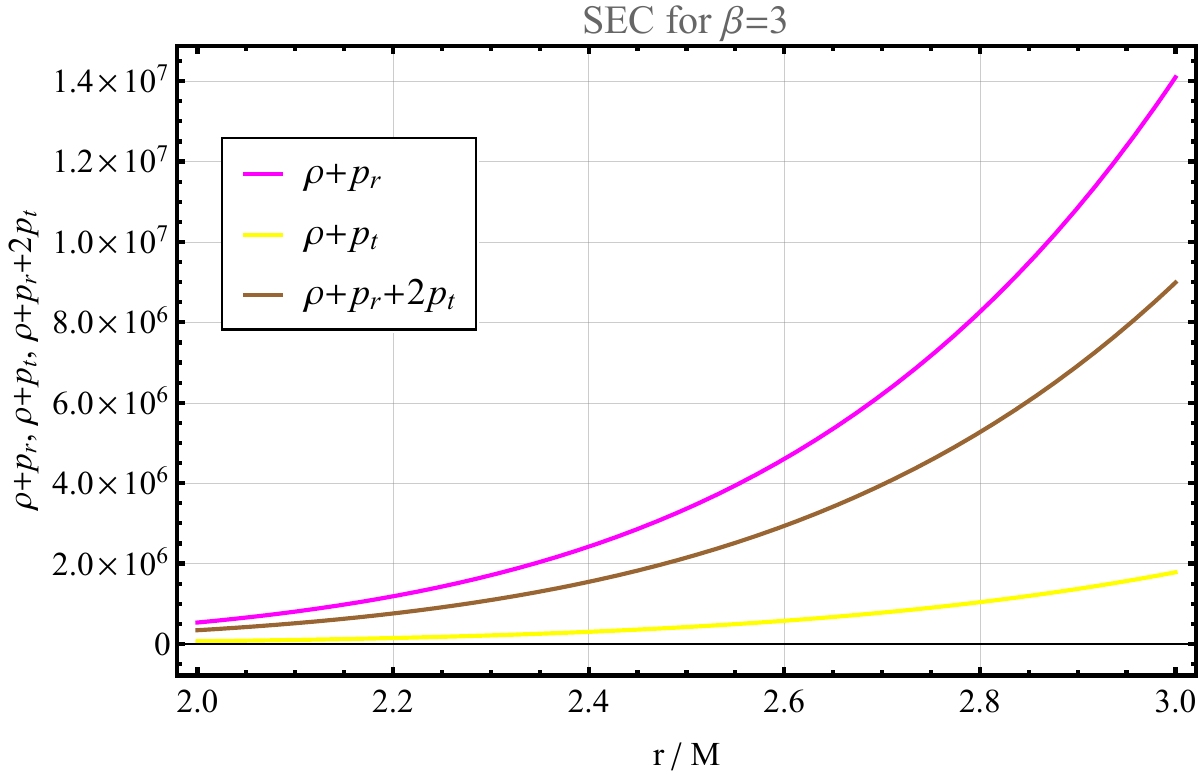}
        \caption*{(b)}
    \end{subfigure}

    \vspace{0.5cm}

    \begin{subfigure}[b]{0.9\textwidth}
        \centering
        \includegraphics[width=9cm]{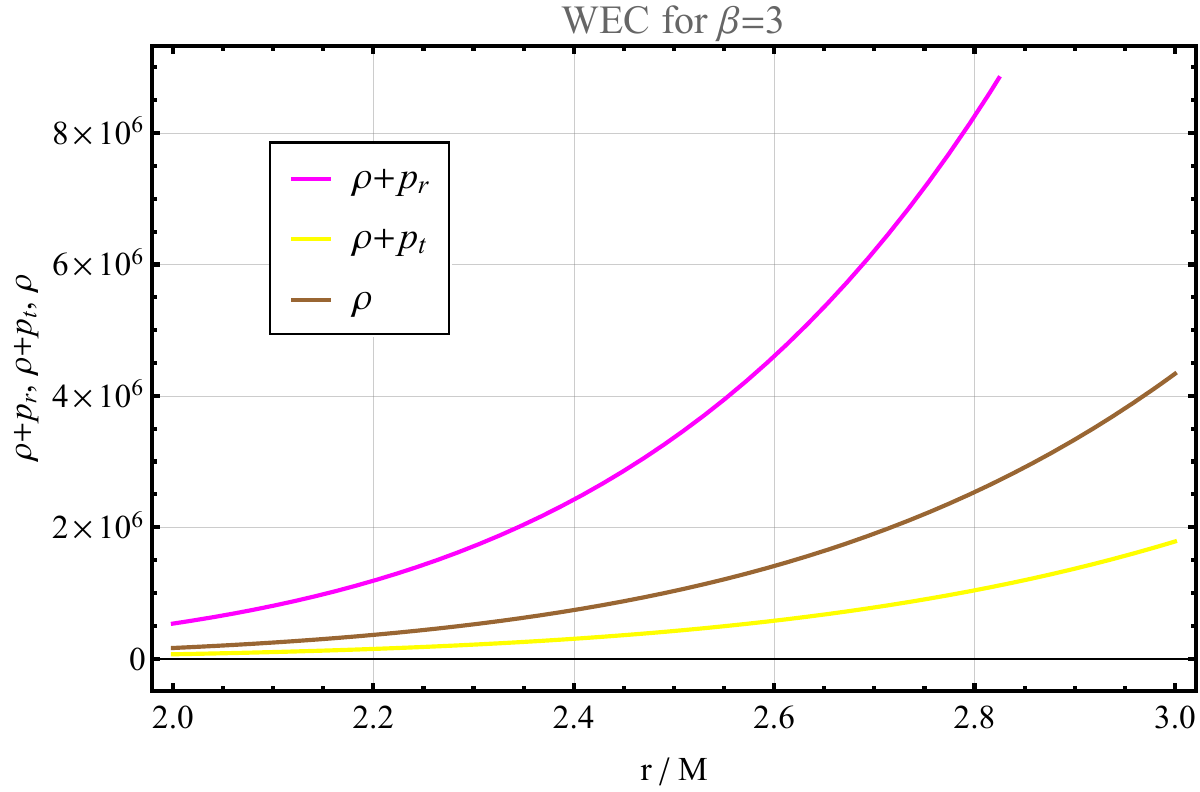}
        \caption*{(c)}
    \end{subfigure}

    \caption{Behaviour of energy conditions of the $R-a_{1}^2/R+a_{2}T$ case for $\beta=3$ and the values $r_{0}=1, \delta_{0}=1, \alpha=1, M=1$ are chosen.}
    \label{fig:decsecwecbeta3}
\end{figure}

\begin{figure}[H]
    \centering
    \begin{subfigure}[b]{0.45\textwidth}
        \centering
        \includegraphics[width=9cm]{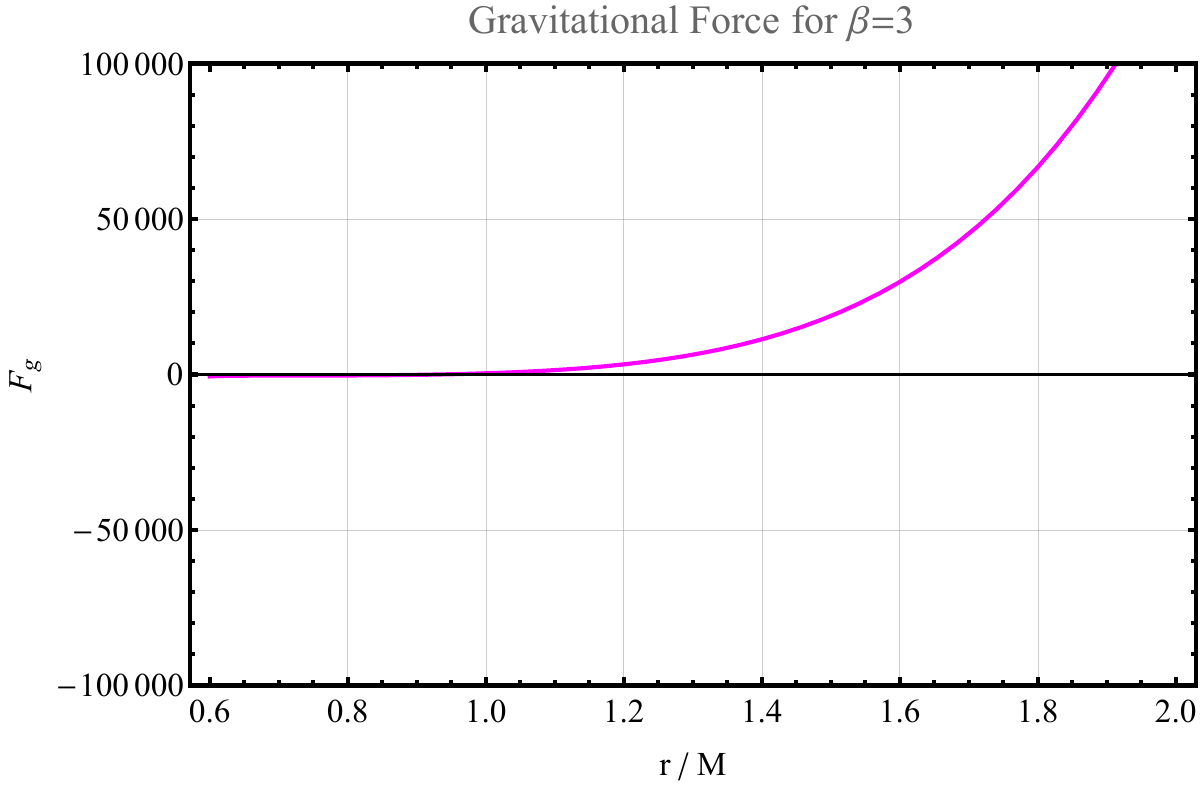}
        \caption*{(a)}
    \end{subfigure}
    \hfill
    \begin{subfigure}[b]{0.45\textwidth}
        \centering
        \includegraphics[width=9cm]{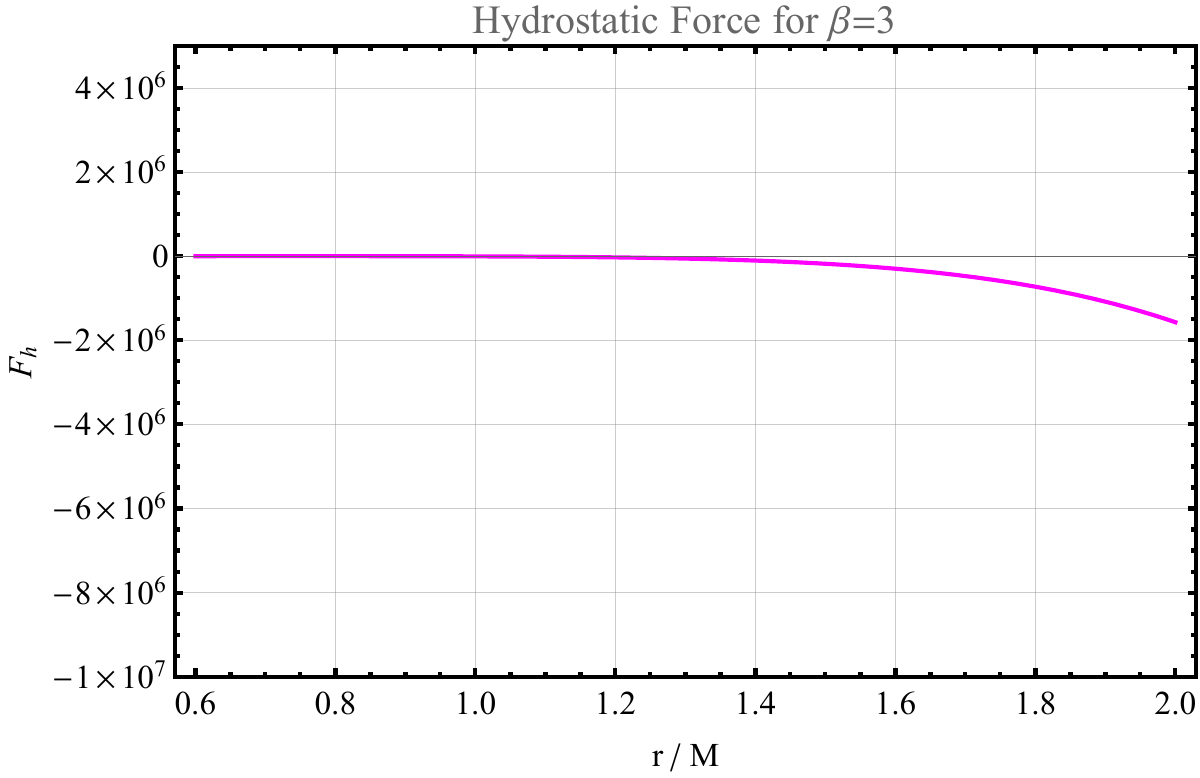}
        \caption*{(b)}
    \end{subfigure}

    \vspace{0.5cm}

    \begin{subfigure}[b]{0.45\textwidth}
        \centering
        \includegraphics[width=9cm]{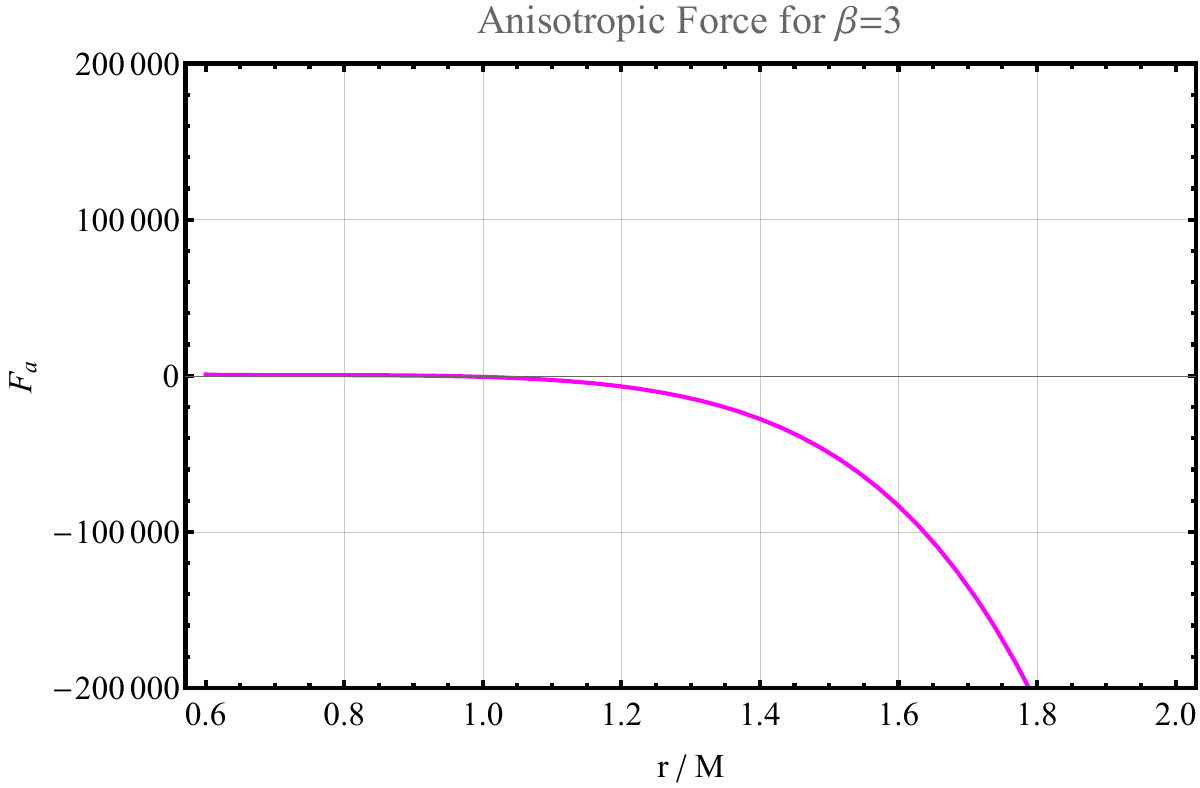}
        \caption*{(c)}
    \end{subfigure}
    \hfill
    \begin{subfigure}[b]{0.45\textwidth}
        \centering
        \includegraphics[width=9cm]{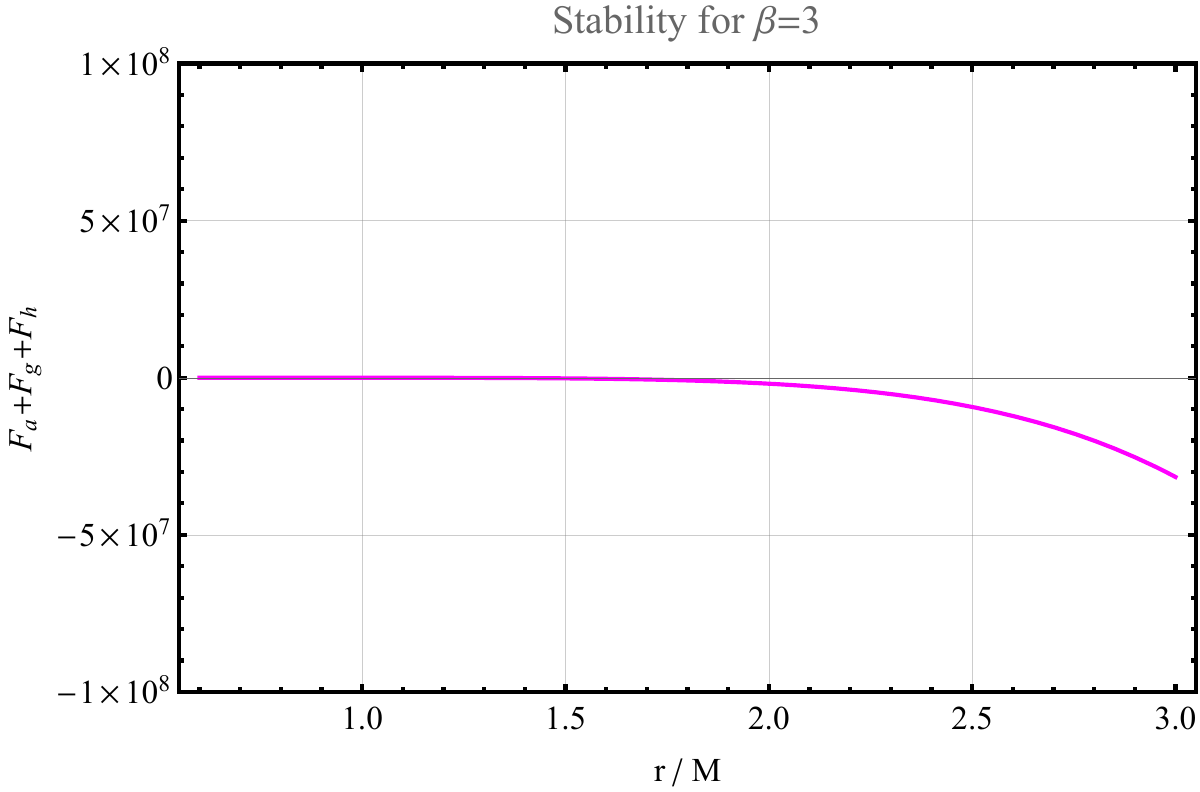}
        \caption*{(d)}
    \end{subfigure}

    \caption{Stability of the $R-a_{1}^2/R+a_{2}T$ case for $\beta=3$ and the values $r_{0}=1$, $\delta_{0}=1$, $\alpha=1$, $M=1$ are chosen. $F_{g}$, $F_{a}$ and $F_{h}$ represent gravitational force, anisotropic force and hydrostatic force, respectively.}
    \label{fig:gravhydroanisbeta3}
\end{figure}

The evolution of the gravitational force, anisotropic force, and hydrostatic force for  $\beta = 1$, $\beta = 2$, and $\beta = 3$ has been examined. However, the three beta values indicate that the geometry possesses an attractive feature, figure \ref{fig:delta1}. We can analyze this behavior using the dimensionless anisotropic parameter, $\Delta=\frac{p_t-p_r}{\rho}$, \cite{Cattoen/2005,Lobo/2013}. In all three cases, since $\rho>0$, if  $p_t<p_r$, then $\Delta <0$, which implies that the geometry is attractive. On the other hand, if $p_t>p_r$, the geometry is repulsive, and if  $p_t=p_r$, the fluid is isotropic. The evolution of the energy and stability conditions in the case of $\beta=1$ is presented in the figures \ref{fig:decsecwecbeta1}, \ref{fig:gravhydroanisbeta1}, in the case of $\beta=2$ is presented in the figures \ref{fig:decsecwecbeta2}, \ref{fig:gravhydroanisbeta2},  and in the case of $\beta=3$ is presented in the figures \ref{fig:decsecwecbeta3}, \ref{fig:gravhydroanisbeta3}.


\subsection{Energy and Stability Conditions for the case  $R+a_{1}^2R^{2}+a_{2}T$ }\label{subsec:sc2}

\begin{figure}[H]
\centering
\includegraphics[width=9cm]{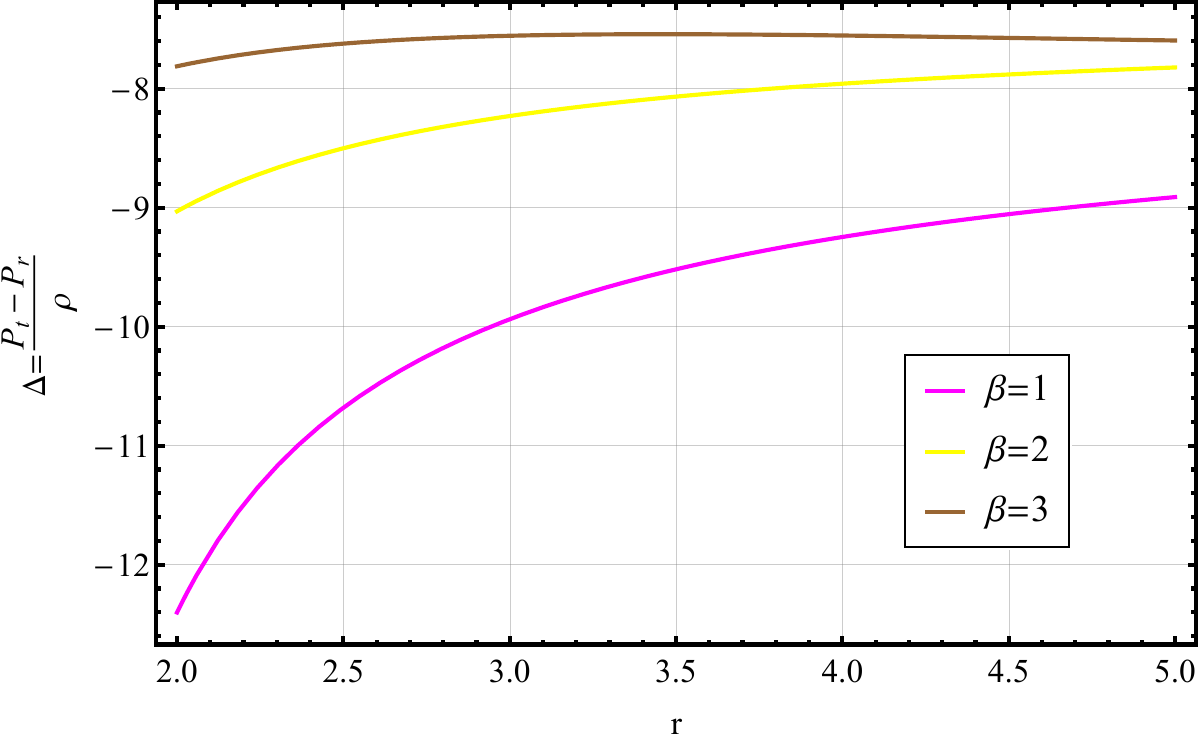}
\caption{Signs of the anisotropy parameter for the case $R+a_{1}^2R^{2}+a_{2}T$.  The curves are plotted for different values of the $\beta$ parameter, $\beta = 1$, $\beta = 2$, $\beta = 3$ and the values $r_{0}=1, \delta_{0}=1, \alpha=1, M=1$ are chosen.} 
\label{del2}
\end{figure}

\begin{figure}[H]
    \centering
    \begin{subfigure}[b]{0.45\textwidth}
        \centering
        \includegraphics[width=9cm]{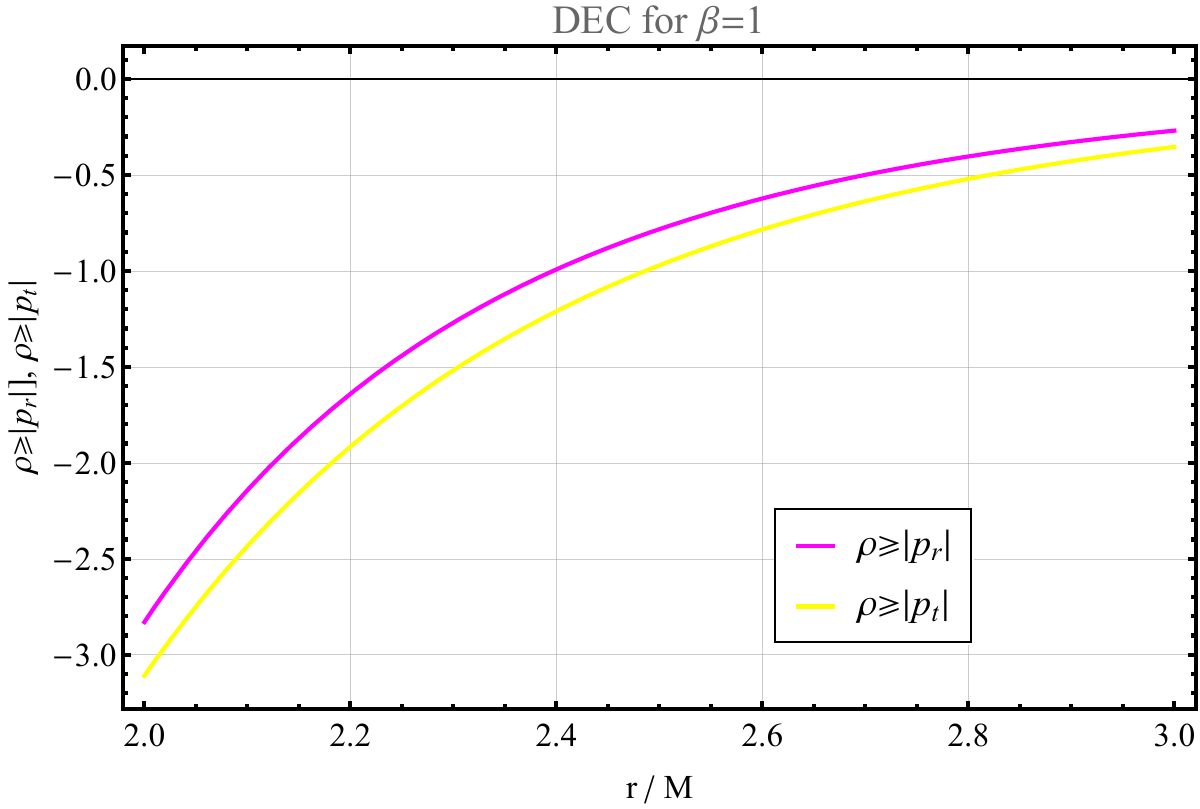}
        \caption*{(a)}
    \end{subfigure}
    \hfill
    \begin{subfigure}[b]{0.45\textwidth}
        \centering
        \includegraphics[width=9cm]{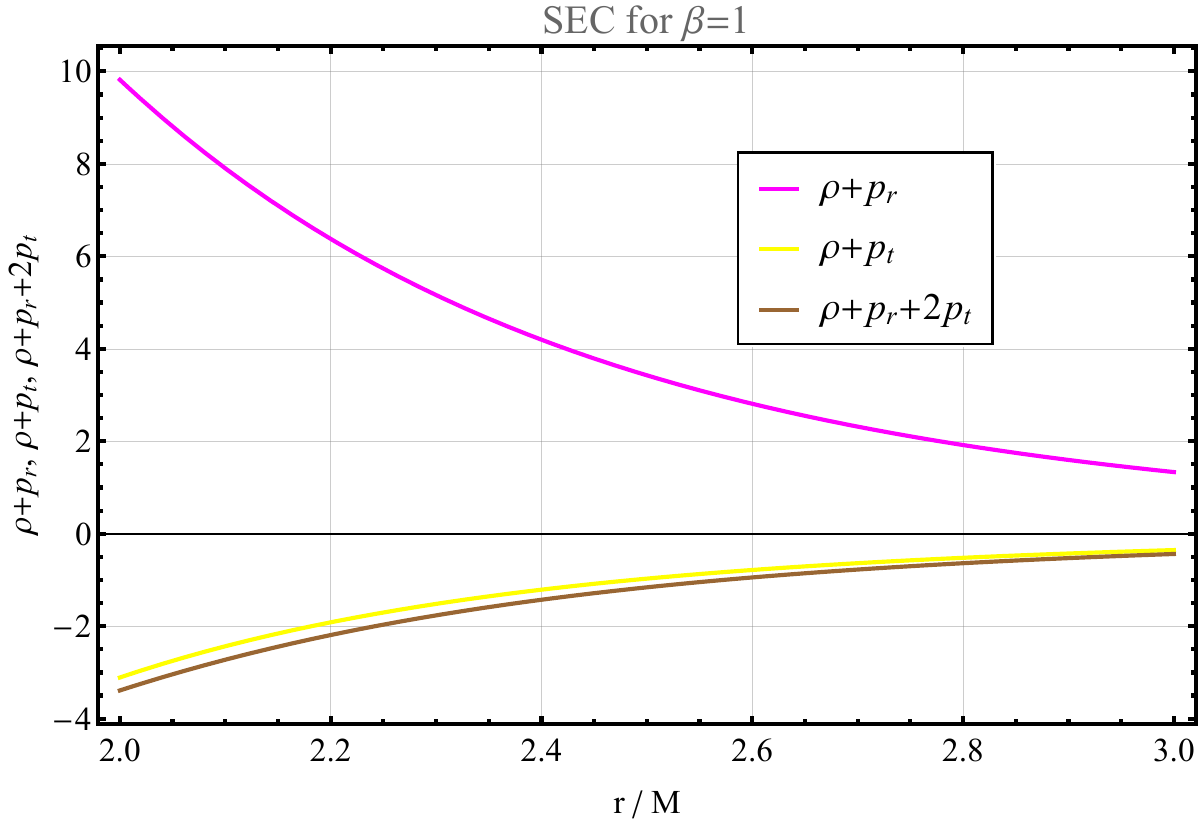}
        \caption*{(b)}
    \end{subfigure}

    \vspace{0.5cm}

    \begin{subfigure}[b]{0.9\textwidth}
        \centering
        \includegraphics[width=9cm]{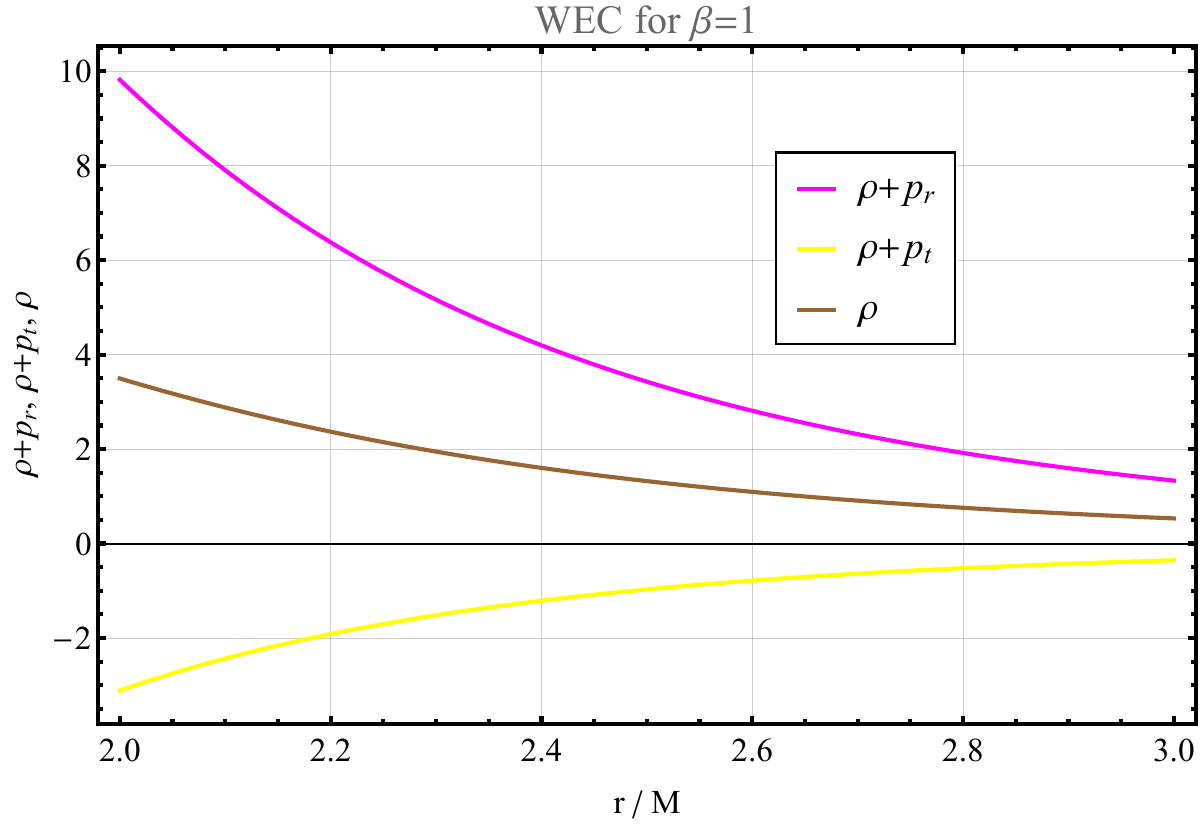}
        \caption*{(c)}
    \end{subfigure}

    \caption{Behaviour of energy conditions of the $R + a_{1}^2 R^{2} + a_{2}T$ case for $\beta=1$ and the values $r_{0}=1$, $\delta_{0}=1$, $\alpha=1$, $M=1$ are chosen.}
    \label{fig:decsecwecbeta22}
\end{figure}

\begin{figure}[H]
    \centering
    \begin{subfigure}[b]{0.45\textwidth}
        \centering
        \includegraphics[width=9cm]{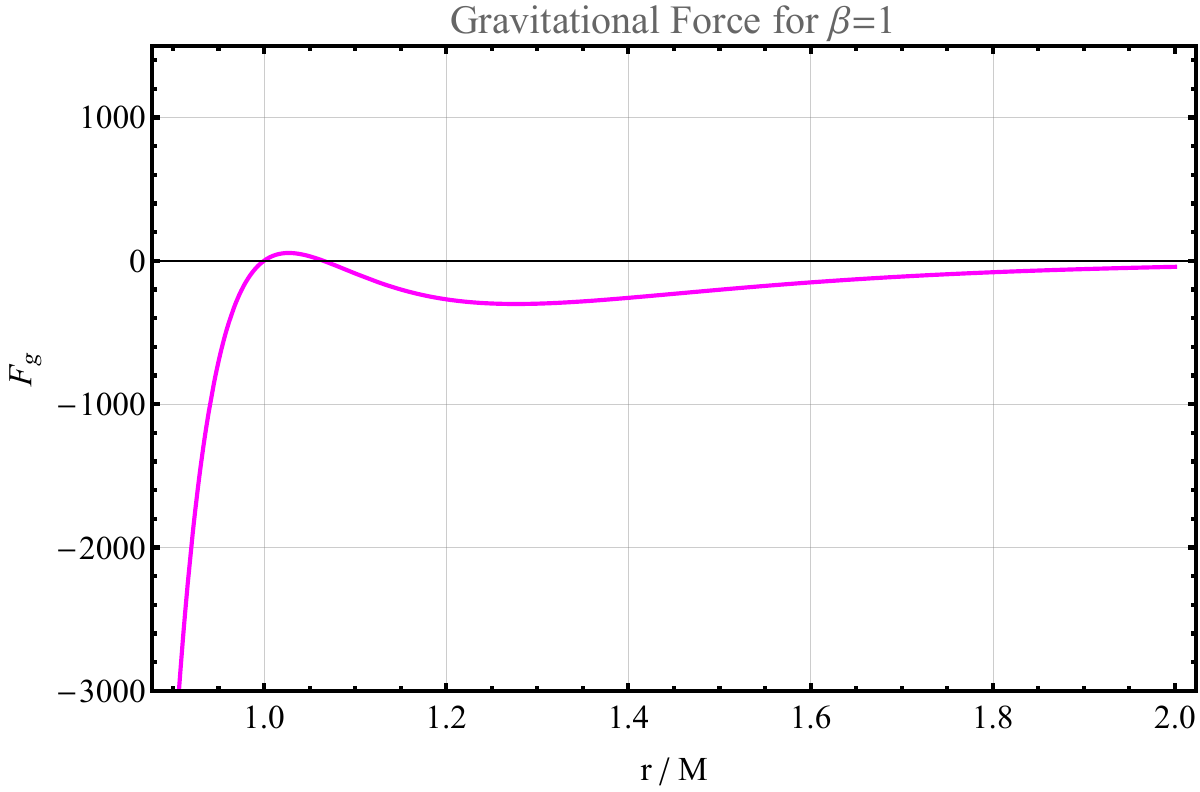}
        \caption*{(a)}
    \end{subfigure}
    \hfill
    \begin{subfigure}[b]{0.45\textwidth}
        \centering
        \includegraphics[width=9cm]{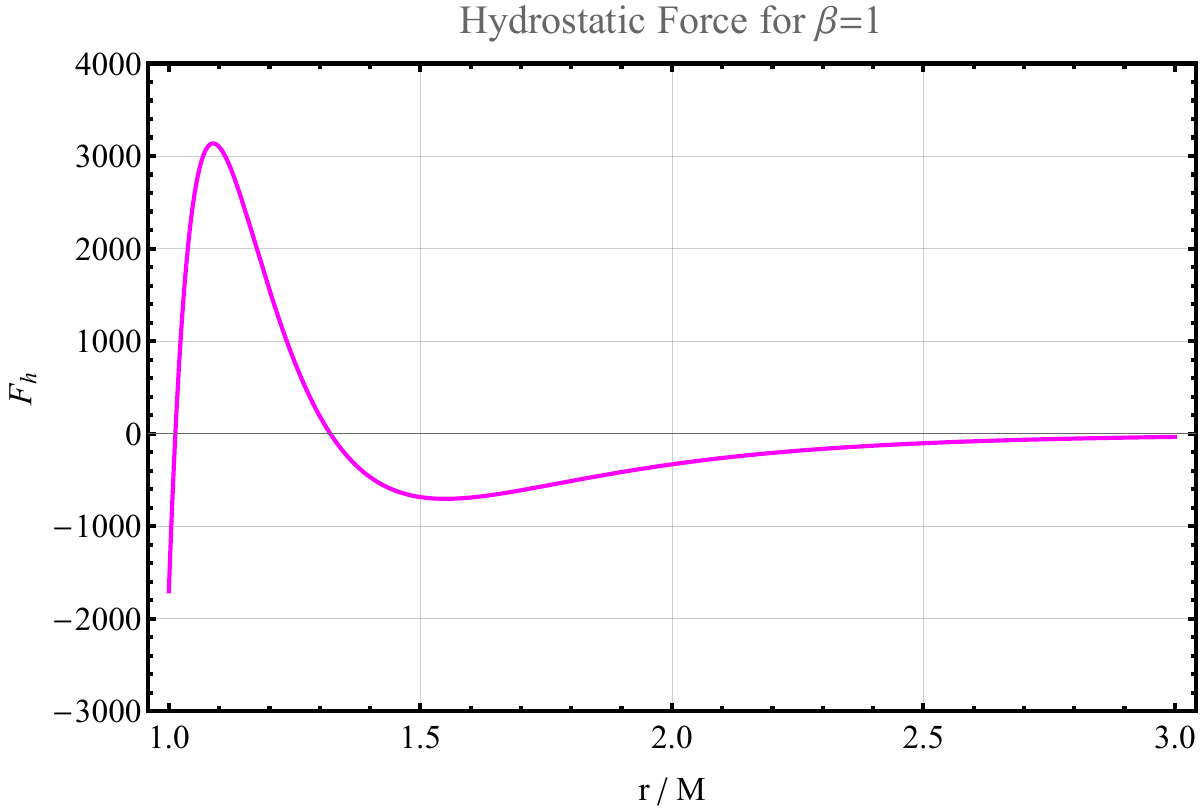}
        \caption*{(b)}
    \end{subfigure}

    \vspace{0.5cm}

    \begin{subfigure}[b]{0.45\textwidth}
        \centering
        \includegraphics[width=9cm]{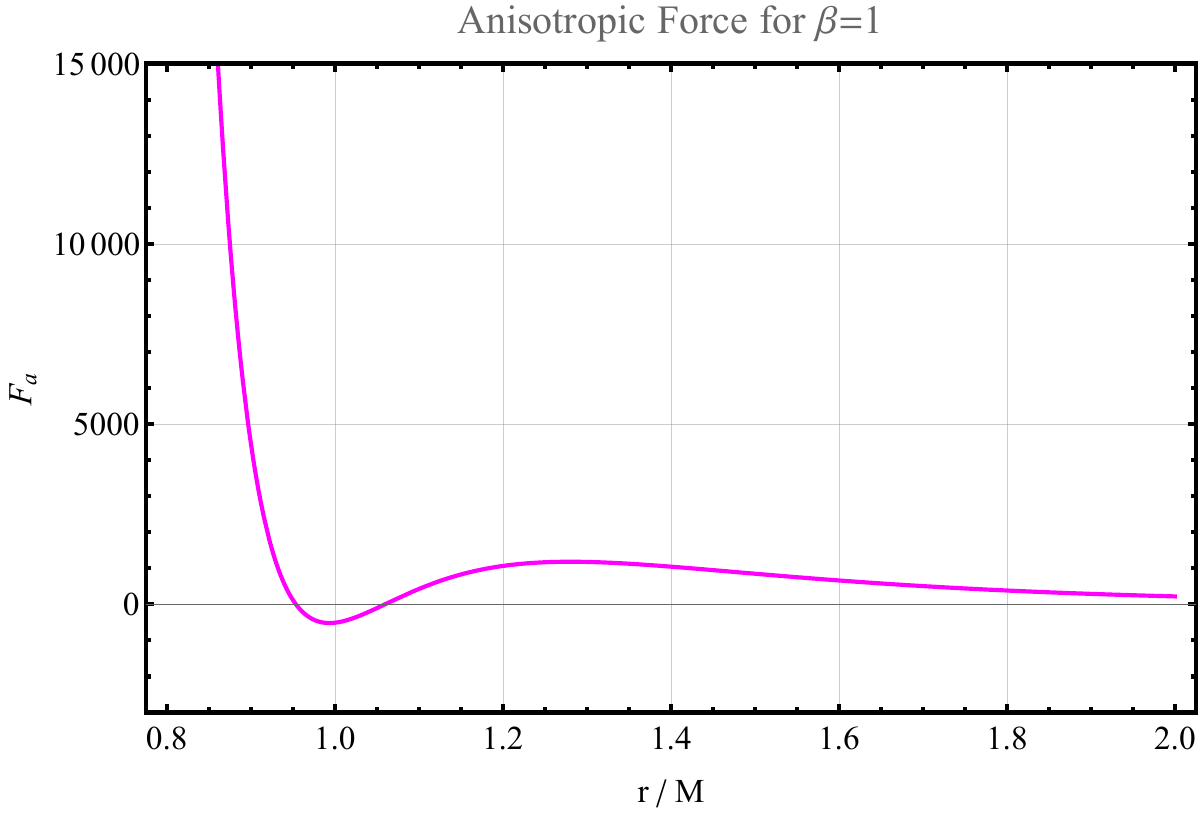}
        \caption*{(c)}
    \end{subfigure}
    \hfill
    \begin{subfigure}[b]{0.45\textwidth}
        \centering
        \includegraphics[width=9cm]{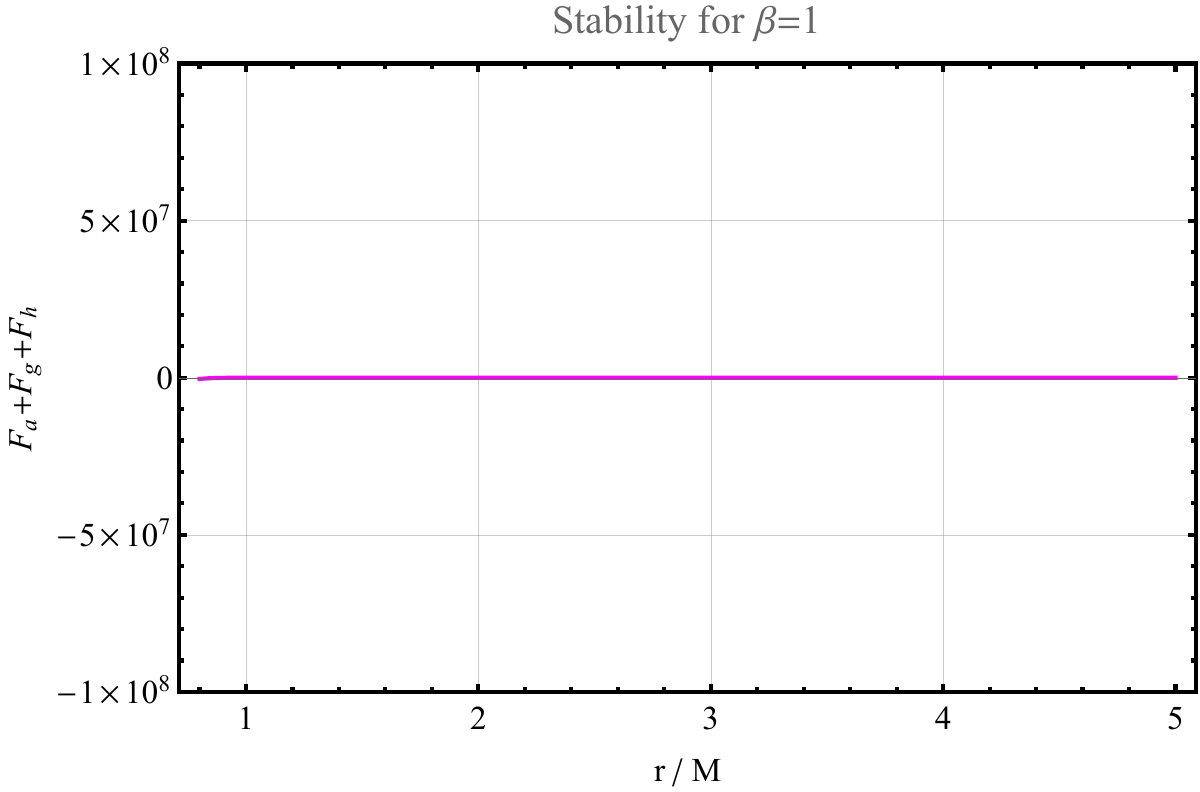}
        \caption*{(d)}
    \end{subfigure}

    \caption{Stability of the $R + a_{1}^2 R^{2} + a_{2}T$ case for $\beta=1$ and the values $r_{0}=1$, $\delta_{0}=1$, $\alpha=1$, $M=1$ are chosen. $F_{g}$, $F_{a}$ and $F_{h}$ represent gravitational force, anisotropic force and hydrostatic force, respectively.}
    \label{fig:gravhydroanisbeta22}
\end{figure}

\begin{figure}[H]
    \centering
    \begin{subfigure}[b]{0.45\textwidth}
        \centering
        \includegraphics[width=9cm]{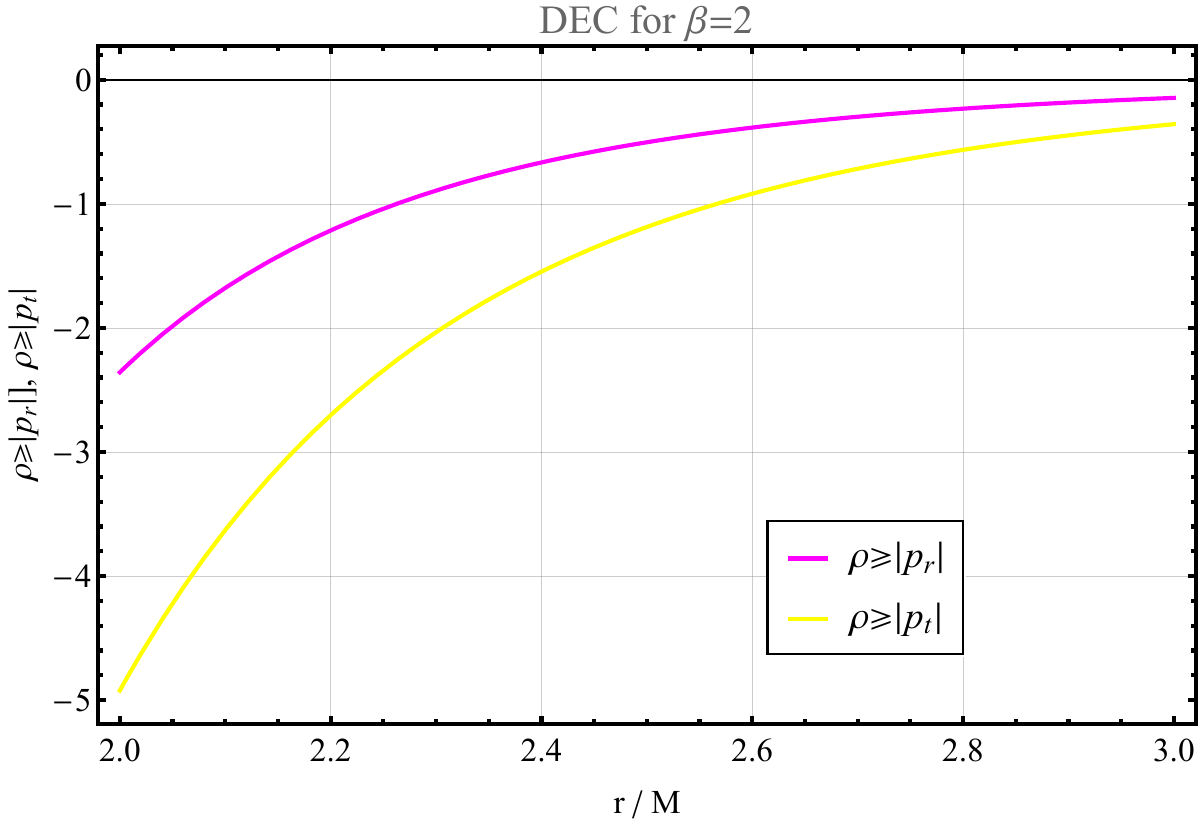}
        \caption*{(a)}
    \end{subfigure}
    \hfill
    \begin{subfigure}[b]{0.45\textwidth}
        \centering
        \includegraphics[width=9cm]{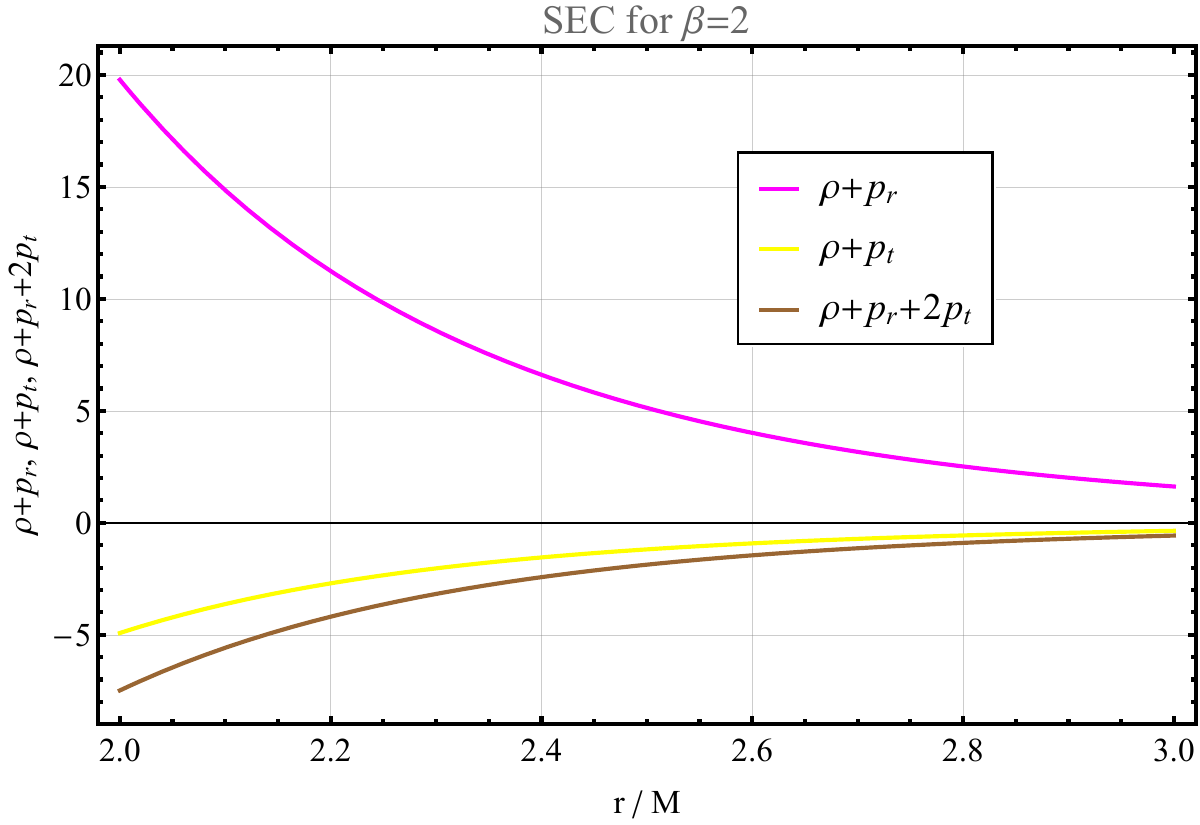}
        \caption*{(b)}
    \end{subfigure}

    \vspace{0.5cm}

    \begin{subfigure}[b]{0.9\textwidth}
        \centering
        \includegraphics[width=9cm]{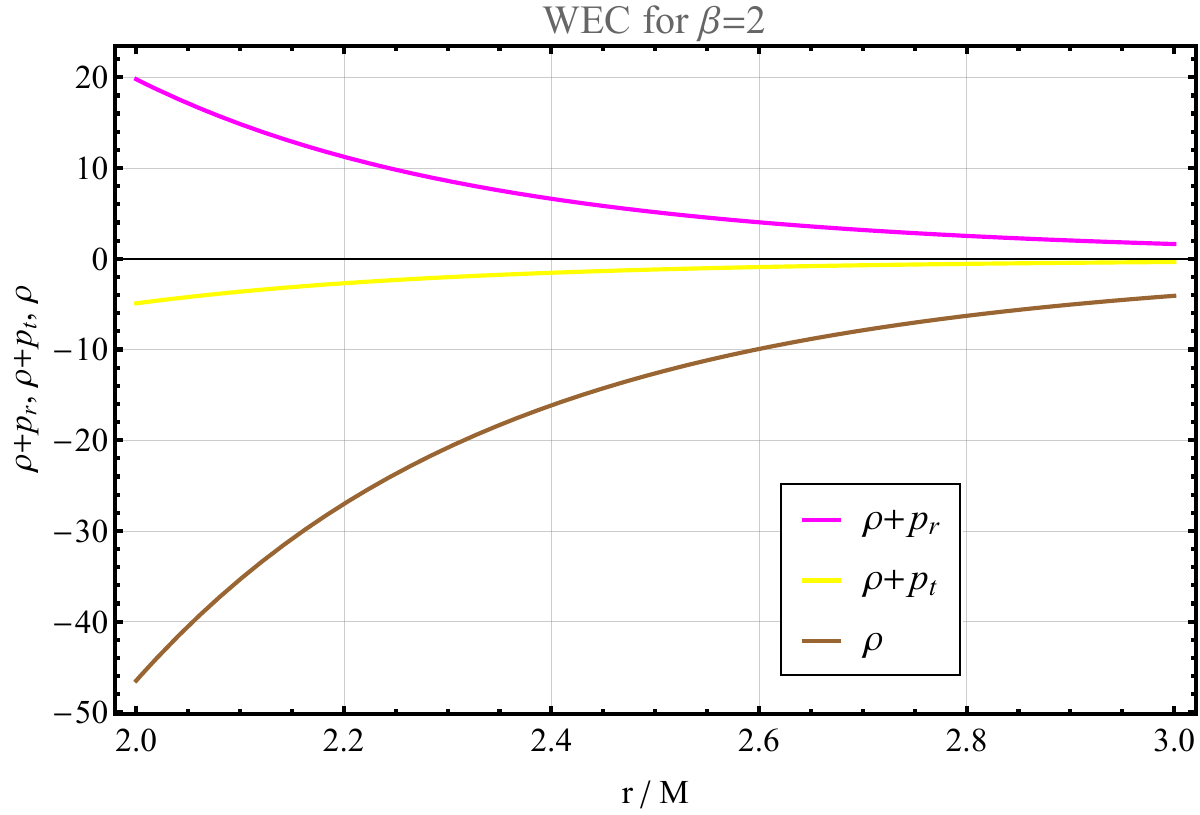}
        \caption*{(c)}
    \end{subfigure}

    \caption{Behaviour of energy conditions of the $R + a_{1}^2 R^{2} + a_{2}T$ case for $\beta=2$ and the values $r_{0}=1$, $\delta_{0}=1$, $\alpha=1$, $M=1$ are chosen.}
    \label{fig:decsecwecbeta222}
\end{figure}

\begin{figure}[H]
    \centering
    \begin{subfigure}[b]{0.45\textwidth}
        \centering
        \includegraphics[width=9cm]{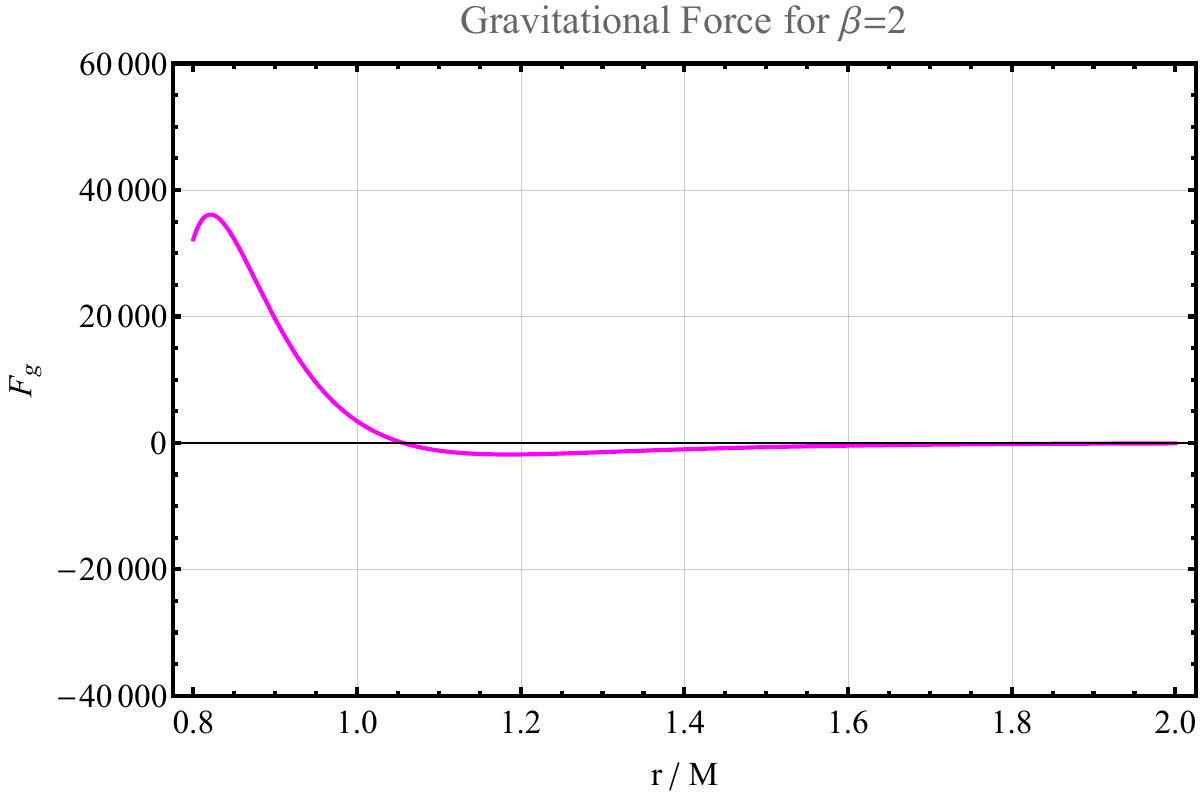}
        \caption*{(a)}
    \end{subfigure}
    \hfill
    \begin{subfigure}[b]{0.45\textwidth}
        \centering
        \includegraphics[width=9cm]{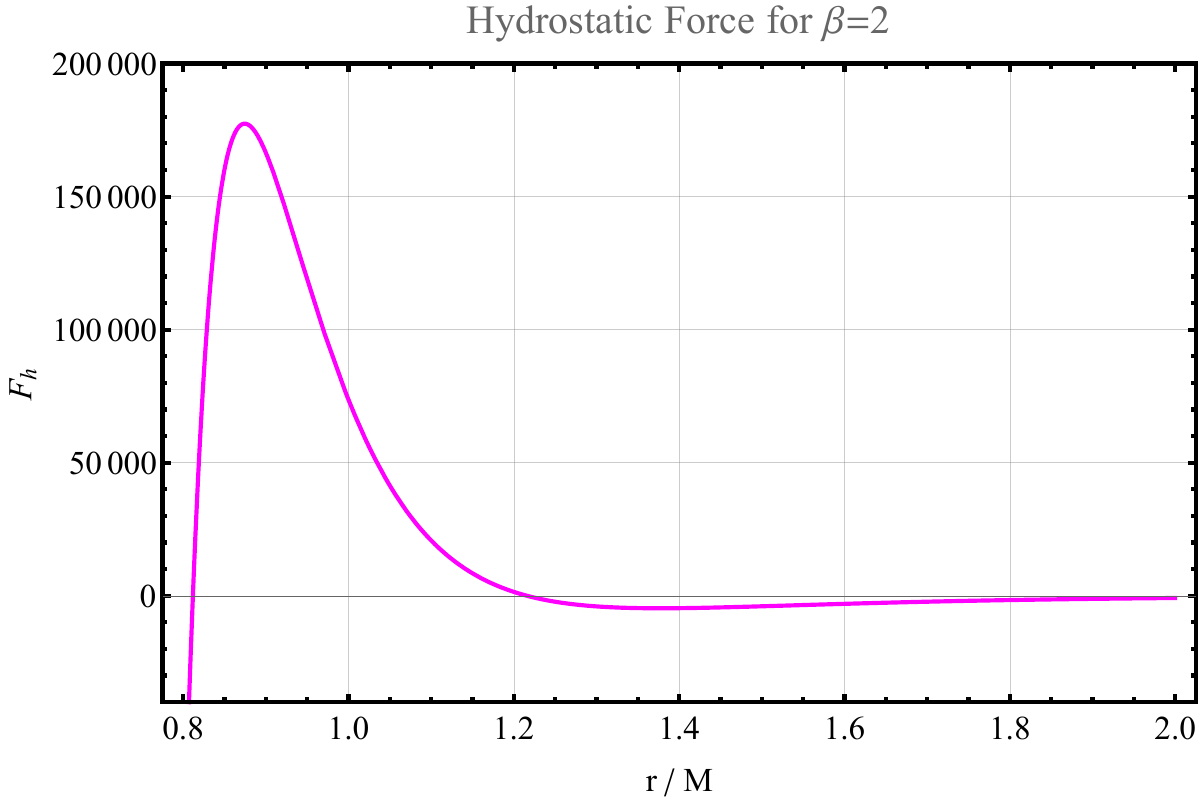}
        \caption*{(b)}
    \end{subfigure}

    \vspace{0.5cm}

    \begin{subfigure}[b]{0.45\textwidth}
        \centering
        \includegraphics[width=9cm]{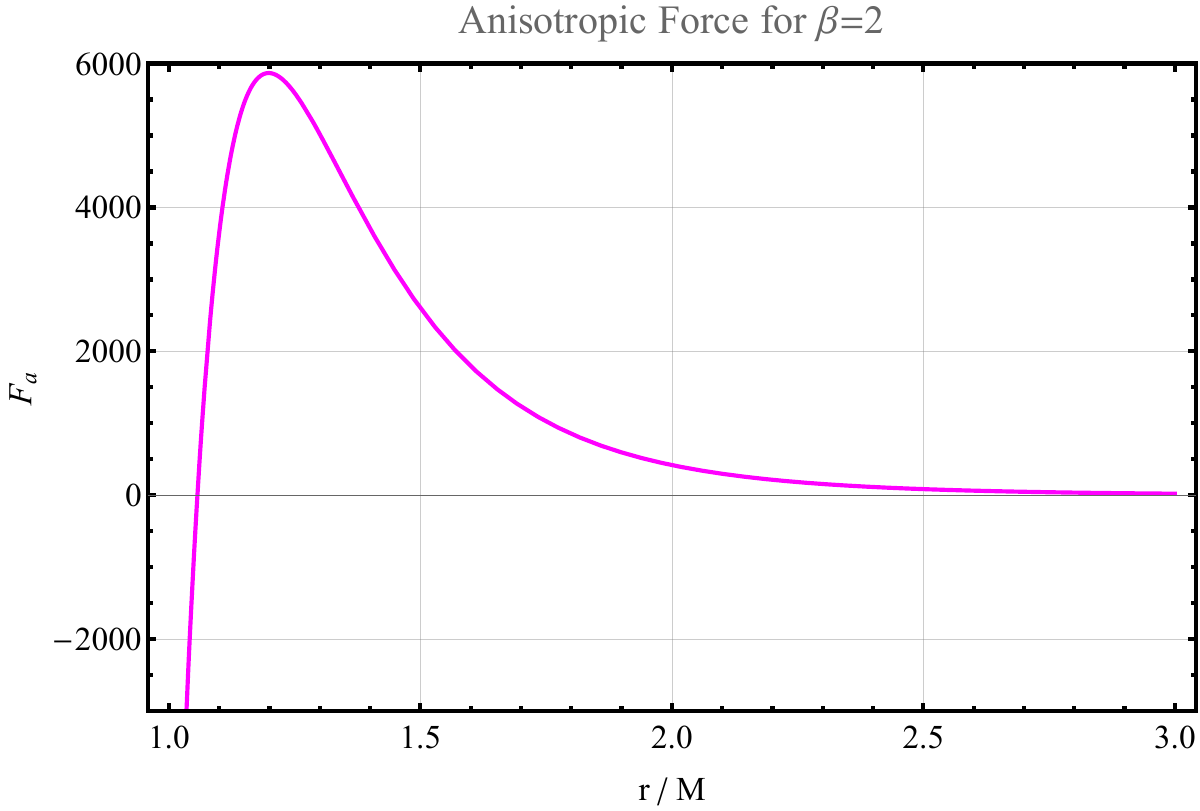}
        \caption*{(c)}
    \end{subfigure}
    \hfill
    \begin{subfigure}[b]{0.45\textwidth}
        \centering
        \includegraphics[width=9cm]{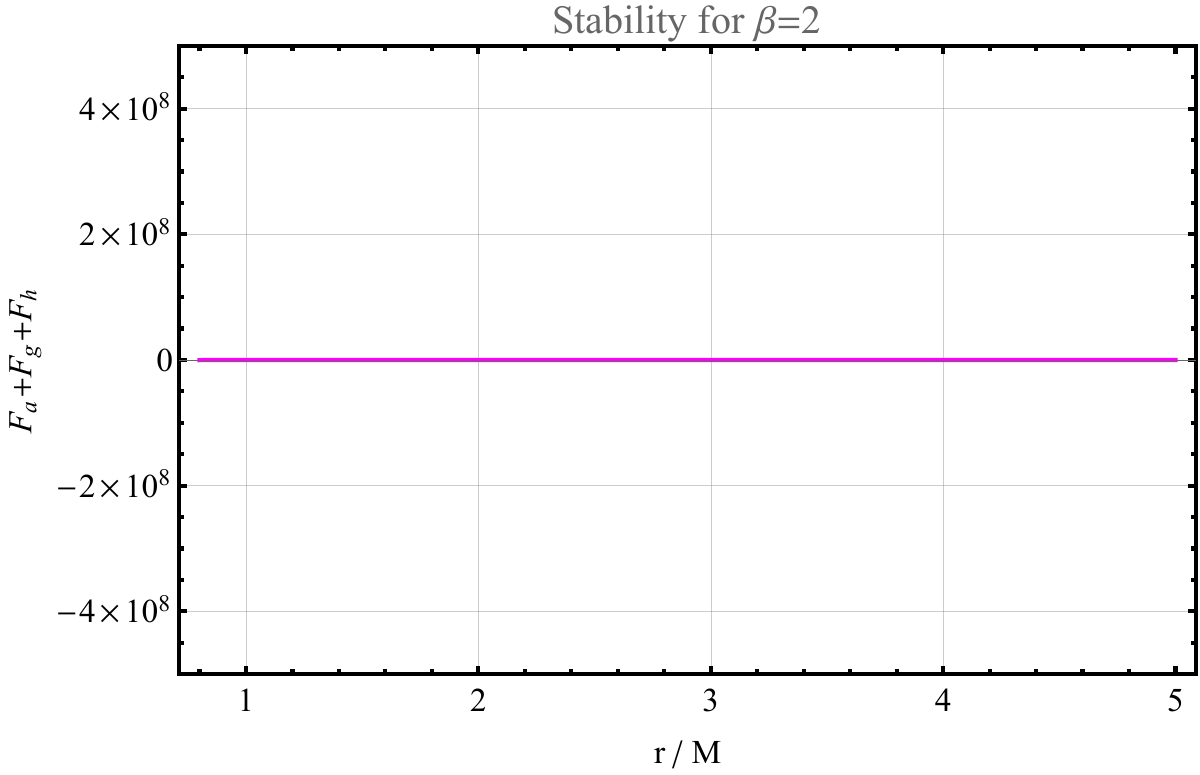}
        \caption*{(d)}
    \end{subfigure}

    \caption{Stability of the $R + a_{1}^2 R^{2} + a_{2}T$ case for $\beta=2$ and the values $r_{0}=1$, $\delta_{0}=1$, $\alpha=1$, $M=1$ are chosen. $F_{g}$, $F_{a}$ and $F_{h}$ represent gravitational force, anisotropic force and hydrostatic force, respectively.}
    \label{fig:gravhydroanisbeta222}
\end{figure}

\begin{figure}[H]
    \centering
    \begin{subfigure}[b]{0.45\textwidth}
        \centering
        \includegraphics[width=9cm]{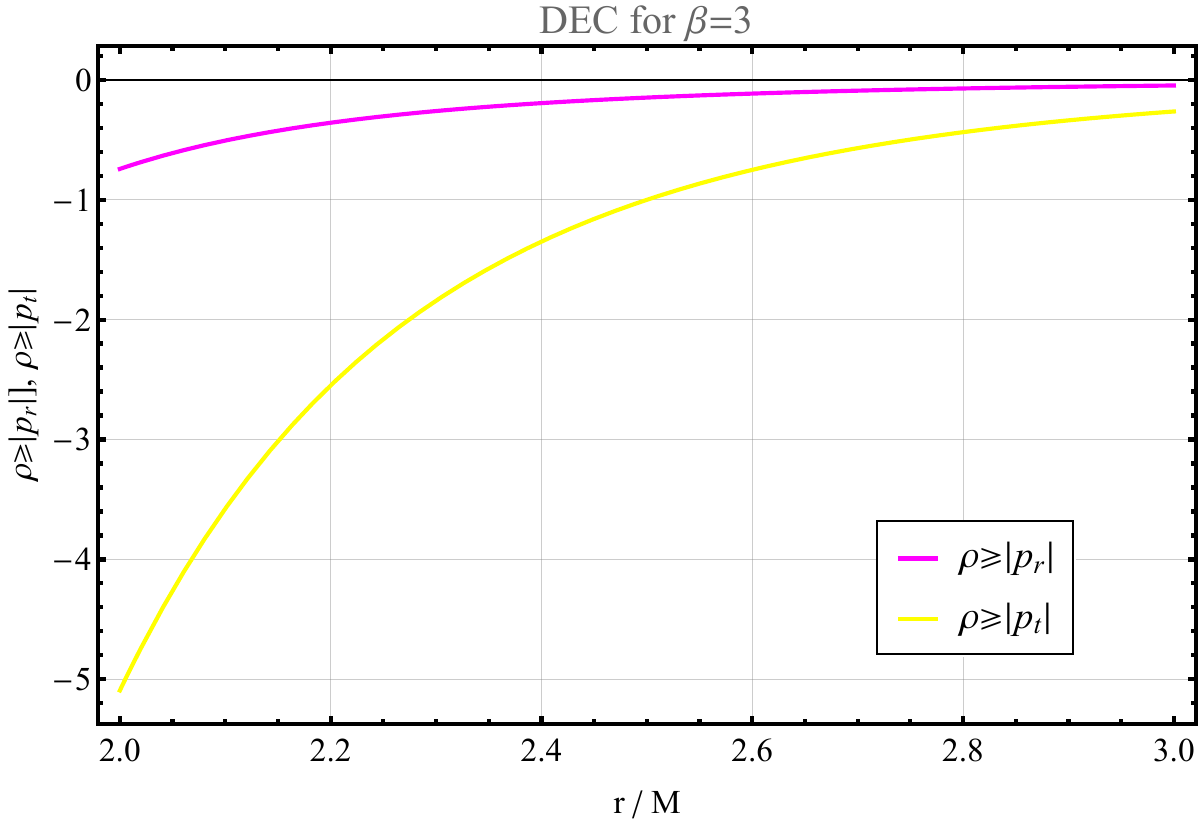}
        \caption*{(a)}
    \end{subfigure}
    \hfill
    \begin{subfigure}[b]{0.45\textwidth}
        \centering
        \includegraphics[width=9cm]{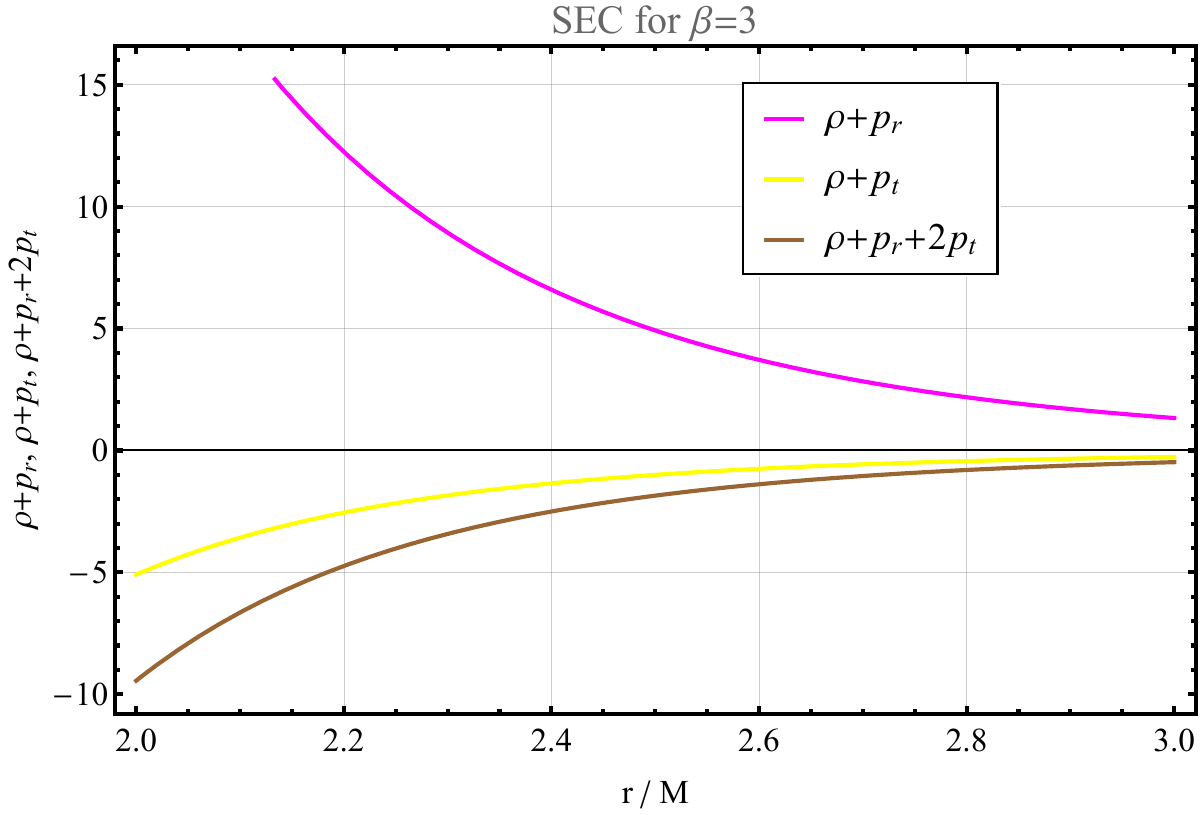}
        \caption*{(b)}
    \end{subfigure}

    \vspace{0.5cm}

    \begin{subfigure}[b]{0.9\textwidth}
        \centering
        \includegraphics[width=9cm]{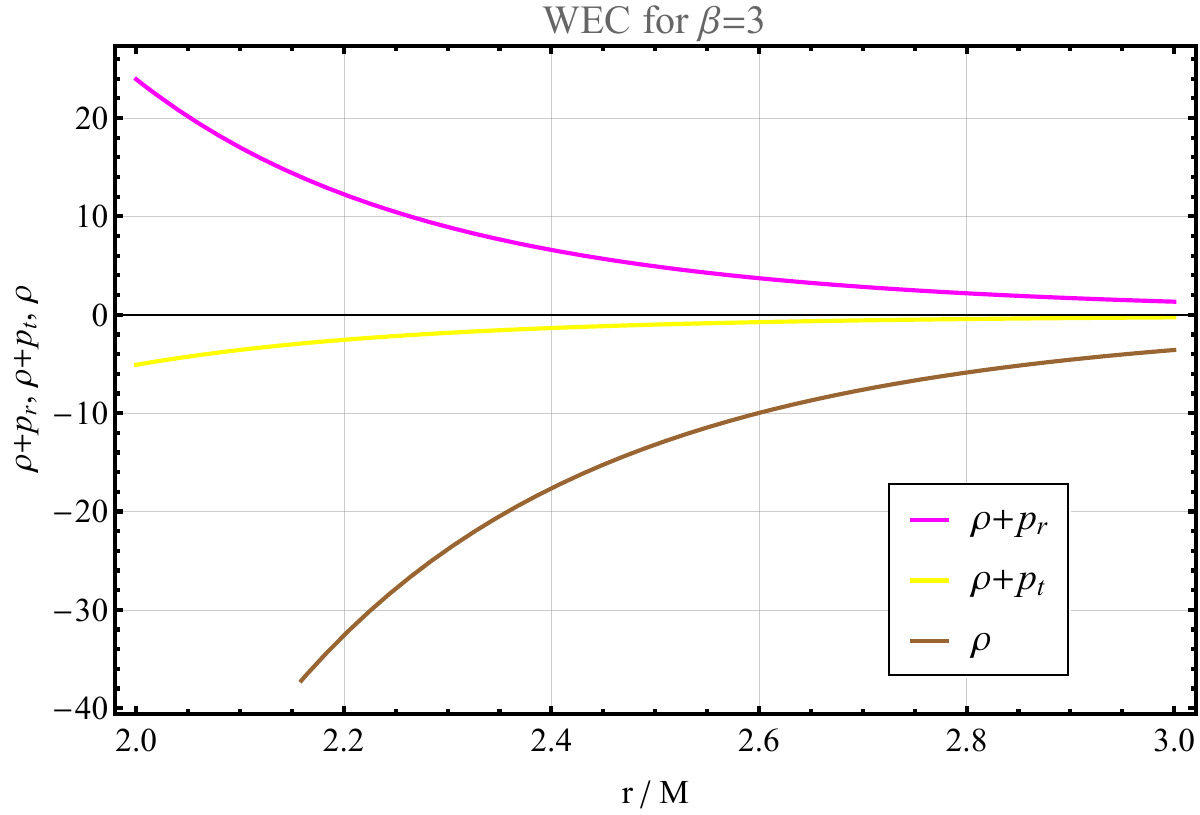}
        \caption*{(c)}
    \end{subfigure}

    \caption{Behaviour of energy conditions of the $R + a_{1}^2 R^{2} + a_{2}T$ case for $\beta=3$ and the values $r_{0}=1$, $\delta_{0}=1$, $\alpha=1$, $M=1$ are chosen.}
    \label{fig:decsecwecbeta3333}
\end{figure}

\begin{figure}[H]
    \centering
    \begin{subfigure}[b]{0.45\textwidth}
        \centering
        \includegraphics[width=9cm]{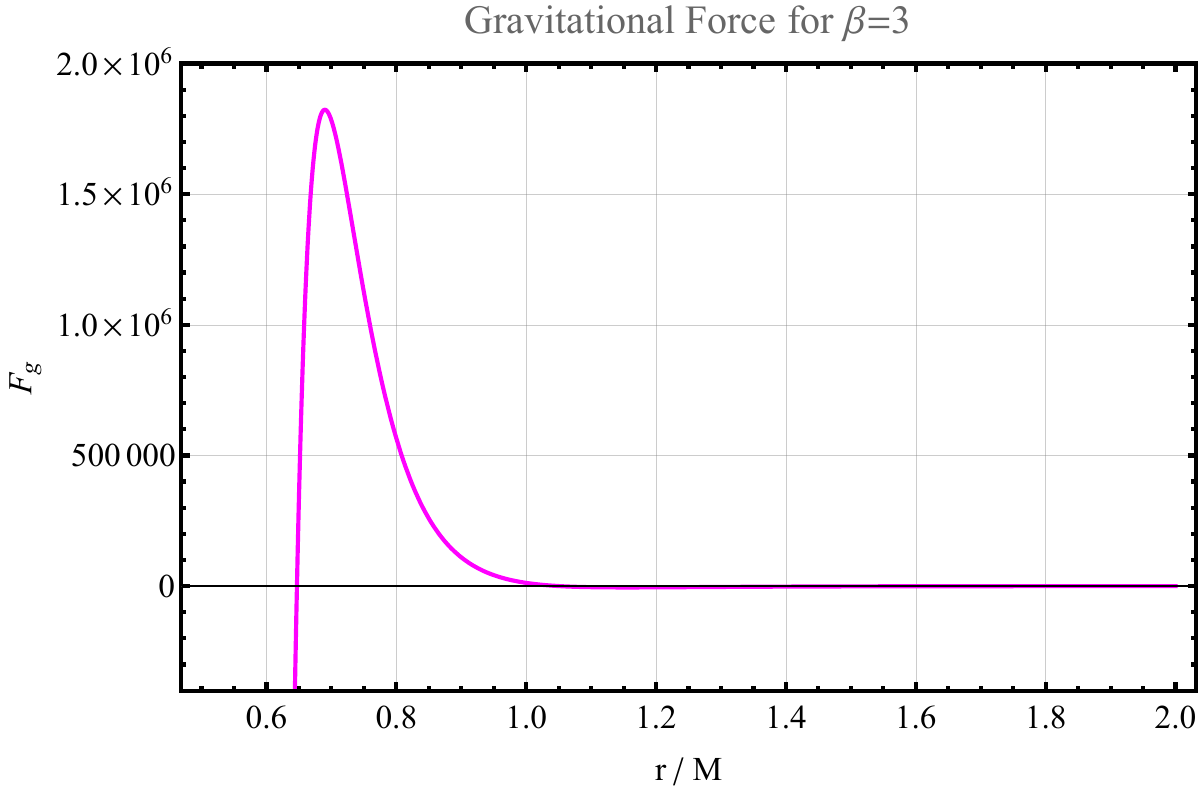}
        \caption*{(a)}
    \end{subfigure}
    \hfill
    \begin{subfigure}[b]{0.45\textwidth}
        \centering
        \includegraphics[width=9cm]{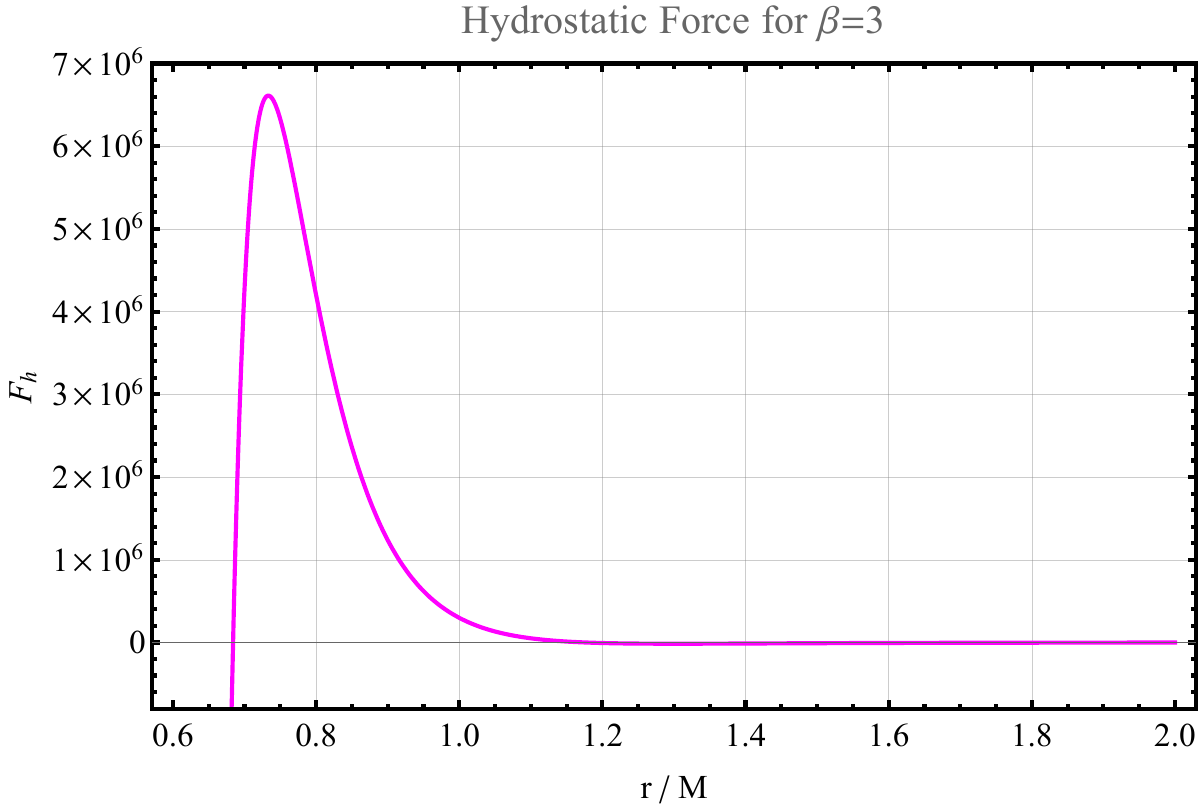}
        \caption*{(b)}
    \end{subfigure}

    \vspace{0.5cm}

    \begin{subfigure}[b]{0.45\textwidth}
        \centering
        \includegraphics[width=9cm]{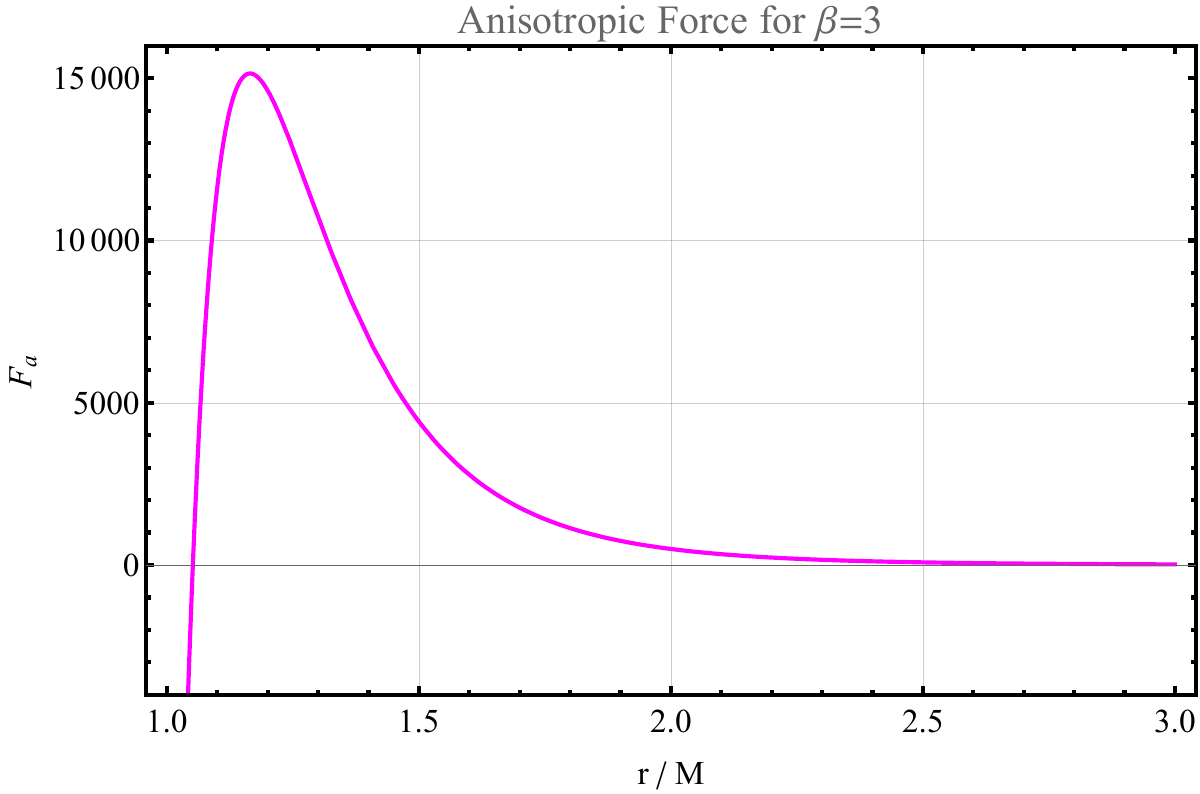}
        \caption*{(c)}
    \end{subfigure}
    \hfill
    \begin{subfigure}[b]{0.45\textwidth}
        \centering
        \includegraphics[width=9cm]{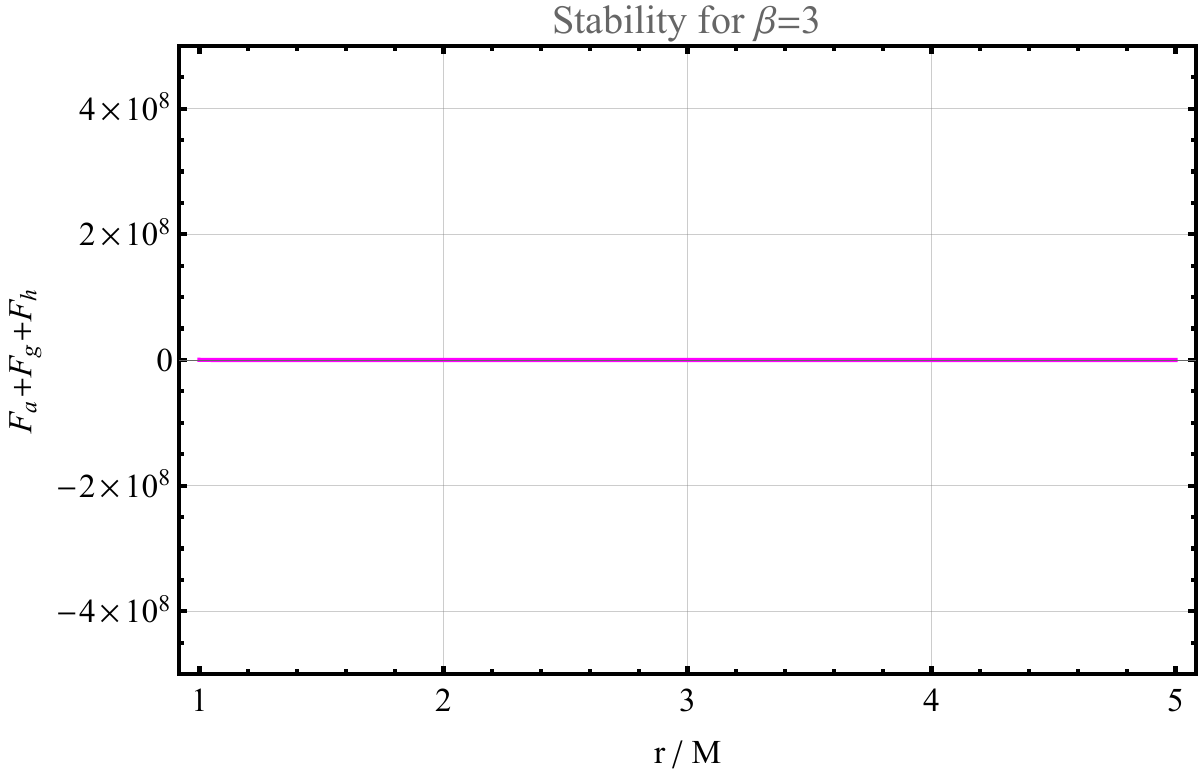}
        \caption*{(d)}
    \end{subfigure}

    \caption{Stability of the $R + a_{1}^2 R^{2} + a_{2}T$ case for $\beta=3$ and the values $r_{0}=1$, $\delta_{0}=1$, $\alpha=1$, $M=1$ are chosen. $F_{g}$, $F_{a}$ and $F_{h}$ represent gravitational force, anisotropic force and hydrostatic force, respectively.}
    \label{fig:gravhydroanisbeta3333}
\end{figure}

Just as in section \ref{subsec:sc1}, the examination of the dimensionless anisotropic parameter demonstrates that gravity possesses an attractive character, see Figure \ref{del2}. The evolution of the energy and stability conditions in the case of $\beta=1$ is presented in the figures \ref{fig:decsecwecbeta22}, \ref{fig:gravhydroanisbeta22},  in the case of $\beta=2$ is presented in the figures \ref{fig:decsecwecbeta222}, \ref{fig:gravhydroanisbeta222},  and in the case of $\beta=3$ is presented in the figures \ref{fig:decsecwecbeta3333}, \ref{fig:gravhydroanisbeta3333}.


\section{Karmarkar Condition}\label{sec:Karma}

In this study, to understand the effect of the chosen redshift function on wormhole geometry, we examine the Karmarkar condition \cite{karm} and investigate whether the presence of exotic matter is necessary under the conditions we have established. In this context,  static spherically symmetric spacetime can be written as;

\begin{equation}\label{kkm}
ds^2=-e^{\delta (r)} dt^2 +e^{\Sigma (r)} dr^2 +r^2 d\theta^2 +r^2\sin^2\theta d\phi^2.
\end{equation}

The non-zero Riemann curvature according to above space-time, equation \ref{kkm}, are;
\begin{equation}\label{kkm1}
R_{1414}=\frac{e^{\delta} (2 \delta'' + \delta'^2-\delta' \Sigma')}{4},
R_{1212}=\frac{r \Sigma'}{2}.
\end{equation}
\begin{equation}\label{kkm2}
R_{2323}=\frac{r^2 Sin^2 \theta (e^\Sigma -1)}{e^\Sigma}, 
R_{3434}=\frac{r Sin^2 \theta \Sigma' e^{\delta-\Sigma}}{2}.
\end{equation}

Karmarkar condition is given as follows;
\begin{equation}\label{kkm3}
R_{1414}=\frac{R_{1212} R_{3434} + R_{1224} R_{1334}}{R_{2323}},
\end{equation}
with $R_{2323}\neq0$. By substituting Riemann curvature in Karmarkar relation, we obtain:
\begin{equation}\label{kkm4}\nonumber
\frac{\delta' \Sigma'}{1-e^{\Sigma}}=\delta' \Sigma'-2 \delta''-\delta'^2,
\end{equation}
The solution of the above differential equation is given as
\begin{equation}\label{kkm5}
e^{\Sigma}=1+ \Gamma e^{\delta} \delta'^2.
\end{equation}
Here, $\Gamma$ is an integrating constant. By comparing equation \ref{eq:metric} and equation \ref{kkm}, we get
\begin{equation}\label{kkm6}
\Sigma(r)= Log [\frac{r}{r-b(r)}].
\end{equation}
Using equations \ref{kkm5}, \ref{eq:def_zeta} and \ref{kkm6}, we find
\begin{equation}\label{kkm7}
b(r)=r- \frac{r}{1+\Gamma e^{\delta} \delta'^2}.
\end{equation}

\begin{figure}[H]
\includegraphics[width=7cm]{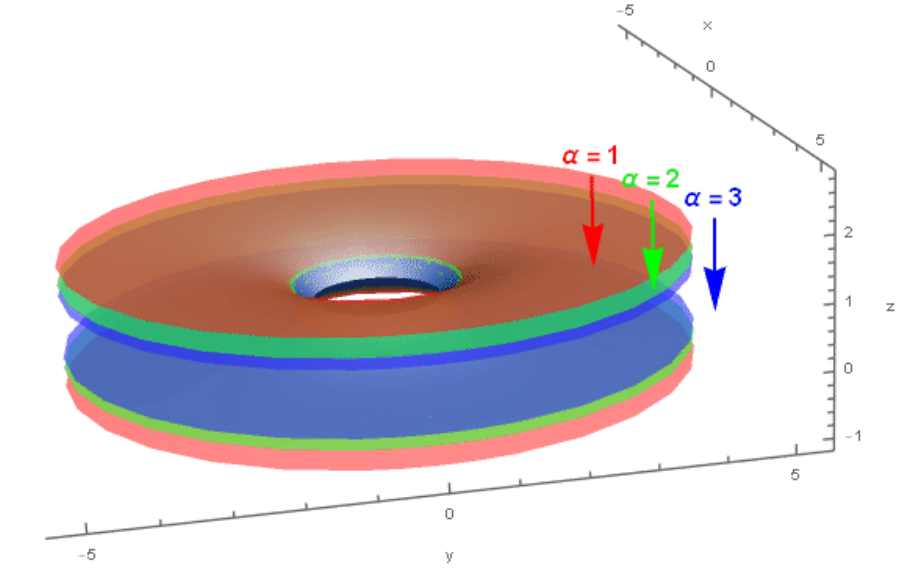}
\centering
\caption{Embedding diagram obtained by applying the Karmakar condition for different $\alpha$ values and $r_{0}=1$, $C=2$.}
\label{karmaembed1}       
\end{figure}

As discussed in Section \ref{sec:WomGeo}, when we apply the condition $b(r_0) - r_0 = 0$ to equation \ref{kkm7}, we arrive at the result $r_0 = 0$. To overcome this problem, we can add a constant $C$ to equation \ref{kkm7}. Then, using the condition $b(r_0) - r_0 = 0$, we obtain the constant as  $\Gamma= \frac{r_{0}^{2}(r_0-C)}{e^{\delta}\alpha^2 b_{0}^2}$. After substituting the value of $\Gamma$ in to equation \ref{kkm7}, we find
\begin{equation}\label{kkmson}
b(r)=r- \frac{r^{2\alpha+3}}{r^{2\alpha+2} + r_{0}^{2\alpha+2}( r_{0}-C)}.
\end{equation}
Using eq. \ref{kkmson}, we draw emmbedding diagrams, figure \ref{karmaembed1}.

\begin{figure}[H]
    \centering
    \begin{subfigure}[b]{0.45\textwidth}
        \centering
        \includegraphics[width=9cm]{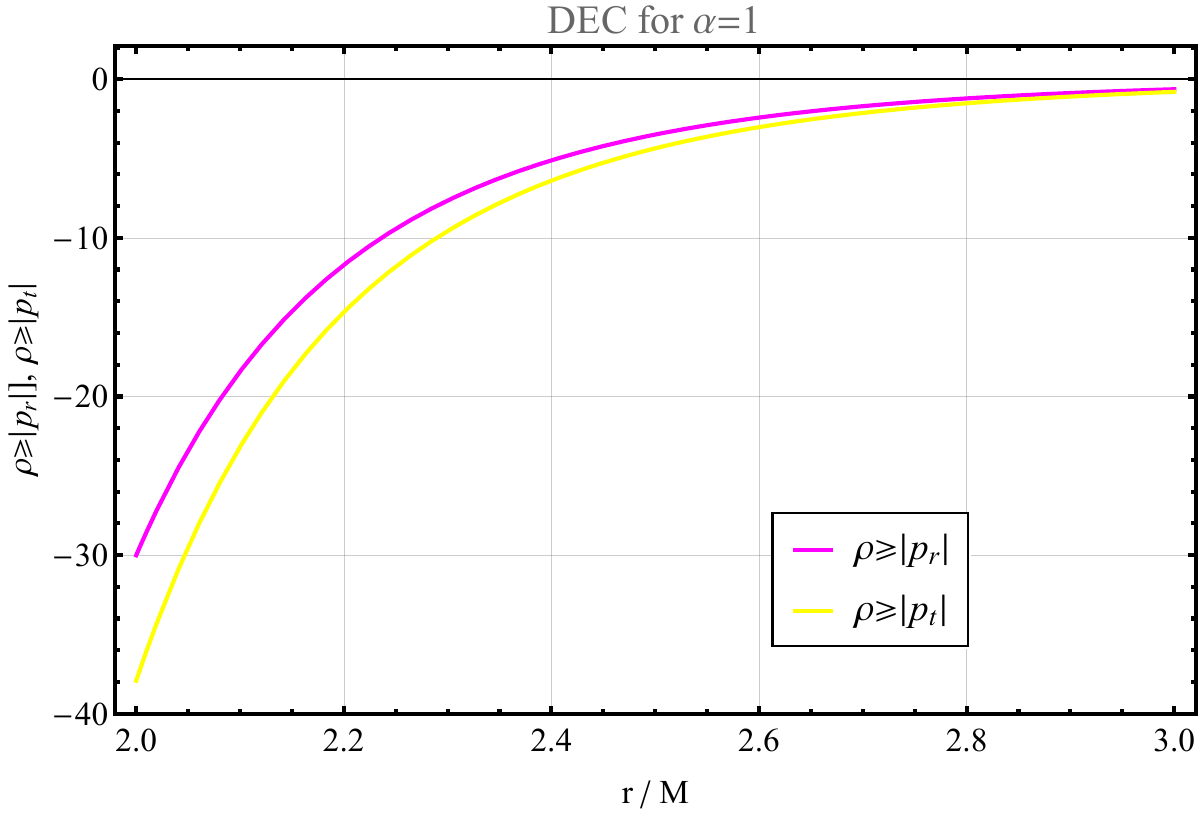}
        \caption*{(a)}
    \end{subfigure}
    \hfill
    \begin{subfigure}[b]{0.45\textwidth}
        \centering
        \includegraphics[width=9cm]{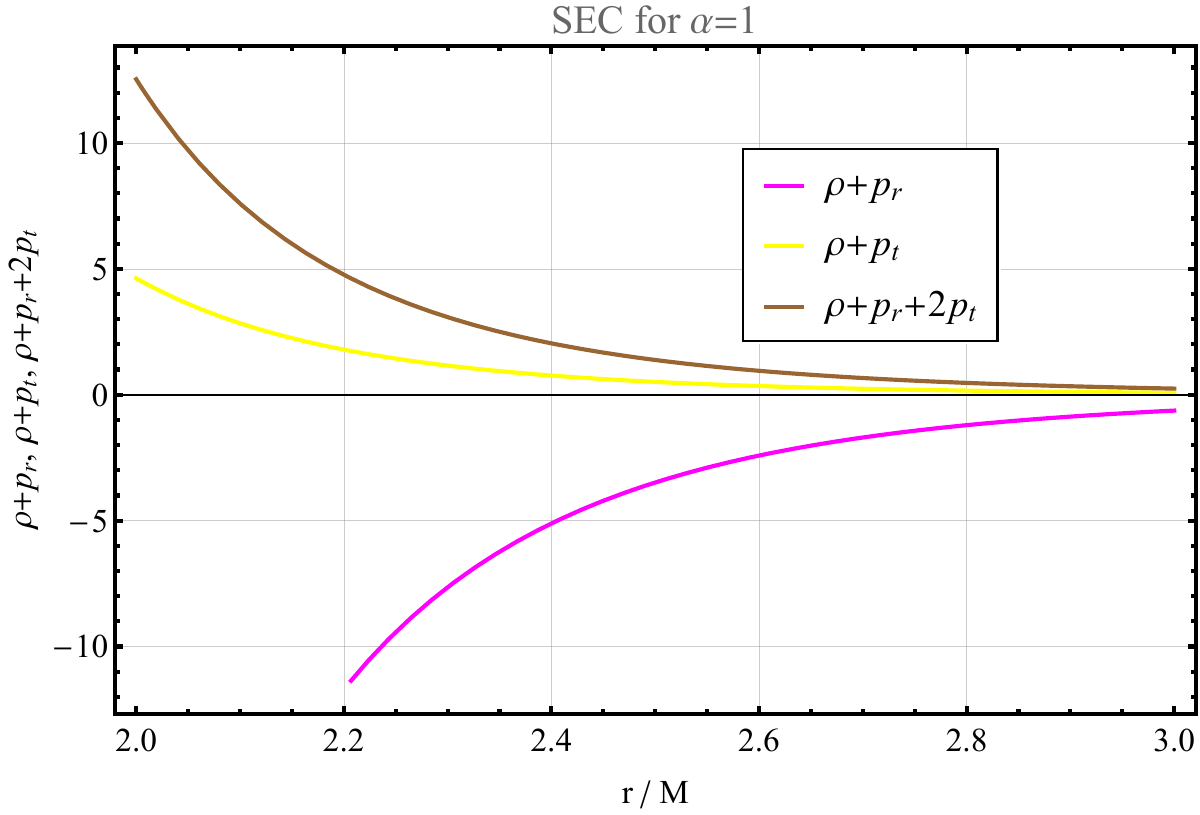}
        \caption*{(b)}
    \end{subfigure}

    \vspace{0.5cm}

    \begin{subfigure}[b]{0.9\textwidth}
        \centering
        \includegraphics[width=9cm]{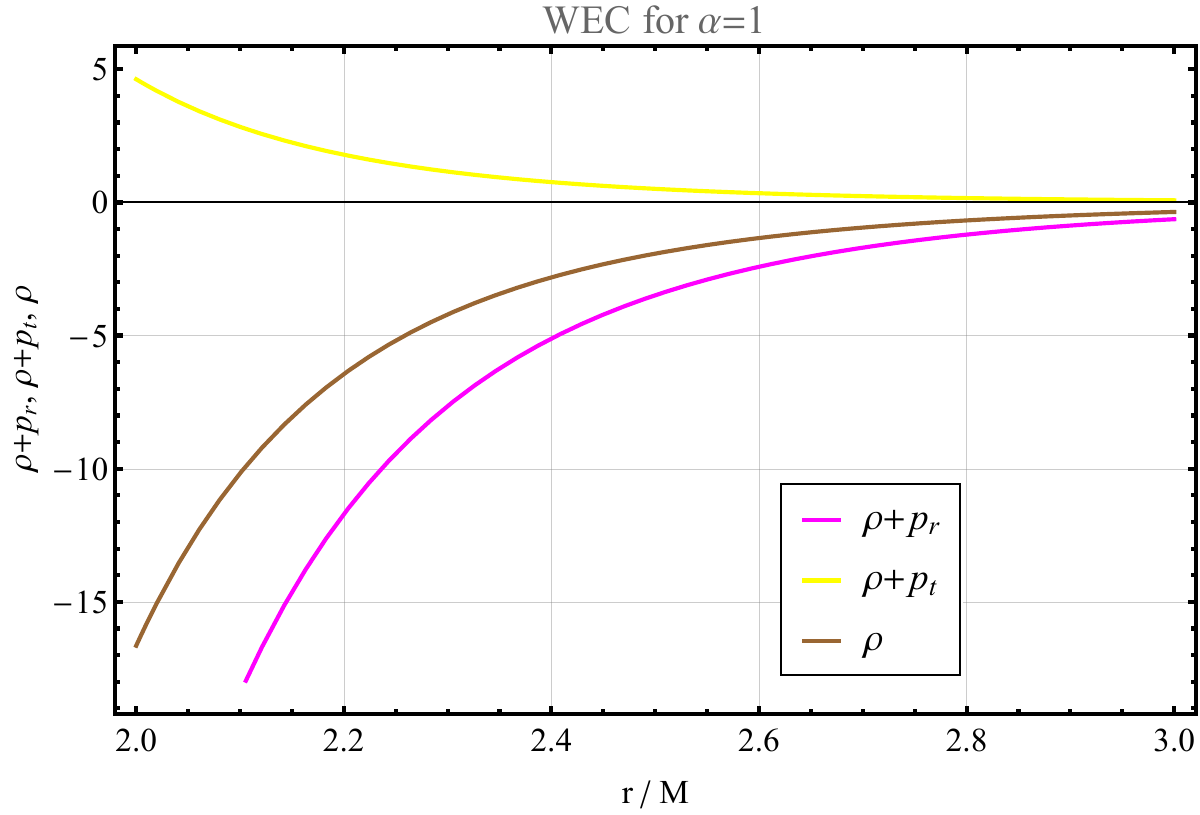}
        \caption*{(c)}
    \end{subfigure}

    \caption{Karmarkar analysis of energy conditions for $\alpha=1$. The values $r_{0}=1$, $\delta_{0}=1$, $\alpha=1$, $M=1$ are chosen.}
    \label{fig:karmakaralpha1}
\end{figure}

\begin{figure}[H]
    \centering
    \begin{subfigure}[b]{0.45\textwidth}
        \centering
        \includegraphics[width=9cm]{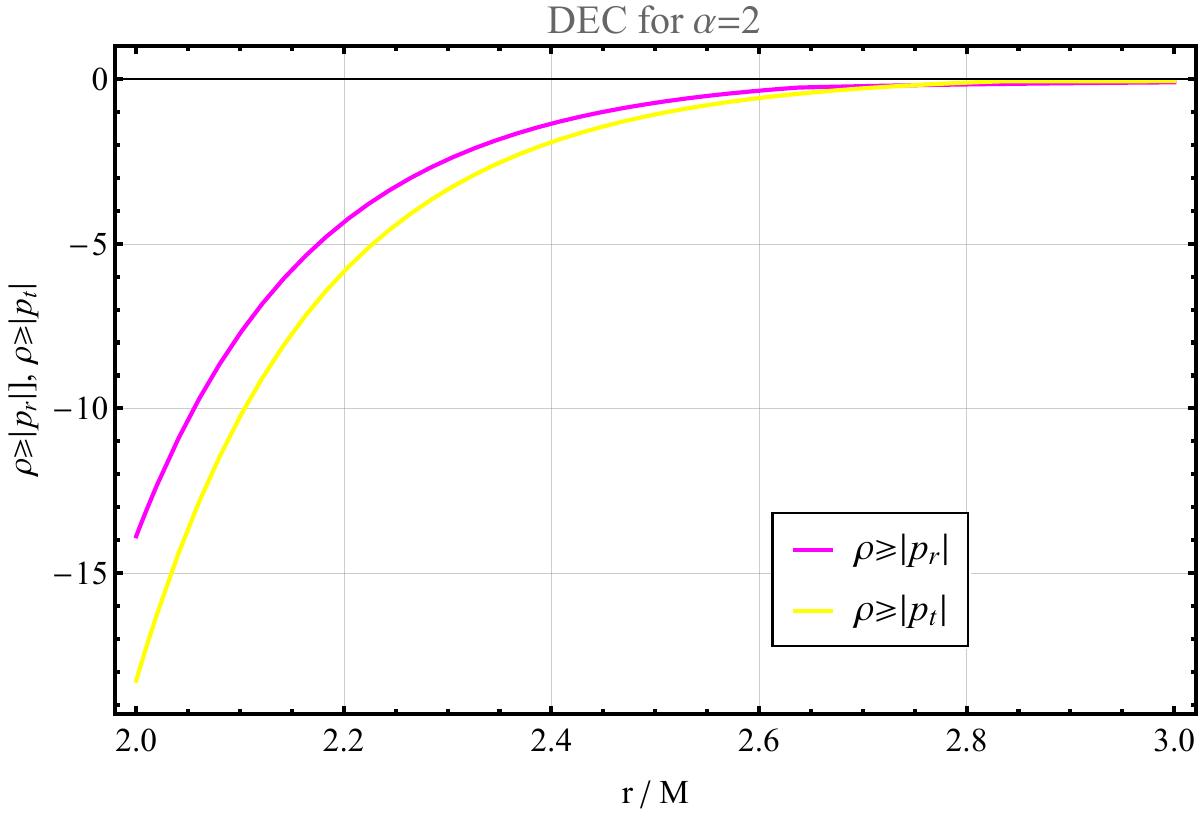}
        \caption*{(a)}
    \end{subfigure}
    \hfill
    \begin{subfigure}[b]{0.45\textwidth}
        \centering
        \includegraphics[width=9cm]{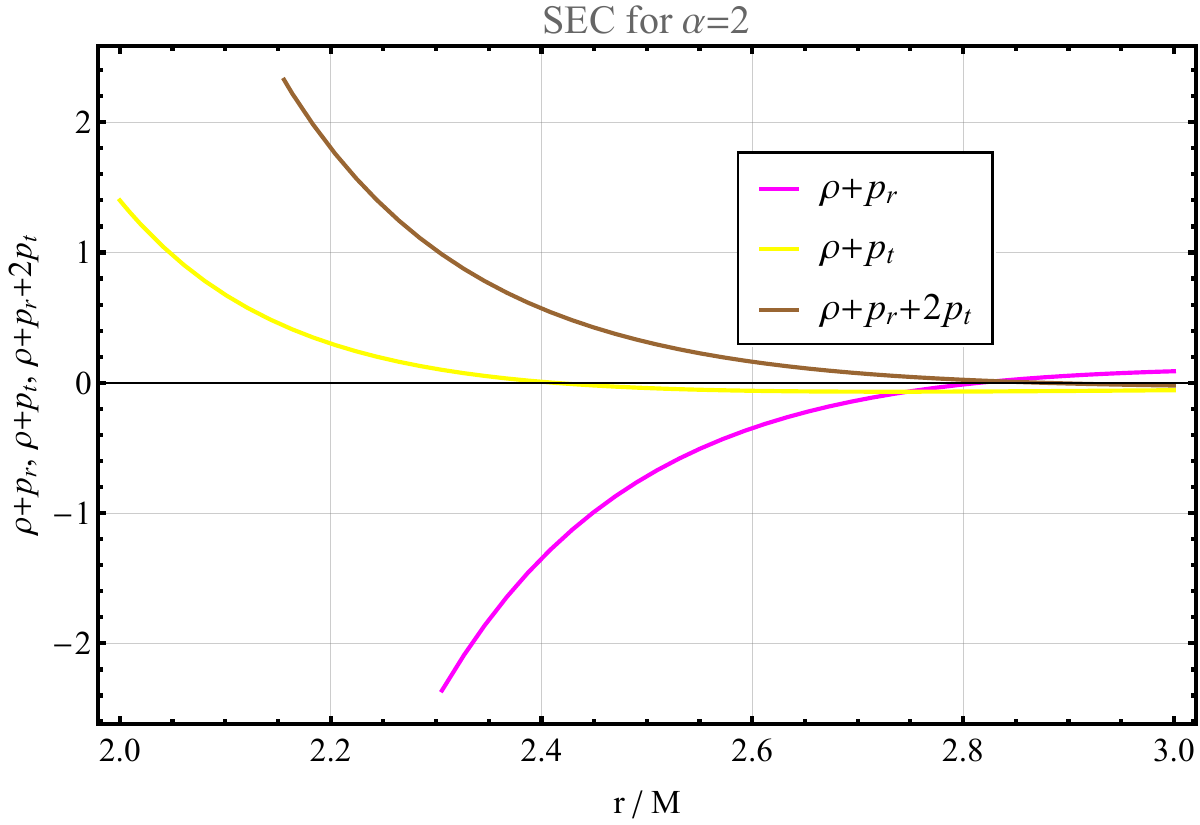}
        \caption*{(b)}
    \end{subfigure}

    \vspace{0.5cm}

    \begin{subfigure}[b]{0.9\textwidth}
        \centering
        \includegraphics[width=9cm]{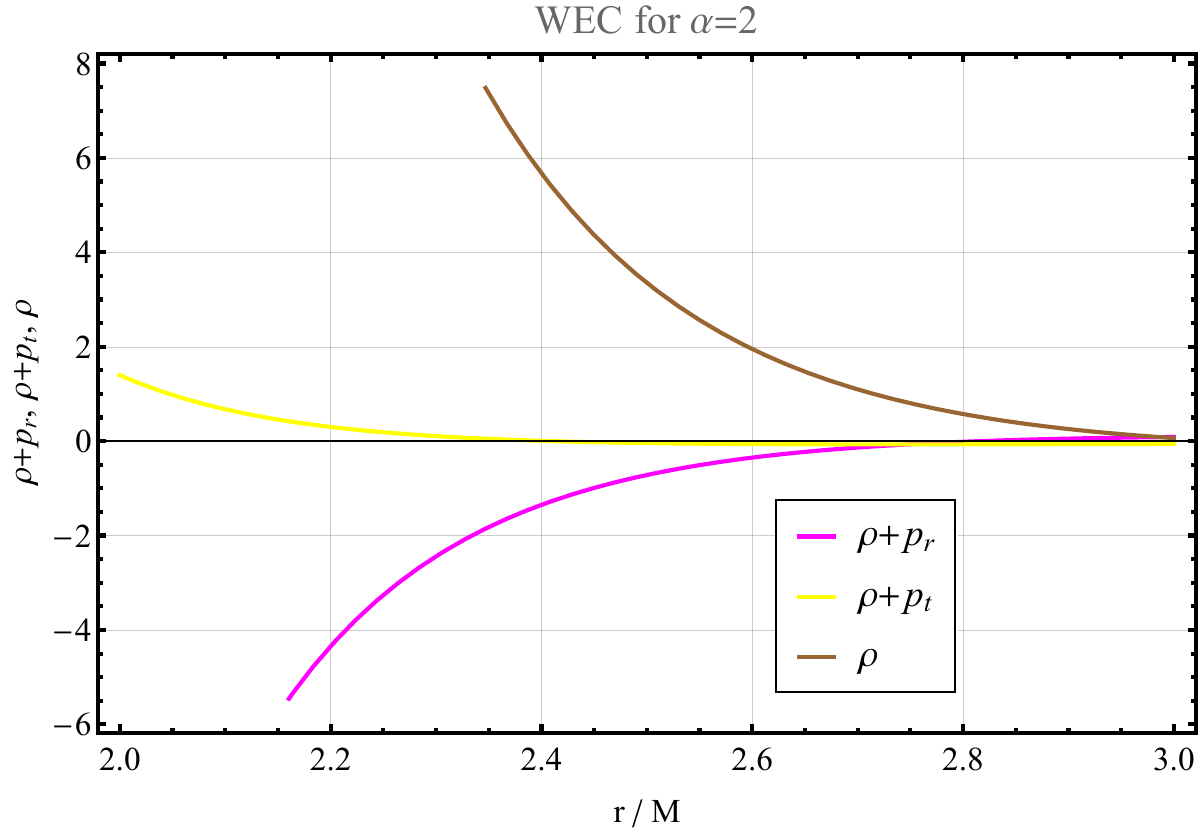}
        \caption*{(c)}
    \end{subfigure}

    \caption{Karmarkar analysis of energy conditions for $\alpha=2$. The values $r_{0}=1$, $\delta_{0}=1$, $\alpha=1$, $M=1$ are chosen.}
    \label{fig:karmakaralpha2}
\end{figure}

\begin{figure}[H]
    \centering
    \begin{subfigure}[b]{0.45\textwidth}
        \centering
        \includegraphics[width=9cm]{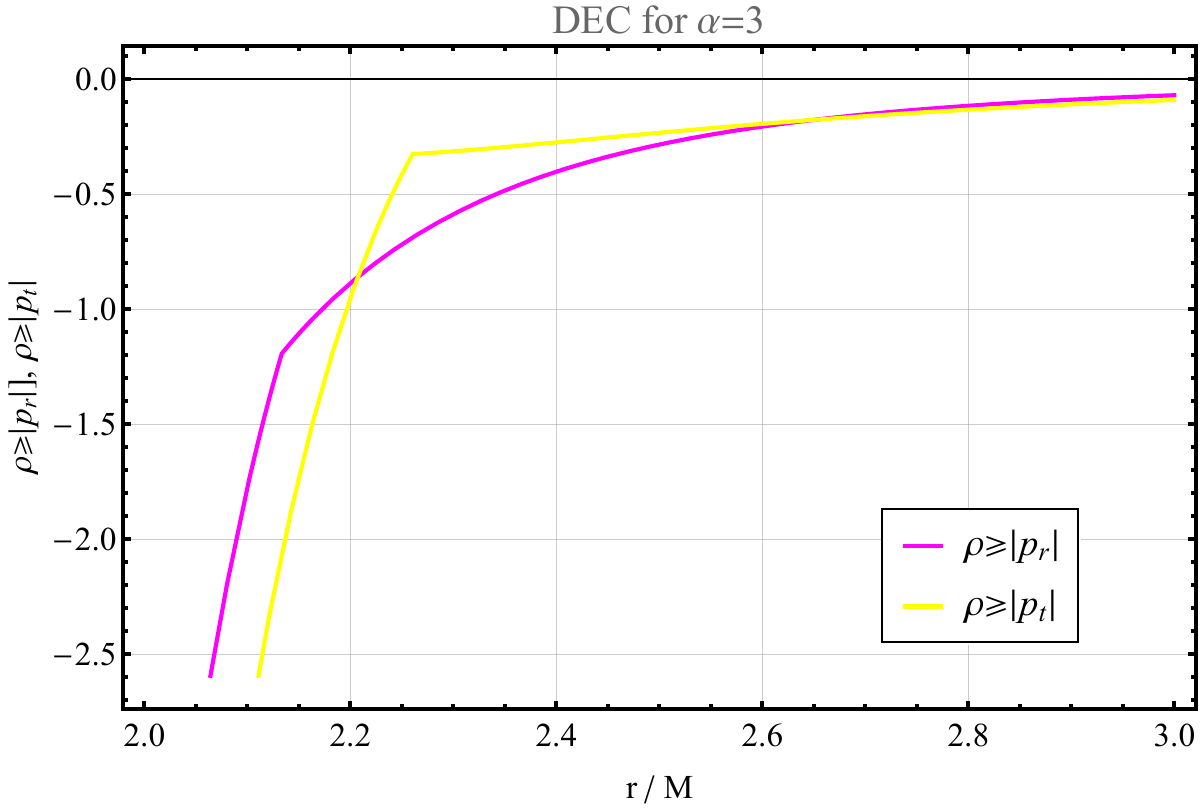}
        \caption*{(a)}
    \end{subfigure}
    \hfill
    \begin{subfigure}[b]{0.45\textwidth}
        \centering
        \includegraphics[width=9cm]{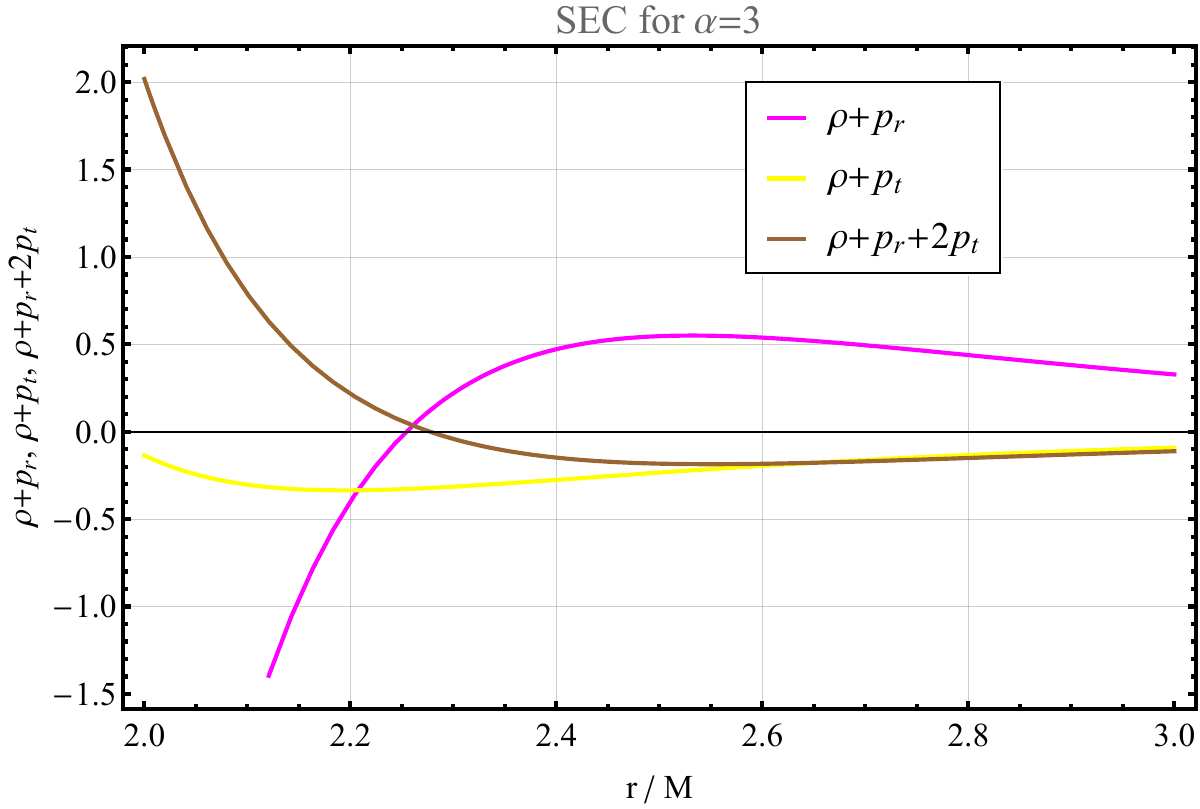}
        \caption*{(b)}
    \end{subfigure}

    \vspace{0.5cm}

    \begin{subfigure}[b]{0.9\textwidth}
        \centering
        \includegraphics[width=9cm]{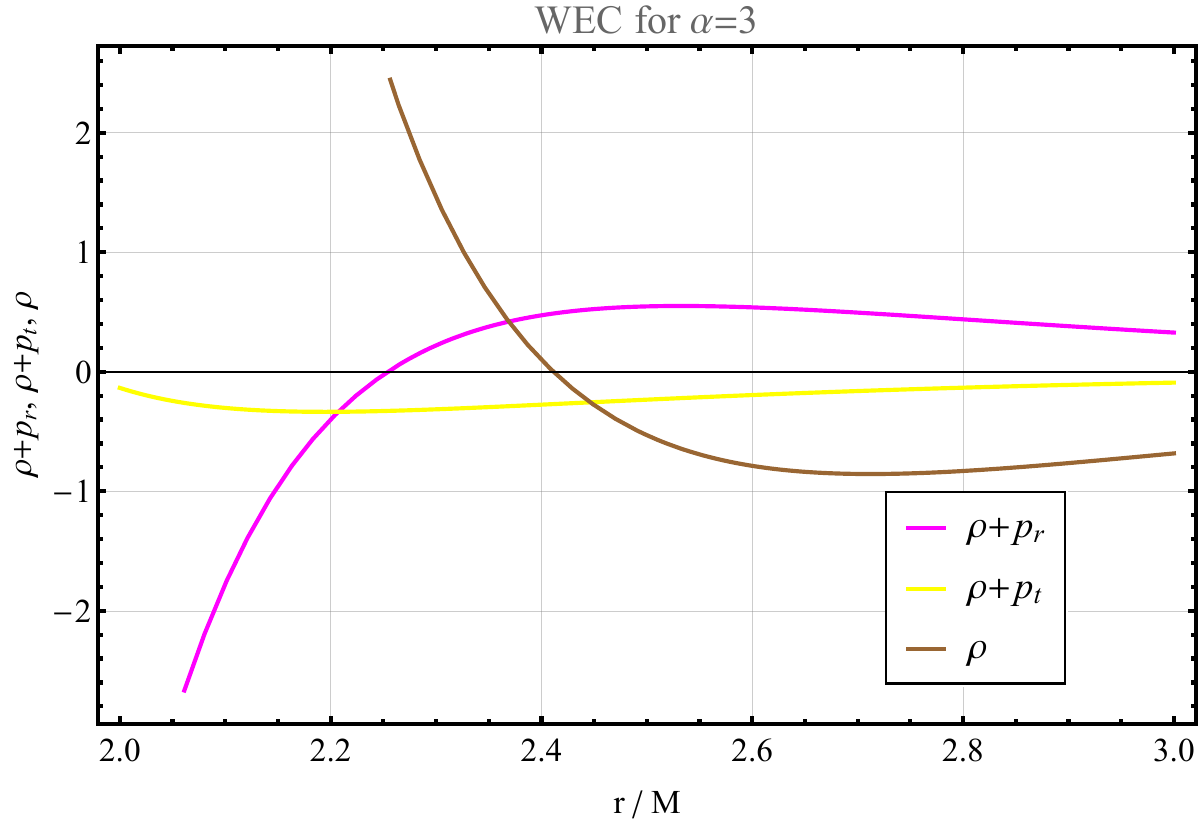}
        \caption*{(c)}
    \end{subfigure}

    \caption{Karmarkar analysis of energy conditions for $\alpha=3$. The values $r_{0}=1$, $\delta_{0}=1$, $\alpha=3$, $M=1$ are chosen.}
    \label{fig:karmakaralpha3}
\end{figure}

Also in this study, based on the solutions of the TOV equations in our two $f(R,T)$ models, we analyzed the geometry and stability of wormholes during the late evolution of the universe and concluded that the wormholes are static and stable. However, in the $f(R,T)$ model we examined in Section \ref{subsec:sc2}, due to the partial preservation of the energy conditions, the presence of exotic matter is required. We aim to investigate whether the application of the Karmarkar condition to our model in section \ref{subsec:sc2} eliminates the need for exotic matter. In this context, we examine the evolution of the energy conditions for different values of $\alpha$. The results obtained are shown in fig. \ref{fig:karmakaralpha1}, fig. \ref{fig:karmakaralpha2}, and fig. \ref{fig:karmakaralpha3} (the energy conditions are examined  using as  $a_{1}=9\pi$ and $a_{2}=-9\pi$, respectively and $C=2$)

\section{Conclusion}\label{sec:conc}

In this study, we have thoroughly analyzed wormhole solutions using two well-known models \cite{wu2018} that generate the accelerating expansion of the universe. The dimensionless anisotropy parameter, $\Delta$, in the wormhole model we analyzed in the presence of anisotropic matter takes negative values in both the $R-a_{1}^2/R+a_{2}T$ case $1$, section \ref{subsec:sc1}, and the $R+a_{1}^2R^{2}+a_{2}T$ case $2$,  section \ref{subsec:sc2}. This is particularly important as it indicates that the wormholes possess an attractive gravitational potential. When examining the energy conditions for case $1$, as shown in figures \ref{fig:decsecwecbeta1}, \ref{fig:decsecwecbeta2}, and \ref{fig:decsecwecbeta3}, it is observed that the WEC, SEC, and Null energy conditions are satisfied, while the DEC is satisfied in the tangential pressure direction. Moreover, we analyzed the forces acting on the wormhole model and the evolution of the TOV equation for various $\beta$ values in figures \ref{fig:gravhydroanisbeta1}, \ref{fig:gravhydroanisbeta2}, and \ref{fig:gravhydroanisbeta3}. As $\beta$ values increase, the gravitational force loses its effectiveness at the throat radius $r_{0}=1$, but it becomes effective again at larger radii. Additionally, as $\beta$ values increase, the hydrostatic force loses its effectiveness near the throat radius and as one moves away from the throat. However, for increasing $\beta$ values, the anisotropic force, though not effective near the throat radius, takes on negative values as one moves away from the throat. According to the TOV equation, our wormhole model maintains its stable structure in all three cases, particularly around the throat region, but moves away from the static stable state as $\beta$ values increase. This is thought to be due to the behavior exhibited by the anisotropic and hydrostatic forces at larger radii and with increasing $\beta$ values.

When examining the energy conditions for the case $2$, as shown in figures \ref{fig:decsecwecbeta22}, \ref{fig:decsecwecbeta222}, and \ref{fig:decsecwecbeta3333}, it is observed that the DEC condition is violated, while the SEC, WEC, and Null energy conditions are partially maintained. Furthermore, the evolution of the effective forces according to the TOV equation for increasing $\beta$ values is examined in figures \ref{fig:gravhydroanisbeta22}, \ref{fig:gravhydroanisbeta222}, and \ref{fig:gravhydroanisbeta3333}. Around the throat radius for $\beta=1$, the gravitational force behaves attractively, but as the distance from the throat increases, it exhibits a repulsive character. Additionally, as the $\beta$ value increases, the fluctuations in the hydrostatic and anisotropic forces become more stable. In all three $\beta$ values, our wormhole model satisfies the TOV equation across all radii. It is also observed that the evolution of the energy density, $\rho$, is highly dependent on the value of $\beta$. As the $\beta$ value increases, the energy density becomes negative (The evolution of energy density, $\rho$, is shown in the figures \ref{fig:decsecwecbeta22}, \ref{fig:decsecwecbeta222}, and \ref{fig:decsecwecbeta3333}).  This is quite an interesting situation. It is known that quantum fluctuations cause the formation of Casimir forces between two parallel conducting plates, and the magnitude of these forces increases monotonically as the distance decreases. As shown in the figure \ref{fig:embed1}, the $\beta$ values significantly alter the wormhole geometry. One possible explanation could be that the changing geometry generates Casimir forces, triggering the transition from positive energy (baryonic matter) to negative energy (dark matter, dark energy, or other exotic matter).

We can take a closer look at the Casimir effect. For this, our first step will be to define the equation of state of the system. The equation of state is defined as $P= w \rho$, where $P =\frac{1}{3}(p_{r}+2p_{t}) $. Garattini (2016) has examined the properties of Casimir wormholes \cite{garat}. In light of Garattini's work, the $w$ parameter for Casimir wormholes, denoted as $w_{casimir}$ (see equation $99$, \cite{garat}), can be defined as follows
\begin{equation}\label{casegu}
w_{casimir}=\frac{1}{3}\left( 3-2\left( \frac{9r+r_{0}}{3r+r_{0}}\right) \right).
\end{equation}
The graphs obtained for the two f(R,T) functions used in the paper are shown below.

\begin{figure}[H]
    \centering
    \begin{subfigure}[b]{0.49\textwidth}
        \centering
        \includegraphics[width=\textwidth]{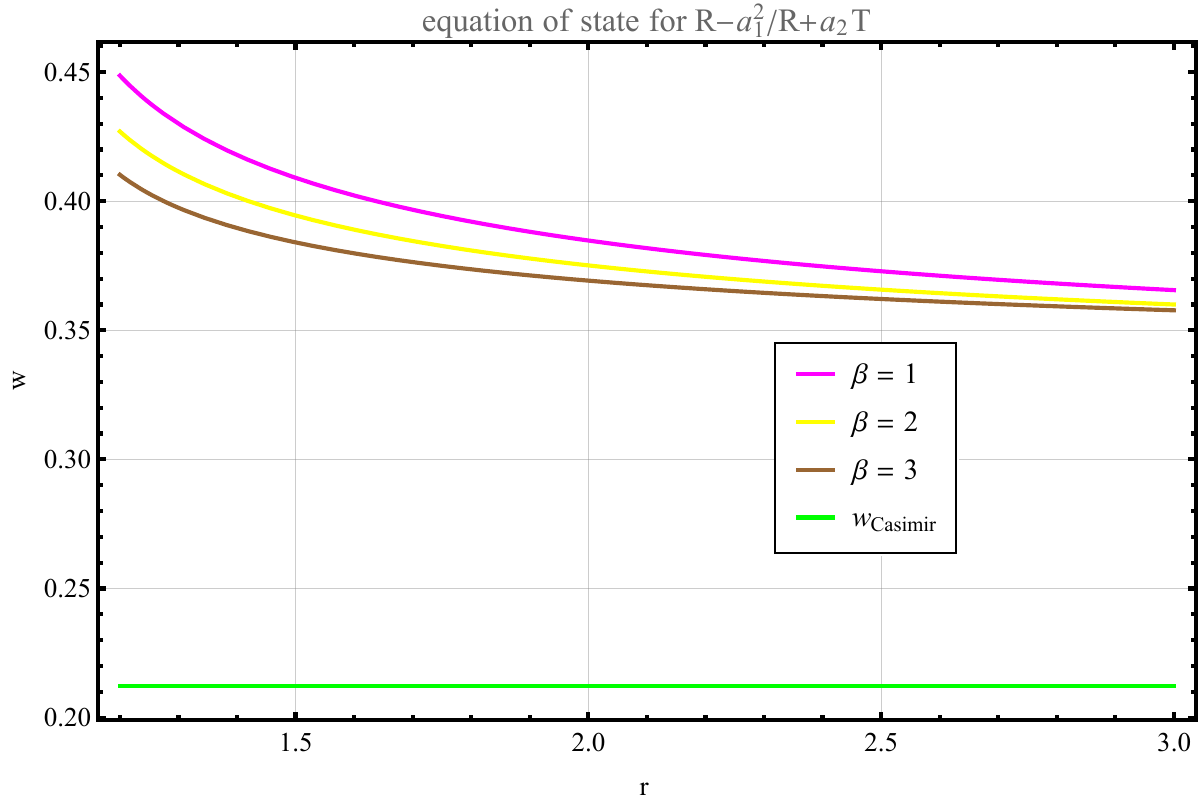}
        \caption*{(a)}
    \end{subfigure}
    \hfill
    \begin{subfigure}[b]{0.49\textwidth}
        \centering
        \includegraphics[width=\textwidth]{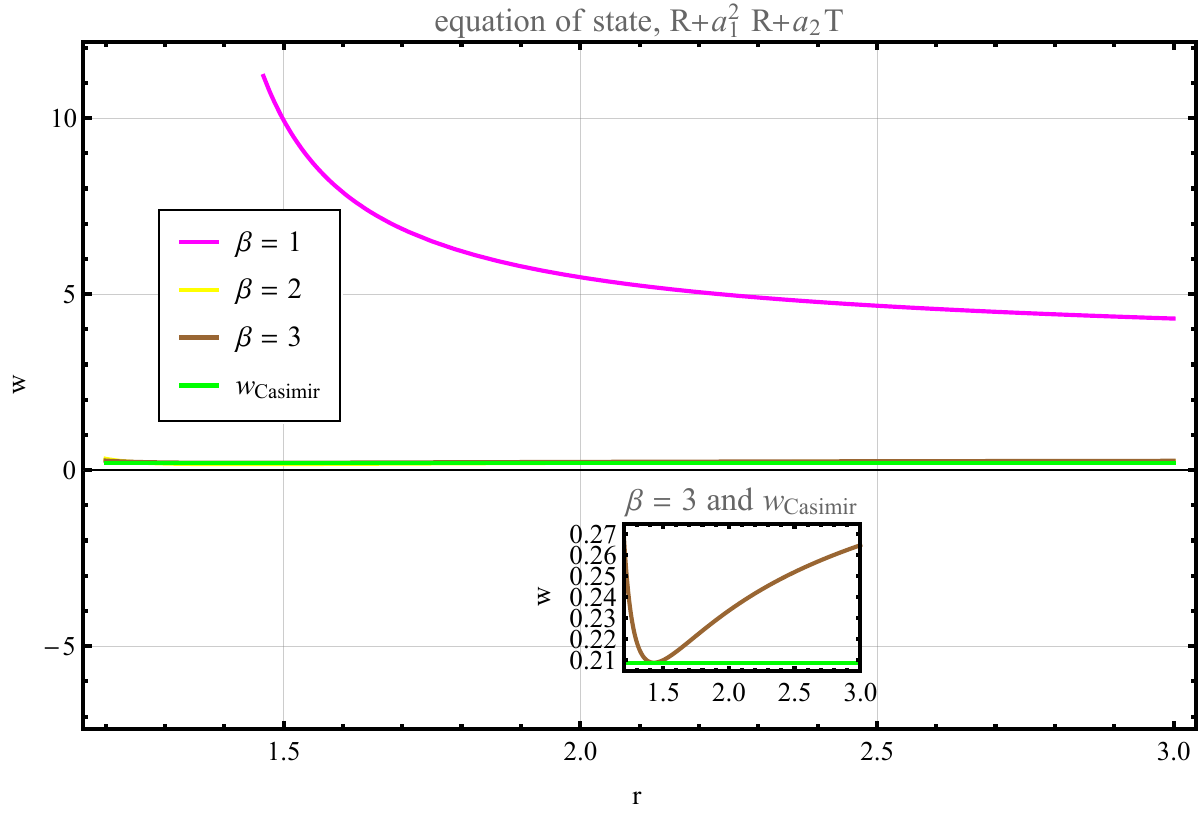}
        \caption*{(b)}
    \end{subfigure}

    \caption{The relationship between the parameter $w$, obtained for different values of $\beta$, and the Casimir wormhole parameter $w_{casimir}$ is shown by $R-a_{1}^2/R+a_{2}T$ (left graph) and $R+a_{1}^2R^{2}+a_{2}T$ (right graph). As can be seen, the $w$ parameter for $R+a_{1}^2R^{2}+a_{2}T$ is in strong agreement with the Casimir wormhole parameter $w_{casimir}$ for a specific value of the $\beta$ parameter ($\beta=3$).}
    \label{fig:casi12}
\end{figure}

For the first function (upper part of figure \ref{fig:casi12}), we observed in our previous analyses that $\rho>0$. Examining the upper part of the graph, also we see that $w > 0$. For this condition to be satisfied, $ P > 0$ must hold. This condition suggests the possibility of warm dark matter. For the second function (right side of figure \ref{fig:casi12}), we know that $\rho < 0$. The lower part of the figure \ref{fig:casi12} shows that the condition $w > 0$ is satisfied for this function (no solution was found in the range given for $\beta=2$ because it does not intersect with the Casimir equation of state). Examining the graph, it is understood that the equation of state parameter, $w$, obtained for $\beta=3$ exhibits behavior similar to the state parameter defined for Casimir wormhole, $w_{casimir}$. Therefore, it can be concluded that specifically for $\beta=3$ the wormhole is supported by Casimir energy around the throat radius.
Here a question immediately comes to mind: can the parameters $a_{1}$ and $a_{2}$ (or any other parameters) that satisfy each energy condition individually be chosen? Theoretically, they can be selected, but in that case, the condition for the generality of the solutions would not be fulfilled.

In Section \ref{subsec:sc2}, the cases where the energy conditions are violated have been reexamined with the Karmarkar condition. Our aim here was also to understand the role of the shape function, which we determined through the redshift function, in the violation of energy conditions. The Karmarkar analysis also predicts the presence of exotic matter for the section \ref{subsec:sc2}. In short, when we re-examine the cases where the energy conditions are violated using the Karmarkar analysis, the need for exotic matter does not disappear, see the figures \ref{fig:karmakaralpha1}, \ref{fig:karmakaralpha2} and \ref{fig:karmakaralpha3}. 

Kavya et al. \cite{Kavya} have demonstrated that the negative energy arising from the Casimir effect can lead to the creation of exotic matter on a small scale, which is necessary to stabilize wormholes, and their calculations have also provided evidence for the presence of positive energy near the throat region. The findings of Kavya et al. are consistent with the results presented in this article, which suggest that the wormhole geometry triggers a transformation from positive to negative energy through the influence of Casimir forces. Furthermore, Sahoo et al. \cite{Sahootwo} have investigated Casimir wormholes with GUP (Generalized Uncertainty Principle) correction and have shown that while almost no exotic matter is required to support a traversable wormhole in the absence of GUP correction, only a small amount of exotic matter is needed when the GUP correction is taken into account. The results obtained by Sahoo et al. have shown that the characteristic length scale introduced through GUP significantly influences the distribution of exotic matter. The assumption proposed in this study—that the variation in the radial length component also affects the matter distribution—can be evaluated on the same footing as the results of Sahoo et al. Moreover, our results are also in agreement with the study by Banerjee et al., which demonstrated that the Casimir stress energy is an ideal candidate for maintaining the stability of wormholes within the framework of $f(R,T)$ theory \cite{banerjeet} (also see, \cite{zubairm}). The study by Santos et al., which demonstrated that quantum vacuum fluctuations associated with the Yang–Mills field within hadrons can provide the appropriate negative energy, is consistent with the hypothesis put forward in the present work \cite{santosetal}.

Demonstrating that Casimir energy can play a role in the stability of wormholes supports the idea that quantum vacuum fluctuations can have a direct impact on gravitational structures. For that reason, the possibility that the wormhole geometry triggers the transition from baryonic matter to non-baryonic matter could provide a new horizon for testing grand unified theories.

\end{document}

%% file: fizik.tex
\newcommand{\bc}{\begin{center}}
\newcommand{\ec}{\end{center}}

\renewcommand{\phi}{\varphi}

%% file: Second_Submit_TASK_arxiv_ready.bbl
\begin{thebibliography}{99}

\bibitem{Flamm}
L.~Flamm, "{Beitrage zur Einsteinschen Gravitationstheorie},''
  {{\em Phys. Z.} {\bfseries 17} (1916)}.
\bibitem{ER}
A.~Einstein and N.~Rosen, "{Constructing `hair' for the three charge hole},"  {{\em Phys. Rev} {\bfseries 48} (1935) 73-77}.
\bibitem{MorrisThorne}
M.~S.~Morris and K.~S.~Thorne, "{Wormholes in spacetime and their use for interstellar travel},"  {{\em Am. J. Phys} {\bfseries 56} (1988)}.
\bibitem{Armendariz}
C.~A.~Picon, "{On a class of stable, traversable Lorentzian wormholes in classical general relativity}," \href{https://doi.org/10.1103/PhysRevD.65.104010} {{\em Phys. Rev. D} {\bfseries 65} (2002) 104010}.
\bibitem{SushkovPha}
S.~Sushkov, "{Wormholes supported by a phantom energy}," \href{https://doi.org/10.1103/PhysRevD.71.043520} {{\em Phys. Rev. D} {\bfseries 71} (2005) 043520}.
\bibitem{SNLobo}
F.~S.~N.~Lobo, "{Phantom energy traversable wormholes}," \href{https://doi.org/10.1103/PhysRevD.71.084011} {{\em Phys. Rev. D} {\bfseries 71} (2005) 084011}.
\bibitem{Zaslavskii}
O.~B.~Zaslavskii, "{ Exactly solvable model of wormhole supported by phantom energy}," \href{https://doi.org/10.1103/PhysRevD.72.061303} {{\em Phys. Rev. D} {\bfseries 72} (2005) 061303}.
\bibitem{Lobo_2006} 
F.~S.~N.~Lobo, "{Chaplygin traversable wormholes}," \href{https://doi.org/10.1103/PhysRevD.73.064028} {{\em Phys. Rev. D} {\bfseries 73} (2006)  064028}.
\bibitem{peter}
M.~Jamil, P.~K.~F.~Kuhfittig, F.~ Rahaman and S.~A.~Rakib, "{Wormholes supported by polytropic phantom energy},"  \href{https://doi.org/10.1140/epjc/s10052-010-1325-3} {{\em Eur. Phys. J. C } {\bfseries 67} (2010)  513}.
\bibitem{jamil1} 
M.~Jamil and  M.~U.~Farooq, "{Phantom Wormholes in (2+1)-dimensions}," \href{https://doi.org/10.1007/s10773-010-0263-z} {{\em  Int. J. Theor. Phys. } {\bfseries 49} (2010)  835}.
\bibitem{jamil2} 
M.~Jamil, "{Evolution of a Schwarzschild black hole in phantom-like Chaplygin gas cosmologies}," \href{https://doi.org/10.1140/epjc/s10052-009-1051-x} {{\em  Int. J. Theor. Phys. } {\bfseries 62} (2009)  609}.
\bibitem{cataldo} 
M.~Cataldo and  F.~Orellana, "{Static phantom wormholes of finite size}," \href{https://doi.org/10.1103/PhysRevD.96.064022} {{\em  Phys. Rev. D} {\bfseries 96} (2017)  064022}.
\bibitem{parsaei} 
F.~Parsaei and S.~Rastgoo, "{Asymptotically flat wormhole solutions with variable equation-of-state parameter}," \href{https://doi.org/10.1103/PhysRevD.99.104037} {{\em  Phys. Rev. D} {\bfseries 99} (2019) 104037}.
\bibitem{Kuhfittig:2018hts} 
P.~K.~F.~Kuhfittig and  V.~D.~Gladney, "{A model for dark energy based on the theory of embedding}," \href{https://doi.org/10.12988/astp.2018.8627} {{\em  Adv. Stud. Theor. Phys.} {\bfseries 12} (2018) 233}.
\bibitem{wang} 
D.~Wang,  X.~-h.~Meng, "{Traversable geometric dark energy wormholes constrained by astrophysical observations}," \href{https://doi.org/10.1140/epjc/s10052-016-4321-4} {{\em  Eur. Phys. J. C} {\bfseries 76} (2016) 484}.
\bibitem{blaz} 
J.~L.~Bl\'azquez-Salcedo, C.~Knoll and  E.~Radu, "{Traversable wormholes in Einstein-Dirac-Maxwell theory}," \href{https://doi.org/10.1103/PhysRevLett.126.101102} {{\em  Phys. Rev. Lett.} {\bfseries 126} (2021) 101102}.
\bibitem{israel} 
W.~Israel, {{\em Nuovo Cim. B} {\bfseries 144} (1966) 1}.
\bibitem{darmois} 
M.~Visser, S.~Kar and  N.~Dadhich, "{Traversable wormholes with arbitrarily small energy condition violations}" \href{https://doi.org/10.1103/PhysRevLett.90.201102} {{\em  Phys. Rev. Lett. } {\bfseries 90} (2003) 201102}.
\bibitem{camera1}
M.~La Camera, "{Wormhole solutions in the Randall-Sundrum scenario}," \href{https://doi.org/10.1016/j.physletb.2003.08.042} {{\em  Phys. Lett. B. } {\bfseries 27} (2003) 573}.
\bibitem{lobo55}
F.~S.~N.~Lobo,  M.~A.~Oliveira, "{Wormhole geometries in f(R) modified theories of gravity}," \href{https://doi.org/10.1103/PhysRevD.80.104012} {{\em  Phys. Rev. D} {\bfseries 80} (2009)  104012}.
\bibitem{capozziello1}
S.~Capozziello, T.~ Harko, T.~S.~Koivisto,  F.~S.~N.~Lobo, and  G.~J.~Olmo, "{Wormholes supported by hybrid metric-Palatini gravity}," \href{https://doi.org/10.1103/PhysRevD.86.127504} {{\em  Phys. Rev. D} {\bfseries 86} (2012) 127504}.
\bibitem{rosa1}
J.~L.~Rosa, J.~P.~S.~Lemos and F.~S.~N.~Lobo, "{Wormholes in generalized hybrid metric-Palatini gravity obeying the matter null energy condition everywhere}," \href{https://doi.org/10.1103/PhysRevD.98.064054} {{\em  Phys. Rev. D} {\bfseries 98} (2018)  064054}.
\bibitem{rosa22}
J.~L.~Rosa and P.~M.~Kull, "{Non-exotic traversable wormhole solutions in linear f(R,T) gravity}," \href{https://doi.org/10.1140/epjc/s10052-022-11135-w} {{\em   Eur. Phys. J. C} {\bfseries 82} (2022)  1154}.
\bibitem{lobo1}
M.~N.~Christiansen (Editor), T.~K.~Rasmussen (Editor), E.~Anderson (Contributor), G.~Basini (Contributor), S.~Capozziello (Contributor) \textit{Classical and Quantum Gravity Research}, (Nova Science Publishers, 2008).
\bibitem{garattini1}
R.~Garattini and F.~S.~N.~Lobo, "{Self sustained phantom wormholes in semi-classical gravity}," \href{https://doi.org/10.1088/0264-9381/24/9/016} {{\em Classical Quantum Gravity } {\bfseries 24} (2007) 2401}.
\bibitem{lobo2}
F.~S.~N.~Lobo, "{General class of wormhole geometries in conformal Weyl gravity}," \href{https://doi.org/10.1088/0264-9381/25/17/175006} {{\em Classical Quantum Gravity } {\bfseries 25} (2008) 175006}.
\bibitem{garattini2}
R.~Garattini and F.~S.~N.~Lobo, "{Self-sustained traversable wormholes in noncommutative geometry}," \href{https://doi.org/10.1016/j.physletb.2008.11.064} {{\em Physics Letters B } {\bfseries 671} (2009) 146}.
\bibitem{lobo3}
F.~S.~N.~Lobo and M.~A.~Oliveira, "{General class of vacuum Brans-Dicke wormholes}," \href{https://doi.org/10.1103/PhysRevD.81.067501} {{\em Phys. Rev. D } {\bfseries 81} (2010) 067501}.
\bibitem{garattini3}
R.~Garattini and  F.~S.~N.~Lobo, "{Self-sustained wormholes in modified dispersion relations}," \href{https://doi.org/10.1103/PhysRevD.85.024043} {{\em Phys. Rev. D } {\bfseries 85} (2012) 024043}.
\bibitem{myrzakulov1}
R.~Myrzakulov, L.~Sebastiani, S.~Vagnozzi and S.~Zerbini, "{Static spherically symmetric solutions in mimetic gravity: rotation curves and wormholes}," \href{https://doi.org/10.1088/0264-9381/33/12/125005} {{\em  Class. Quant. Grav. } {\bfseries 33} (2016) 125005}.
\bibitem{harko1}
T.~Harko, F.~S.~N.~Lobo, M.~K.~Mak and S.~V.~Sushkov, "{Modified-gravity wormholes without exotic matter}," \href{https://doi.org/10.1103/PhysRevD.87.067504} {{\em Phys. Rev. D} {\bfseries 87} (2013) 067504}.
\bibitem{Anchordoqu1997}
L.~ A.~ Anchordoqui, S.~ .P.~ Bergliaffa and D.~ F.~Torres, "{Brans-Dicke wormholes in nonvacuum spacetime}," \href{ https://doi.org/10.1103/PhysRevD.55.5226} {{\em Phys. Rev. D} {\bfseries 55} (1997) 5226}.
\bibitem{Bhawal1992}
B.~ Bhawal and S.~ Kar, "{Lorentzian wormholes in Einstein-Gauss-Bonnet theory}," \href{https://doi.org/10.1103/PhysRevD.46.2464} {{\em Phys. Rev. D} {\bfseries 46} (1992) 2464}.
\bibitem{dotti1}
J.~Oliva and R.~Troncoso, "{Static wormholes in vacuum for conformal gravity}," \href{https://doi.org/10.1142/S0217751X09044930}  {{\em Int. Jour. Mod. Phys. A} {\bfseries 24} (2009) 1528-1532}.
\bibitem{bronnikov1}
K.~A.~Bronnikov and S.~W.~Kim, "{Possible wormholes in a brane world}," \href{https://doi.org/10.1103/PhysRevD.67.064027} {{\em Phys. Rev. D} {\bfseries 67} (2003) 064027}.
\bibitem{lobo6}
F.~S.~N.~Lobo, "{General class of braneworld wormholes}", \href{https://doi.org/10.1103/PhysRevD.75.064027}  {{\em Phys. Rev. D} {\bfseries 75} (2007) 064027}.
\bibitem{agnese1}
A.~G.~Agnese and M.~La Camera, "{Wormholes in the Brans-Dicke theory of gravitation}," \href{ https://doi.org/10.1103/PhysRevD.51.2011} {{\em Phys. Rev. D} {\bfseries 51} (1995) 2011}.
\bibitem{nandi1}
K.~K.~Nandi, B.~Bhattacharjee, S~ M.~K.~Alam and J.~Evans, "{Brans-Dicke wormholes in the Jordan and Einstein frames}," \href{https://doi.org/10.1103/PhysRevD.57.823} {{\em Phys. Rev. D} {\bfseries 57} (1998) 823}.
\bibitem{benedictis} 
A.~De Benedictis and D.~Horvat, "{On wormhole throats in f (R) gravity theory}," \href{https://doi.org/10.1007/s10714-012-1412-x} {{\em Gen. Relativ. Gravit.} {\bfseries 44} (2012) 2711}.
\bibitem{eiroa} 
E.~F.~Eiroa and G.~F.~Aguirre, "{Thin-shell wormholes with charge in F(R) gravity}," \href{https://doi.org/10.1140/epjc/s10052-016-3984-1} {{\em Eur. Phys. J. C} {\bfseries 76} (2016) 132}.
\bibitem{mazhar} 
S.~H.~Mazharimousavi and M.~Halilsoy, "{Necessary Conditions for Having Wormholes in f(R) Gravity}," \href{https://doi.org/10.1142/S0217732316502035} {{\em Mod. Phys. Lett. A} {\bfseries 31} (2016) 1650192}.
\bibitem{godani}
N.~Godani and G.~C.~Samanta, "{Traversable Wormholes in f(R) with Constant and Variable Redshift Functions}," \href{https://doi.org/10.1016/j.newast.2020.101399} {{\em  New Astron.} {\bfseries 80} (2020) 101399}.
\bibitem{pavlovic}
P.~Pavlovic and M.~Sossich, "{Wormholes in viable f(R) modified theories of gravity and Weak Energy Condition}," \href{https://doi.org/10.1140/epjc/s10052-015-3331-y} {{\em Eur. Phys. J. C} {\bfseries 75} (2015) 117}.
\bibitem{zubair0}
M.~Zubair, R.~Saleem, Y.~Ahmad and G.~Abbas, "{Exact wormholes solutions without exotic matter in f(R,T) gravity}," \href{https://doi.org/10.1016/j.cjph.2019.11.018} {{\em Int.J.Geom.Meth.Mod.Phys.} {\bfseries 16} (2019) 1950046}.
\bibitem{zubair}
 M.~Zubair, S.~Waheed and Y.~Ahmad, "{Static Spherically Symmetric Wormholes in f(R,T) Gravity}," \href{https://doi.org/10.1140/epjc/s10052-016-4288-1} {{\em Eur. Phys. J. C.} {\bfseries 76} (2016) 444}.
\bibitem{garciaa} 
N.~M.~Garcia and F.~S.~N.~Lobo, "{Exact solutions of Brans-Dicke wormholes in the presence of matter}," \href{https://doi.org/10.1142/S021773231103739X} {{\em Modern Physics Letters A.} {\bfseries 26} (2011) 3067}.
\bibitem{papan} 
E.~Papantonopoulos, C.~Vlachos, "{Wormhole solutions in modified Brans-Dicke theory}," \href{https://doi.org/10.1103/PhysRevD.101.064025} {{\em Phys. Rev. D.} {\bfseries 10} (2020) 064025 }.
\bibitem{Chanda:2021dvc}
A.~Chanda, S.~Dey and  B.~C.~Paul, "{Study of Gravastars in Rastall Gravity}," \href{ https://doi.org/10.1088/1475-7516/2021/07/004} {{\em JCAP} {\bfseries 2021} (2021) 004}.
\bibitem{Harko:2011kv}
T.~Harko, F.~S.~N.~Lobo, S.~Nojiri and S.~D.~Odintsov, "{f(R,T) gravity}," \href{https://doi.org/10.1103/PhysRevD.84.024020} {{\em Phys. Rev. D.} {\bfseries 84} (2011) 024020}.
\bibitem{BarrientosO:2014mys}
O.~J.~Barrientos and  F.~R.~Guillermo, "{Comment on f(R,T)}," \href{https://doi.org/10.1103/PhysRevD.90.028501}  {{\em Phys. Rev. D.} {\bfseries 90} (2014) 028501}.
\bibitem{moraes1}
P.~.H~ R.~S.~Moraes and  P.~K.~Sahoo, "{Modeling wormholes in f(R,T) gravity}," \href{https://doi.org/10.1103/PhysRevD.96.044038} {{\em Phys. Rev. D.} {\bfseries 96} (2017) 044038}.
\bibitem{mishra1}
A.~K.~Mishra, U.~K.~Sharma, V.~C.~Dubey and A.~Pradhan, "{Traversable wormholes in f(R,T) gravity}," \href{https://doi.org/10.1007/s10509-020-3743-5} {{\em Astrophys. Space Sci.} {\bfseries 365} (2020) 34}.
\bibitem{sahoo1}
P.~Sahoo, P.~H.~R.~S.~Moraes, M.~M.~Lapola and P.~K.~Sahoo, "{Traversable wormholes in the traceless f(R,T) gravity}," \href{https://doi.org/10.1142/S0218271821501005} {{\em Int. Journ. Mod. Phys. D} {\bfseries 30} (2021) 2150100}.
\bibitem{SALEEM}
R.~Saleem and M.~I.~Aslam, "{Traversable wormholes in f(R,T) gravity with vanishing speed of sound}," \href{https://doi.org/10.1016/j.cjph.2023.06.004} {{\em Chinese Journal of Physics} {\bfseries 85} (2023) 741-751}.
\bibitem{CHANDRA2021}
A.~Chandra, S.~Dey and B.~C.~Paul, "{Anisotropic compact objects in f(T) gravity with Finch–Skea geometry}," \href{https://link.springer.com/article/10.1140/epjc/s10052-019-7020-0} {{\em Eur. Phys. Jour. C.} {\bfseries 53} (2021) 78}.
\bibitem{wu2018}
J.~Wu, G.~Li, T.~Harko and S.~D.~Liang, "{Palatini formulation of f(R, T) gravity theory, and its cosmological implications}," \href{https://link.springer.com/article/10.1140/epjc/s10052-018-5923-9} {{\em Eur. Phys. Jour. C.} {\bfseries 78} (2018) 430}.
\bibitem{harko-2014} 
T.~Harko, "{Thermodynamic interpretation of the generalized gravity models with geometry-matter coupling}," \href{https://doi.org/10.1103/PhysRevD.90.044067} {{\em Phys. Rev. D} {\bfseries 90} (2014)  044067}.
\bibitem{bertolami2009}
O.~ Bertolami and M.~C.~Sequeira, "{Energy conditions and stability in f(R) theories of gravity with nonminimal coupling to matter}," \href{https://doi.org/10.1103/PhysRevD.79.104010} {{\em Phys. Rev. D} {\bfseries 79} (2009)  104010}.
\bibitem{capozziello2018}
S.~Capozziello, S.~Nojiri and S.~D.~Odintsov, "{The role of energy conditions in f(R) cosmology}," \href{10.1016/j.physletb.2018.03.064} {{\em Physics Letters B} {\bfseries 781} (2018) 99-106}.
\bibitem{santos2017}
C.~S.~Santos, J.~Santos, S.~Capozziello and J. S. Alcaniz, "{Strong energy condition and the repulsive character of f(R) gravity}," \href{https://doi.org/10.1007/s10714-017-2212-0} {{\em Gen. Rel. Gravit} {\bfseries 49} (2017) 50}.
\bibitem{zubair2015}
M.~Zubai and S.~Waheed, "{Energy Conditions in f(T) Gravity with Non- Minimal Torsion-Matter Coupling}," \href{https://doi.org/10.1007/s10509-014-2181-7} {{\em  Eur. Phys. J. Plus} {\bfseries 137} (2022) 755}.
\bibitem{ganiyeva}
N.~Ganiyeva, J.~L.~ Rosa and F.~ S.~ N.~Lobo "{Wormhole geometries in $f\left(R,T^2\right)$ gravity satisfying the energy conditions}," \href{https://arxiv.org/abs/2502.19323} {{\em Contribution to the proceedings of the $17th$ Marcel Grossmann Meeting} (2025) }.
\bibitem{Royet}
D.~Roy, A.~Dutta, B.~ Ghosh and S.~Chakraborty, "{Investigating Evolving Wormholes in $f(R,T)$ Gravity}," \href{https://doi.org/10.48550/arXiv.2501.12237} {{\em accepted paper IJMPA} (2025)}.
\bibitem{yousaf}
M.~Yousaf, and H.~Asad, "{Impact of modified Chaplygin gas on electrically charged thin-shell wormhole models}, "\href{https://doi.org/10.1016/j.dark.2025.101841} {{\em  Physics of the Dark Universe} {\bfseries 48} (2025) 101841}.
\bibitem{Bhatti}
M.~Z.~Bhatti, M.~Yousaf, and Z.~Yousaf, "{Construction of thin-shell wormhole models in the geometric representation of f(R,T) gravity}, "\href{https://doi.org/10.1016/j.newast.2023.102132} {{\em  New Astronomy} {\bfseries 106} (2024) 102132}.
\bibitem{Rastpars}
S.~Rastgoo, and F.~Parsaei, "{Wormholes in f(R,T) gravity with variable equation of state}, "\href{https://doi.org/10.1016/j.nuclphysb.2025.116797} {{\em  Nuclear Physics B} {\bfseries 1011} (2025) 116797}.
\bibitem{chaudaryet}
S.~Chaudhary, S.~K.~Maurya, J.~ Kumar and S.~Kiroriwal, "{Physically viable travsersable wormhole solutions and energy conditions in F(R,T) gravity within $R^{2}$ formalism via specific form of shape functions}, "\href{https://doi.org/10.1016/j.dark.2024.101565} {{\em Physics of the Dark Universe} {\bfseries 46} (2024) 101565}.
\bibitem{Lu}
J.~Lu, M.~Xu, J.~Guo, and R.~Li, "{Investigating the physical properties of traversable wormholes in the modified f(R, T) gravity}, "\href{https://doi.org/10.1007/s10714-024-03223-x} {{\em Gen. Rel. Grav.} {\bfseries 56} (2024) 37}.
\bibitem{Tangphati}
T.~Tangphati, A.~Banerjee, and A.~Pradhan, "{Wormholes and energy conditions in f(R,T) gravity}, "\href{https://doi.org/10.1142/S0219887824501093} {{\em International Journal of Geometric Methods in Modern Physics} {\bfseries 21} (2024) 2450109}.
\bibitem{Yashwanth}
B.~R.~Yashwanth, S.~K.~ Narasimhamurthy, and Z.~Nekouee, "{Generalized Finslerian Wormhole Models in f(R,T) Gravity}, "\href{https://doi.org/10.3390/particles7030043} {{\em Particles} {\bfseries 7} (2024) 747-767}.
\bibitem{Mondal}
M.~Mondal, and F.~Rahaman, "{Possible existence of galactic wormholes in f(R,T) gravity}, "\href{https://doi.org/10.1140/epjp/s13360-023-04688-6} {{\em Eur. Phys. J. Plus } {\bfseries 139} (2024) 39}.
\bibitem{Azmat}
H.~Azmat, Q.~Muneer, M.~Zubair, E.~Gudekli, I.~Ahmad, and S.~Waheed, "{Class of charged traversable Casimir wormholes in f(R,T) gravity}, "\href{https://doi.org/10.1016/j.nuclphysb.2023.116396} {{\em Nuclear Physics B} {\bfseries 998} (2024) 116396}.
\bibitem{Chadtwo}
S.~Chaudhary, S.~K.~Maurya, J.~ Kumar, and S.~Kiroriwal,  "{Traversable wormhole solutions with phantom fluid in modified f(R, T) gravity}, "\href{https://doi.org/10.1007/s12043-024-02811-5} {{\em Pramana J. Phys.} {\bfseries 98} (2024) 139}.
\bibitem{zubair2022}
M.~Zubair, Q.~Muneer and S.~Waheed, "{Energy Constraints for Evolving Spherical and Hyperbolic Wormholes in f(R,T) Gravity}," \href{https://doi.org/10.1140/epjp/s13360-022-02946-7} {{\em Eur. Phys. J. Plus} {\bfseries 355} (2015) 361-369}.
\bibitem{Mandal2022}
S.~ Mandal, P.~K.~Sahoo and J.~R.~L.~Santos, "{Energy conditions in f(Q) gravity}," \href{https://doi.org/10.1103/PhysRevD.102.024057} {{\em Phys. Rev. D} {\bfseries 102} (2020)  024027 }.
\bibitem{Curiel}
E.~Curiel, "{A Primer on Energy Conditions.}," \href{https://doi.org/10.1007/978-1-4939-3210-8_3} {{\em Einstein Studies} {\bfseries 13} (2017)}.
\bibitem{Cattoen/2005} 
C.~Cattoen, F.~Tristan and  M.~Visser, "{Gravastars must have anisotropic pressures}," \href{https://doi.org/10.1088/0264-9381/22/20/002} {{\em  Classical Quantum Gravity} {\bfseries 22} (2005) 22}.
\bibitem{Lobo/2013}  
F.~S.~N.~Lobo, F.~Parsaei and F.~N.~Riazi, "{New asymptotically flat phantom wormhole solutions}," \href{https://doi.org/10.1103/PhysRevD.87.084030} {{\em Phy. Rev. D} {\bfseries 87} (2013) 084030}.
\bibitem{karm} 
K.~R.~Karmarkar, {{\em Proc. Indian Acad. Sci. A} {\bfseries 27} (1948) 56}.
\bibitem{garat} 
G.~ Remo, "{Casimir wormholes}," \href{https://doi.org/10.1140/epjc/s10052-019-7468-y} {{\em The European Physical Journal C} {\bfseries 79} (2019) 11}.
\bibitem{Kavya} 
N.~S.~Kavya, C.~S.~Varsha, L.~Sudharani, and V.~Venkatesha, "{Unifying non-commutative geometry with Casimir energy: A novel f(R) wormhole solution}," \href{https://www.sciencedirect.com/science/article/pii/S0550321325000045} {{\em Nuclear Physics B} {\bfseries 1011} (2025) 116794}.
\bibitem{Sahootwo} 
A.~Sahoo, S.~K.~Tripathy, B.~Mishra, S.~Ray "{Casimir wormhole with GUP correction in extended symmetric teleparallel gravity}," \href{https://doi.org/10.1140/epjc/s10052-024-12638-4} {{\em The European Physical Journal C} {\bfseries 84} (2024) 325}.
\bibitem{banerjeet}
A.~Banerjee, S.~ Hansraj, and A.~Pradhan,  "{ Wormholes In f(R,T) Gravity with Casimir Stress Energy}," \href{https://ssrn.com/abstract=4604823} {{SSRN, ISSN: 1556-5068}  (2023)}.
\bibitem{zubairm}
M.~Zubair, S.~Waheed, M.~ Farooq, A.~H.~Alkhakdi, and A.~Ali, "{ New Casimir wormholes in f(R, T) gravity admitting conformal killing vectors}," \href{https://doi.org/10.1140/epjp/s13360-023-04539-4} {{\em The European Physical Journal C} {\bfseries 138} (2023) 902}.
\bibitem{santosetal}
A.~C.~L.~Santos, R.~V.~Maluf, and C.~R.~Muniz, "{Generating 4-dimensional wormholes with Yang–Mills Casimir sources}," \href{https://doi.org/10.1016/j.aop.2024.169775.} {{\em Annals of Physics} {\bfseries 469} (2024) 169775}
\end{thebibliography}
